\documentclass[journal]{IEEEtran}
\usepackage{amsmath,amssymb,amsfonts,mathrsfs}
\usepackage{epsfig,epstopdf}
\usepackage{graphicx,graphics}
\usepackage{color}
\usepackage{float}
\usepackage{ifpdf}

\newtheorem{theorem}{Theorem}[section]
\newtheorem{corollary}{Corollary}[section]
\newtheorem{definition}{Definition}[section]
\newtheorem{lemma}{Lemma}[section]
\newtheorem{proposition}{Proposition}[section]
\newtheorem{remark}{Remark}[section]
\newtheorem{example}{Example}[section]

\begin{document}

\title{The Kalman Decomposition for Linear Quantum Systems \thanks{%
This research is supported in part by a National Natural Science Foundation
of China grant (No. 61374057), Hong Kong RGC grant (No. 531213, 15206915),
the Australian Research Council under grant FL110100020 and the Air Force Office of Scientific
Research (AFOSR) under agreement number FA2386-16-1-4065.}}
\date{\today }
\author{G. Zhang,\thanks{
G. Zhang is with the Department of Applied Mathematics, The Hong Kong
Polytechnic University, Hong Kong. (e-mail: Guofeng.Zhang@polyu.edu.hk).} S.
Grivopoulos, \thanks{
S.~Grivopoulos was with the School of Engineering and
Information Technology, University of New South Wales, Canberra, ACT, 2600,
Australia. (e-mail: symeon.grivopoulos@gmail.com).}  
 I. R. Petersen, \textit{Fellow, IEEE,} \thanks{
I.~R. Petersen is with the Research School of Engineering, Australian National University, Canberra ACT 2601, Australia, (e-mail: i.r.petersen@gmail.com).}
J. E. Gough\thanks{
J. E. Gough is with Department of Physics, Aberystwyth University, Wales,
SY23 2BZ, Aberystwyth, UK. (e-mail: jug@aber.ac.uk).} }
\maketitle

\begin{abstract}
This paper studies the Kalman decomposition for linear quantum systems.
Contrary to the classical case, the coordinate transformation used for the
decomposition must belong to a specific class of transformations as a
consequence of the laws of quantum mechanics. We propose a construction
method for such transformations that put the system in a Kalman canonical
form. Furthermore, we uncover an interesting structure for the obtained
decomposition. In the case of passive systems, it is shown that there exist
only controllable/observable and uncontrollable/unobservable subsystems. In
the general case, controllable/unobservable and uncontrollable/observable
subsystems may also be present, but their respective system variables must
be conjugate variables of each other. This decomposition naturally exposes
decoherence-free modes, quantum-nondemolition modes, quantum-mechanics-free
subsystems, and back-action evasion measurements in the quantum system, which
are useful resources for quantum information processing, and quantum
measurements. The theory developed is applied to physical examples.

\end{abstract}

%\tableofcontents

%%%%%%%%%%%%%%%%%%%%%%%%%%%
%%%%%%%%%%%%%%%%%%%%%%%%%%%
%%%%%%%%%%%%%%%%%%%%%%%%%%%

\section{Introduction}

\label{sec:intro}

Over the past few decades, great progress has been made in the theoretical
investigation and experimental realization of controlled quantum systems. In
particular, a multitude of control methods have been proposed and tested;
see e.g. \cite{WM08,WM10,DP10, SW10, AT12, DFK+12, ZJ12, KJ14, PR14, ZLW+14,ZZZ+15}%
. Linear quantum systems play a prominent role in these developments. In
quantum optics, linear models are commonly used because they are often
adequate approximations for more general dynamics. Furthermore, control
problems for linear systems often enjoy analytical or computationally
tractable solutions. In addition to their wide applications in quantum
optics \cite{GZ00, WM08, M08, WM10}, linear quantum systems have also found
useful applications in many other quantum-mechanical systems, including
circuit quantum electro-dynamical (circuit QED) systems \cite{MJP+11}, \cite%
{KAK13}, cavity QED systems \cite{DJ99}, quantum
opto-mechanical systems \cite{TC10}, \cite{MHP+11}, \cite{HM12}, \cite%
{DFK+12}, \cite{MCP+12}, \cite{NY13}, \cite{NY14}, \cite{ODP+16}, atomic
ensembles \cite{SvHM04}, \cite{NY13}, and quantum memories \cite{HCH+13}, 
\cite{YJ14}.

Controllability and observability are two fundamental notions in modern
control theory \cite{ZDG96}, \cite{HK97}, \cite{CF}. Roughly speaking,
controllability describes the external input's ability to steer internal
system states, while observability refers to the capability of
reconstructing the state-space trajectories of a dynamical system based on
its input--output data. Recently, these two fundamental notions have been
investigated for linear quantum systems. In the study of optimal
measurement-based linear quadratic Gaussian (LQG) control, Wiseman and
Doherty showed the equivalence between detectability and stabilizability 
\cite{WD05}. Yamamoto and Guta proved that controllability and observability
are equivalent for passive linear quantum systems \cite[ Lemma 3.1]{GY13}
and they imply Hurwitz stability \cite[Lemma 3.2]{GY13}. Gough and Zhang
showed that the equivalence between controllability and observability holds
for general (namely, not necessarily passive) linear quantum systems \cite[%
Proposition 1]{GZ15}. Moreover, in the passive case, it is proved that
Hurwitz stability, controllability and observability are all equivalent \cite%
[Lemma 2]{GZ15}. The controllability and observability of passive linear
quantum systems have been studied by Maalouf and Petersen \cite{MP11a};
using these notions the authors established a complex-domain bounded real
lemma for passive linear quantum systems \cite[Theorem 6.5]{MP11a}; see also 
\cite{JNP08}, \cite{JG10}, \cite{GZ15}. Nurdin \cite{HIN13} studied model
reduction for linear quantum systems based on controllability and
observability decompositions; see also \cite{IRP11}. Interestingly,
controllability and observability are closely related to the so-called
decoherence-free subsystems (DFSs),  \cite{WC12a}, \cite{WC12b}, \cite{DFK+12},  \cite{NY13}, \cite{NY14}, \cite{GZ15}, and references therein, quantum-nondemolition (QND) variables 
\cite{HMW95}, \cite{TC12},  \cite%
{NY13}, \cite{NY14}, and back-action evasion (BAE) measurements 
\cite{TC10}, \cite{WC13}, \cite{ODP+16}, \cite{NY14}, which are useful for
quantum information processing  \cite{TC10}, \cite%
{DFK+12}, \cite{NY14}, \cite{ZZZ+15}. 

 Of course, realistic quantum information processing applications such as quantum computers will require going beyond linear quantum systems. Nevertheless, having the theoretical tools to identify all of these useful resources in linear quantum systems is a necessary step in this direction. Moreover, an improved understanding of quantum linear systems may aid in the construction of a quantum computer such as for example in proposed approaches to quantum computing involving cluster states and quantum measurements \cite{MvLGWRN06}. Also, the theory of quantum linear systems has many other potential applications in quantum technology including quantum measurements \cite{KJ14} and quantum communications \cite{DLCZ01}.

Notwithstanding the above advances, a result corresponding to the classical
Kalman decomposition (e.g., see \cite[Chapter 2]{HK97}, \cite[Chapter 3]%
{ZDG96}) is still lacking for linear quantum systems. The critical issue is
that, quantum-mechanical laws allow only specific types of coordinate
transformations for linear quantum systems. More specifically, in the real quadrature operator representation where the two quadrature operators can be position and momentum operators respectively, the allowed transformations on quantum linear systems are orthogonal symplectic transformations for passive systems and  symplectic transformations for general (non-passive) systems. In the annihilation-creation operator
representation, which is unitarily equivalent to the real quadrature operator representation, the allowed transformations are unitary transformations for passive systems and Bogoliubov transformations for
general (non-passive) systems. It is not a priori obvious that
transformations to a Kalman canonical form obtained by the standard methods
of linear systems theory will satisfy these requirements for linear \textit{%
quantum} systems. The main purpose of this work is to show that there do
exist unitary, Bogoliubov and symplectic transformations, for the
corresponding cases, that decompose linear quantum systems into
controllable/observable ($co$), controllable/unobservable ($c\bar{o}$),
uncontrollable/observable ($\bar{c}o$), and uncontrollable/unobservable ($%
\bar{c}\bar{o}$) subsystems. More specifically, in Section \ref{sec:passive}%
, we study the Kalman decomposition for passive linear quantum systems. In
particular, we show that in this case, the uncontrollable subspace is
identical to the unobservable subspace, Theorem \ref{thm:passive_Kalman}; we
also give a characterization of these subspaces, Theorem \ref%
{thm:passive_DFS_characterization}. The general non-passive case is studied
in Section \ref{sec:general}. First, we construct the Kalman decomposition
for general linear quantum systems in the annihilation-creation operator
representation, Theorems \ref{thm:general_Kalman_1} and \ref%
{thm:general_Kalman_2}. Then, we translate these theorems into the real
quadrature operator representation for linear quantum systems, Theorems \ref%
{thm:general_Kalman_3} and \ref{thm:general_Kalman_4}. As a by-product, the
real quadrature operator representation of the Kalman canonical form of
passive linear systems is given in Corollary \ref{cor:passive_Kalman}. It is
worth noting that the Kalman decomposition is achieved in a constructive
way, as in the classical case. Moreover, all the transformations involved
are unitary and thus the decomposition can be performed in a numerically
stable way.

The Kalman decomposition of a linear quantum system proposed in this paper
exhibits the following features: 1) The $co$ and $\bar{c}\bar{o}$ subsystems
are linear quantum systems in their own right, as is to be expected from a
physics perspective; see Remark \ref{rem:sept4} for details. 2) The system
variables of the $c\bar{o}$ subsystem are conjugate to those of the $\bar{c}%
o $ subsystem. This fact has already been noticed in \cite{NY13}. An
immediate consequence of this is that, a $c\bar{o}$ subsystem exists if and
only if a $\bar{c}o$ subsystem does, and they always have the same
dimension. Indeed, the question of how to handle the $c\bar{o}$ and $\bar{c}%
o $ subsystems properly is the major technicality involved in the quantum
Kalman decomposition theory proposed in this work, see Lemmas \ref{lem:Delta}%
-\ref{lem:basis_Rh}. 3) The quantum-mechanical notions of Decoherence-Free
subsystems (DFSs), Quantum Non-Demolition (QND) variables, Quantum
Mechanics-Free subsystems (QMFS) and Back-Action Evasion (BAE) measurements,
which are important in quantum information science and measurement theory,
have natural connections with the subspace decomposition. In particular, the 
$\bar{c}\bar{o}$ subsystem of a linear quantum system (if it exists) is a
DFS, and the $\bar{c}o$ subsystem (if it exists) is a QMFS, whose variables
are QND variables; see Theorem \ref{thm:general_Kalman_4}, and Remarks \ref%
{rem:sept22_1} and \ref{rem:BAE}.

The main result of this paper thus shows how methods of classical linear systems theory can be applied to gain a new understanding of the structure of quantum linear systems. In particular, the results which are presented can be applied in analyzing the structure of a given quantum linear system model. These results will also pave the way for future research involving the synthesis of quantum feedback control systems to achieve a desired closed loop structure such as the existence of a DFS or QMFS.

The rest of the paper is organized as follows: In Section \ref{sec:lqs}, we
briefly review linear quantum systems and several physical concepts. In
Section \ref{sec:passive}, we study the Kalman decomposition for passive
linear quantum systems. The general case is studied in Section \ref%
{sec:general}. The proposed methodology is applied to two physical systems
in Section \ref{sec:apps}. Section \ref{sec:con} concludes the paper.

\subsection*{Notation}

\label{subsec:notation}

\begin{enumerate}
\item $x^{\ast}$ denotes the complex conjugate of a complex number $x$ or
the adjoint of an operator $x$. The commutator of two operators $X$ and $Y$
is defined as $[X,Y] \triangleq XY-YX$.

\item For a matrix $X=[x_{ij}]$ with number or operator entries, $%
X^{\#}=[x_{ij}^{\ast}]$, $X^{\top}=[x_{ji}]$ is the usual transpose, and $%
X^{\dag}=(X^{\#})^{\top}$. For a vector $x$, we define $\breve{x}=\bigl[%
\begin{smallmatrix}
x \\ 
x^{\#}%
\end{smallmatrix}%
\bigr]$.

\item $I_{k}$ is the identity matrix, and $0_{k}$ the zero matrix in $%
\mathbb{C}^{k \times k}$. $\delta_{ij}$ denotes the Kronecker delta symbol;
i.e., $I_k=[\delta_{ij}]$. $\mathrm{Ker}\left( X\right)$, $\mathrm{Im}\left(
X\right)$, and $\sigma\left( X\right)$ denote the null space, the range
space, and the spectrum of a matrix $X$, respectively.

\item Let $J_{k} \triangleq \mathrm{diag}(I_k,-I_k)$. For a matrix $X\in 
\mathbb{C}^{2k\times 2r}$, define its $\flat$-adjoint by $X^{\flat }
\triangleq J_{r}X^{\dag }J_{k}$. The $\flat$-adjoint satisfies properties
similar to the usual adjoint, namely $(x_1 A + x_2 B)^{\flat}=x_1^*
A^{\flat} + x_2^* B^{\flat}$, $(AB)^{\flat}=B^{\flat} A^{\flat}$, and $%
(A^{\flat})^{\flat}=A$.

\item Given two matrices $U$, $V\in \mathbb{C}^{k\times r}$, define $\Delta
(U,V) \triangleq [U~V;V^{\#}~U^{\#}]$. A matrix with this structure will be
called \emph{doubled-up} \cite{GJN10}. It is immediate to see that the set
of doubled-up matrices is closed under addition, multiplication and taking ($%
\flat$-) adjoints.

\item A matrix $T \in \mathbb{C}^{2k\times 2k}$ is called \emph{Bogoliubov}
if it is doubled-up and satisfies $TJ_{k}T^{\dag}=T^{\dag}J_{k}T=J_{k}
\Leftrightarrow TT^{\flat}=T^{\flat}T=I_{2k}$. The set of these matrices
forms a complex non-compact Lie group known as the Bogoliubov group.

\item Let $\mathbb{J}_{k} \triangleq \bigl[%
\begin{smallmatrix}
0_{k} & I_k \\ 
-I_k & 0_{k}%
\end{smallmatrix}%
\bigr]$. For a matrix $X\in \mathbb{C}^{2k\times 2r}$, define its $\sharp$-%
\emph{adjoint} $X^{\sharp}$ by $X^{\sharp} \triangleq -\mathbb{J}_{r}X^{\dag}%
\mathbb{J}_{k}$. The $\sharp$-\emph{adjoint} satisfies properties similar to
the usual adjoint, namely $(x_1 A + x_2 B)^{\sharp}=x_1^* A^{\sharp} + x_2^*
B^{\sharp}$, $(AB)^{\sharp}=B^{\sharp} A^{\sharp}$, and $(A^{\sharp})^{%
\sharp}=A$.

\item A matrix $S \in \mathbb{C}^{2k\times 2k}$ is called \emph{symplectic},
if it satisfies $S\mathbb{J}_{k}S^{\dag}=S^{\dag}\mathbb{J}_{k}S=\mathbb{J}%
_{k} \Leftrightarrow SS^{\sharp}=S^{\sharp}S=I_{2k}$. The set of these
matrices forms a complex non-compact group known as the symplectic group.
The subgroup of real symplectic matrices is one-to-one homomorphic to the
Bogoliubov group.
\end{enumerate}

%%%%%%%%%%%%%%%%%%%%%%%%%%%
%%%%%%%%%%%%%%%%%%%%%%%%%%%
%%%%%%%%%%%%%%%%%%%%%%%%%%%

\section{Linear quantum systems}

\label{sec:lqs}

In this section, we briefly introduce linear quantum systems; more details
can be found in, e.g., \cite{KRP92}, \cite{GZ00}, \cite{YK03a}, \cite{WM08}, 
\cite{WM10}, \cite{JNP08}, \cite{GJ09}, \cite{TNP+11}, \cite{ZJ12}, \cite{Z14}. 
\begin{figure}[tbph]
\centering
\includegraphics[width=0.30\textwidth]{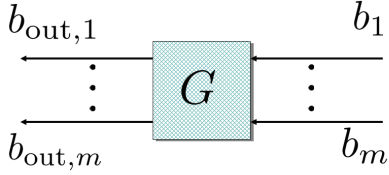}
\caption{A linear quantum system.}
\label{fig:sys}
\end{figure}
The linear quantum system, as shown in Fig. \ref{fig:sys}, is a collection
of $n$ quantum harmonic oscillators driven by $m$ input boson fields. The
mode of oscillator $j$, $j=1,\ldots, n$, is described in terms of its
annihilation operator $\boldsymbol{a} _{j}$, and its creation operator $%
\boldsymbol{a}_{j}^{\ast }$, the adjoint operator of $\boldsymbol{a}_{j}$.
These are operators in an infinite-dimensional Hilbert space. The operators $%
\boldsymbol{a}_{j},\boldsymbol{a}_{k}^{\ast }$ satisfy the \emph{canonical
commutation relations} $[\boldsymbol{a}_{j}(t),~\boldsymbol{a}_{k}(t)]=0$, $[%
\boldsymbol{a}_{j}^{\ast}(t),~\boldsymbol{a}_{k}^{\ast}(t)]=0$, and $[%
\boldsymbol{a}_{j}(t),~\boldsymbol{a}_{k}^{\ast}(t)]=\delta _{jk}$, $\forall
j,k=1,\ldots n, \forall t \in \mathbb{R}{\color{red}^+}$. Let $\boldsymbol{a}=[ \boldsymbol{%
a}_{1}~\cdots ~\boldsymbol{a}_{n}]^{\top}$. The system Hamiltonian $%
\boldsymbol{H}$ is given by $\boldsymbol{H}=(1/2)\boldsymbol{\breve{a}}
^{\dag }\Omega \boldsymbol{\breve{a}}$, where $\boldsymbol{\breve{a}}=[ 
\boldsymbol{a}^{\top}~(\boldsymbol{a}^{\#})^{\top}]^{\top}$, and $\Omega
=\Delta (\Omega _{-},\Omega _{+})\in \mathbb{C}^{2n\times 2n}$ is a
Hermitian matrix with $\Omega _{-},\Omega _{+}\in \mathbb{C}^{n\times n}$.
The coupling of the system to the input fields is described by the operator $%
\boldsymbol{L}=[C_{-}\ C_{+}] \boldsymbol{\breve{a}}$, with $C_{-},C_{+}\in 
\mathbb{C}^{m\times n}$. The input boson field $k$, $k=1,\ldots, m$, is
described in terms of an annihilation operator $\boldsymbol{b}_{k}(t)$ and a
creation operator $\boldsymbol{b}_{k}^{\ast }(t)$, the adjoint operator of $%
\boldsymbol{b}_{k}(t)$. These are operators on a symmetric Fock space (a
special kind of infinite-dimensional Hilbert space). The operators $%
\boldsymbol{b}_{k}(t)$ and $\boldsymbol{b} _{k}^{\ast }(t)$ satisfy the
singular commutation relations $[\boldsymbol{b} _{j}(t),~\boldsymbol{b}%
_{k}(r)]=0$, $[\boldsymbol{b} _{j}^\ast(t),~ \boldsymbol{b}_{k}^\ast(r)]=0$,
and $[\boldsymbol{b} _{j}(t),~\boldsymbol{b}_{k}^{\ast }(r)]=\delta
_{jk}\delta (t-r),~~\forall j,k=1,\ldots ,m,~\forall t,r\in \mathbb{R}$. Let 
$\boldsymbol{b}(t)=[ \boldsymbol{b}_{1}(t)~\cdots ~\boldsymbol{b}%
_{m}(t)]^{\top}$ and $\boldsymbol{\ \breve{b}}(t)=[\boldsymbol{b(t)}^{\top}~(%
\boldsymbol{b}(t)^{\#})^{\top}]^{\top}$.

The dynamics of the open linear quantum system in Fig. \ref{fig:sys} is
described by the following quantum stochastic differential equations (QSDEs) 
\begin{eqnarray}
\boldsymbol{\dot{\breve{a}}}(t) &=&\mathcal{A}\boldsymbol{\breve{a}}(t)+ 
\mathcal{B}\boldsymbol{\breve{b}}(t),  \label{eq:sys_a} \\
\boldsymbol{\breve{b}}_{\mathrm{out}}(t) &=&\mathcal{C}\boldsymbol{\breve{a}}%
(t)+ \mathcal{D}\boldsymbol{\breve{b}}(t),  \ \ t\geq 0,
 \label{eq:sys_b}
\end{eqnarray}
where the system matrices are given by 
\begin{eqnarray}
\mathcal{D}&=&I_{2m}, ~\mathcal{C}=\Delta(C_{-},\ C_{+}), ~\mathcal{B}=-%
\mathcal{C}^{\flat },\ \mathrm{and}  \notag \\
\mathcal{A}&=&-\imath J_{n}\Omega-\frac{1}{2}\mathcal{C}^{\flat }\mathcal{C}.
\label{eq:sys_ABCD}
\end{eqnarray}
An equivalent way to characterize the structure of (\ref{eq:sys_ABCD})
(given that all matrices are doubled-up) is by the following \textit{%
physical realizability} conditions \cite{JNP08}, \cite{NJP09}, \cite{SP12}, 
\cite{ZJ12}: 
\begin{equation}  \label{eq:PR}
\mathcal{A}+\mathcal{A}^{\flat }+\mathcal{B}\mathcal{B}^{\flat }=0,~
\mathcal{B}=-\mathcal{C}^{\flat }.
\end{equation}
It can be shown that \cite{DEMPUJ16}, the above forms of system matrices are
the only ones with the property that the temporal evolution of (\ref%
{eq:sys_a})-(\ref{eq:sys_b}) preserves the fundamental commutation relations 
\begin{eqnarray*}
\lbrack\boldsymbol{\breve{a}}(t),~\boldsymbol{\breve{a}}^{\dag }(t)] &=&[ 
\boldsymbol{\breve{a}}(0),~\boldsymbol{\breve{a}}^{\dag }(0)],~~~t\geq 0, \\
\lbrack \boldsymbol{\breve{a}}(t),\boldsymbol{\breve{b}}_{\mathrm{out}
}^{\dag }(r)] &=&0,~0\leq r<t.
\end{eqnarray*}
Only under the condition that the above physical realizability conditions
are satisfied, do the QSDEs (\ref{eq:sys_a})-(\ref{eq:sys_b}) represent the
dynamics of a linear quantum system that can be practically implemented,
say, with optical devices, \cite{Leo03,NJD09,ZLHZ12}.

A very important issue for the purpose of this work is the kind of
coordinate transformations $\boldsymbol{\breve{a}}_{\mathrm{new}}=T%
\boldsymbol{\breve{a}}$ allowed in the QSDEs (\ref{eq:sys_a})-(\ref{eq:sys_b}%
). It is straightforward to show that the form of (\ref{eq:sys_ABCD}) is
preserved (with $\mathcal{C}_{\mathrm{new}} = \mathcal{C}T^{-1}$ and $%
\Omega_{\mathrm{new}} = (T^{-1})^{\dag} \Omega T^{-1})$) only if $T$ is
Bogoliubov. This is a system-theoretic re-statement of the quantum
mechanical requirement that $T$ must be Bogoliubov so that the new
annihilation and creation operators also satisfy the canonical commutation
relations. It is this additional constraint on the allowed coordinate
transformations of linear quantum systems that forces us to re-examine the
classical method for constructing the Kalman decomposition for such systems.

Linear quantum systems that do not require an external source of energy for
their operation are called \emph{passive}. For this important class of
systems, $C_+=0$ and $\Omega_+=0$. This results in the QSDEs for system and
field annihilation operators to decouple from those for the creation
operators of either type. Then, a description of the system in terms of
annihilation operators only is possible. The QSDEs for a passive linear
quantum system are (e.g., see \cite[Sec. 3.1]{GZ15}), 
\begin{eqnarray}
\boldsymbol{\dot{a}}(t) &=& \mathcal{A}\boldsymbol{a}(t)+\mathcal{B}%
\boldsymbol{b}(t),  \label{eq:passive_sys_a} \\
\boldsymbol{b }_{\mathrm{out}}(t) &=& \mathcal{C}\boldsymbol{a}(t)+\mathcal{D%
}\boldsymbol{b}(t),  \label{eq:passive_sys_b}
\end{eqnarray}
where 
\begin{equation}  \label{eq:passive_sys_ABCD}
\mathcal{A}=-\imath\Omega _{-}-\frac{1}{2}C_{-}^{\dagger }C_{-},\ \mathcal{B}%
=-C_{-}^{\dagger },\ \mathcal{C}=C_{-},\ \mathcal{D}=I_{m}
\end{equation}
(although we use the same symbols for the system matrices in the passive and
the general cases, it should be clear from the context which case we are
referring to). An equivalent way to characterize the structure of (\ref%
{eq:passive_sys_ABCD}), is by the physical realizability conditions 
\begin{equation}  \label{eq:passive_PR}
\mathcal{A}+\mathcal{A}^{\dagger }+\mathcal{B}\mathcal{B}^{\dagger }=0, ~%
\mathcal{B}=-\mathcal{C}^{\dagger }.
\end{equation}
The restriction that the allowed coordinate transformations of a general
linear quantum system must be Bogoliubov reduces in the passive case to the
requirement that the allowed coordinate transformations of a passive linear
quantum system must be unitary. This can be deduced from the result for the
general case, or directly from (\ref{eq:passive_sys_ABCD}).

So far, we have used the so-called complex annihilation-creation operator
representation to describe the linear quantum system (\ref{eq:sys_a})-(\ref%
{eq:sys_b}). There is another useful representation of this system, the
so-called real quadrature operator representation \cite[Sec. II.E]{ZJ11}. It
can be obtained from the annihilation-creation operator representation
through the following transformations: 
\begin{eqnarray}
\left[ 
\begin{array}{c}
\boldsymbol{q} \\ 
\boldsymbol{p}%
\end{array}%
\right] &\equiv& \boldsymbol{x} \triangleq V_{n}\boldsymbol{\breve{a}}, 
\notag \\
\left[ 
\begin{array}{c}
\boldsymbol{q}_{\mathrm{in}} \\ 
\boldsymbol{p}_{\mathrm{in}}%
\end{array}%
\right] &\equiv& \boldsymbol{u} \triangleq V_{m}\boldsymbol{\breve{b}},\ %
\left[ 
\begin{array}{c}
\boldsymbol{q}_{\mathrm{out}} \\ 
\boldsymbol{p}_{\mathrm{out}}%
\end{array}%
\right] \equiv \boldsymbol{y} \triangleq V_{m}\boldsymbol{\breve{b}}_{%
\mathrm{out}},  \label{complex_to_real_trans}
\end{eqnarray}
where the unitary matrices $V$ are defined by 
\begin{equation*}
V_{k}\triangleq \frac{1}{\sqrt{2}}\left[ 
\begin{array}{cc}
I_{k} & I_{k} \\ 
-\imath I_{k} & \imath I_{k}%
\end{array}%
\right].
\end{equation*}%
The operators $\boldsymbol{q}_{i}$ and $\boldsymbol{p}_{i}$, $i=1,\ldots,n$,
of the real quadrature operator representation are called \emph{conjugate}
variables, and they are self-adjoint operators, that is, observables.
Moreover, they satisfy the canonical commutation relations $[\boldsymbol{q}%
_{j}(t), \boldsymbol{q}_{k}(t)]=0$, $[\boldsymbol{p}_{j}(t), \boldsymbol{p}%
_{k}(t)]=0$, and $[\boldsymbol{q}_{j}(t), \boldsymbol{p}_{k}(t)]=\imath
\delta_{jk} $, $\forall j,k=1,\ldots,n$, $\forall t \in \mathbb{R}$. The
QSDEs that describe the dynamics of the linear quantum system in Fig. \ref%
{fig:sys} in the real quadrature operator representation are the following: 
\begin{eqnarray}
\boldsymbol{\dot{x}} &=& A \boldsymbol{x} + B \boldsymbol{u},
\label{eq:real_sys_a} \\
\boldsymbol{y} &=& C \boldsymbol{x} + D \boldsymbol{u},
\label{eq:real_sys_b}
\end{eqnarray}
where 
\begin{eqnarray}
D &=& V_{m} \mathcal{D} V_{m}^{\dag} = I_{2m}, ~C = V_{m} \mathcal{C}
V_{n}^{\dag},  \notag \\
B &=& V_{n} \mathcal{B} V_{m}^{\dag} = -C^{\sharp},  \notag \\
A &=& V_{n} \mathcal{A} V_{n}^{\dag} = \mathbb{J} H -\frac{1}{2}C^{\sharp}C.
\label{eq:real_sys_ABCD}
\end{eqnarray}
The matrix $H$ in Eq. (\ref{eq:real_sys_ABCD}) is defined by $H \triangleq
V_{n} \Omega V_{n}^{\dag}$ (hence, $\boldsymbol{H}=(1/2)\boldsymbol{x}%
^{\top} H \boldsymbol{x})$, and is real symmetric. In the above, the useful
identities 
\begin{eqnarray*}
V_{k} J_{k} V_{k}^{\dag} = \imath \mathbb{J}_{k}, \ \ V_{k} X^{\flat}
V_{j}^{\dag} = (V_{j} X V_{k}^{\dag})^{\sharp},
\end{eqnarray*}
for $X \in \mathbb{C}^{j \times k}$, were used. The matrices $A$, $B$, $C$, $%
D$, and $H$ are all real due to the fact that $V_{k} X V_{j}^{\dag}$ is real
if and only if $X \in \mathbb{C}^{2k \times 2j}$ is doubled-up.

In the real quadrature operator representation, the physical realizability
conditions (\ref{eq:PR}) take the form 
\begin{equation*}
A + A^{\sharp} + BB^{\sharp}=0,~B=-C^{\sharp}.
\end{equation*}
Finally, the only coordinate transformations that preserve the structure of (%
\ref{eq:real_sys_ABCD}) are real symplectic. This can be deduced from the
fact that only Bogoliubov transformations preserve the structure of (\ref%
{eq:sys_ABCD}), and that $S=V_{n}TV_{n}^{\dag}$ is real symplectic if and
only if $T$ is Bogoliubov. Finally, since $S$ is symplectic, it preserves
the commutation relations.

We end this section by introducing some important notions from quantum
information science and quantum measurement theory. We will show later that
these notions are naturally exposed by the Kalman decomposition of linear
quantum systems. We begin with two well-known notions in linear systems
theory.

The controllability and observability matrices for the linear quantum system
(\ref{eq:sys_a})-(\ref{eq:sys_b}) are defined respectively by (e.g., see 
\cite[Sec. III-B]{NY13} and \cite[Proposition 2]{GZ15}) 
\begin{eqnarray*}
C_{G}&\triangleq& \left[%
\begin{array}{cccc}
\mathcal{B} & \mathcal{A}\mathcal{B} & \cdots & \mathcal{A}^{2n-1}\mathcal{B}%
\end{array}
\right],  \notag \\
O_{G}&\triangleq& \left[%
\begin{array}{c}
\mathcal{C} \\ 
\mathcal{C}\mathcal{A} \\ 
\vdots \\ 
\mathcal{C}\mathcal{A}^{2n-1}%
\end{array}
\right] .
\end{eqnarray*}

%Therefore, the controllable subspace is $\mathrm{Im}(C _{G})$,
%the uncontrollable subspace is $\mathrm{Ker}(C_{G}^{\dagger })$, the observable
%subspace is $\mathrm{Im}(O_{G}^{\dagger })$, and the unobservable subspace is  $\mathrm{Ker}\left( O_{G}\right)$.

$\mathrm{Im}(C _{G})$ and $\mathrm{Ker}\left(O_{G}\right)$ are the
controllable and unobservable subspaces of the space of system variables $%
\mathbb{C}^{2n}$. We define the uncontrollable and observable subspaces to
be their \emph{orthogonal complements} in $\mathbb{C}^{2n}$, that is $%
\mathrm{Ker}(C_{G}^{\dagger })$, and $\mathrm{Im}(O_{G}^{\dagger })$,
respectively.

\begin{definition}
\label{def:DFS} The linear span of the system variables related to the uncontrollable/unobservable subspace of a linear quantum
system is called its \textit{decoherence-free subsystem (DFS)}.
\end{definition}

Decoherence-free subsystems for linear quantum systems have recently been
studied in e.g., \cite{WC12a}, \cite{WC12b},   \cite{DFK+12},  \cite{NY13}, \cite{NY14}, \cite{GZ15}, and references therein.
\begin{definition}
\label{def:QND} An observable $F$ is called a continuous-time \emph{%
quantum-nondemolition (QND) variable} if 
\begin{equation}  \label{eq:QND}
[F(t_1),F(t_2)]=0
\end{equation}
for all time instants $t_1,t_2 \in \mathbb{R}^+$.
\end{definition}

The physical meaning of Eq. (\ref{eq:QND}) is that $F$ may be measured an
arbitrary number of times (in fact, continuously) during the evolution of
the quantum system, with no quantum limit on the predictability of these
measurements \cite{HMW95}, \cite{TC12}, \cite{NY14}.

A natural extension of the notion of a QND variable is the following concept 
\cite{TC12}.

\begin{definition}
\label{QMFS} The span of a set of observables $F_{i}$, $i=1,\ldots,r$, is
called a \emph{quantum mechanics-free subsystem} (QMFS) if 
\begin{equation}  \label{eq:QMFS}
[F_{i}(t_1),F_{j}(t_2)]=0
\end{equation}
for all time instants $t_1,t_2 \in \mathbb{R}^+$, and $i,j=1,\ldots,r$.
\end{definition}

The transfer function of the linear system (\ref{eq:real_sys_a})-(\ref%
{eq:real_sys_b}) is 
\begin{equation*}  \label{tf_uy}
\Xi_{\boldsymbol{u}\to\mathbf{\boldsymbol{y}}} (s) \triangleq D-C(sI-A)^{-1}B.
\end{equation*}
This transfer function relates the overall input $\boldsymbol{u}$ to the
overall output $\boldsymbol{y}$. However, in many applications, we are
interested in a particular subvector $\boldsymbol{u}^\prime$ of the input
vector $\boldsymbol{u}$ and a particular subvector $\boldsymbol{y}^\prime$
of the output vector $\boldsymbol{y}$. This motivates us to introduce the
following concept.

\begin{definition}
\label{def:BAE} For the linear quantum system (\ref{eq:real_sys_a})-(\ref%
{eq:real_sys_b}), let $\Xi_{\boldsymbol{u}^\prime \to\mathbf{\boldsymbol{y}}%
^\prime}(s)$ be the transfer function from a subvector $\boldsymbol{u}^\prime$
of the input vector $\boldsymbol{u}$ and a subvector $\boldsymbol{y}^\prime$
of the output vector $\boldsymbol{y}$. We say that system (\ref%
{eq:real_sys_a})-(\ref{eq:real_sys_b}) realizes the \textit{back-action
evasion (BAE) measurement} of the output $\boldsymbol{y}^\prime$ with
respect to the input $\boldsymbol{u}^\prime$ if $\Xi_{\boldsymbol{u}^\prime
\to\mathbf{\boldsymbol{y}}^\prime}(s)=0$ for all $s$.
\end{definition}

More discussions on BAE measurements can be found in, e.g., \cite{HMW95},  \cite{TC10}, 
\cite{WC13}, \cite{ODP+16}, \cite{NY14} and the references therein.

We shall see that all of these notions emerge naturally from the study of
the Kalman decomposition of a linear quantum system, see Remarks \ref%
{rem:sept22_1} and \ref{rem:BAE}.

%%%%%%%%%%%%%%%%%%%%%%%%%%
%%%%%%%%%%%%%%%%%%%%%%%%%%
%%%%%%%%%%%%%%%%%%%%%%%%%%

\section{The Kalman decomposition for passive linear quantum systems}

\label{sec:passive}

In this section, we study the Kalman decomposition for passive linear
quantum systems. First, we show that their uncontrollable subspace is
identical to their unobservable subspace.

Let us define the controllability and observability matrices of system (\ref%
{eq:passive_sys_a})-(\ref{eq:passive_sys_b}), respectively, by 
\begin{eqnarray*}
C_{G} &\triangleq& \left[%
\begin{array}{cccc}
\mathcal{B} & \mathcal{A}\mathcal{B} & \cdots & \mathcal{A}^{n-1}\mathcal{B}%
\end{array}
\right],  \notag \\
O_{G} &\triangleq& \left[%
\begin{array}{c}
\mathcal{C} \\ 
\mathcal{C}\mathcal{A} \\ 
\vdots \\ 
\mathcal{C}\mathcal{A}^{n-1}%
\end{array}
\right] .
\end{eqnarray*}
$\mathrm{Im}(C _{G})$ and $\mathrm{Ker}\left(O_{G}\right)$ are the
controllable and unobservable subspaces of the space of system variables $%
\mathbb{C}^{n}$. As in the general case, we define the uncontrollable and
observable subspaces to be their \emph{orthogonal complements} in $\mathbb{C}%
^{n}$, that is $\mathrm{Ker}(C_{G}^{\dagger })$, and $\mathrm{Im}%
(O_{G}^{\dagger })$, respectively. We have the following theorem.

%%%%%%%%%%%%%%%%%%%%

\begin{theorem}
\label{thm:passive_Kalman} The uncontrollable and unobservable subspaces of
the passive linear quantum system (\ref{eq:passive_sys_a})-(\ref%
{eq:passive_sys_b}) are identical. That is, 
\begin{equation}
\mathrm{Ker}( C_{G}^{\dagger }) = \mathrm{Ker}\left( O_{G}\right).
\label{key_2}
\end{equation}
\end{theorem}

\textit{Proof}: Let us define the auxiliary matrices 
\begin{eqnarray*}
C_{s} &\triangleq& [ \mathcal{B} \ (-\imath \Omega_{-})\mathcal{B} \ \cdots
\ (-\imath \Omega_{-})^{n-1}\mathcal{B} ],  \notag \\
O_{s} &\triangleq& \left[ 
\begin{array}{c}
\mathcal{C} \\ 
\mathcal{C}(-\imath \Omega_{-}) \\ 
\vdots \\ 
\mathcal{C}(-\imath \Omega_{-})^{n-1}%
\end{array}
\right].
\end{eqnarray*}
It can be readily shown that 
\begin{eqnarray*}
C_{s}^{\dagger } = -\left[ 
\begin{array}{cccc}
I_{m} &  &  &  \\ 
& -I_{m} &  &  \\ 
&  & \ddots &  \\ 
&  &  & (-1)^{n-1}I_{m}%
\end{array}
\right] O_{s}.
\end{eqnarray*}
%\begin{eqnarray*}
%C_{s}^{\dagger } &=&\left[
%\begin{array}{c}
%\mathcal{B}^{\dagger } \\
%\mathcal{B}^{\dagger }(-\imath \Omega_{-})^{\dagger } \\
%\vdots \\
%\mathcal{B}^{\dagger }((-\imath \Omega_{-})^{\dagger })^{n-1}%
%\end{array}
%\right] \\
%&=&-\left[
%\begin{array}{c}
%\mathcal{C} \\
%-\mathcal{C}(-\imath \Omega_{-}) \\
%\vdots \\
%(-1)^{n-1}\mathcal{C}(-\imath \Omega_{-})^{n-1}%
%\end{array}
%\right] \\
%&=& -\left[
%\begin{array}{cccc}
%I_{m} &  &  &  \\
%& -I_{m} &  &  \\
%&  & \ddots &  \\
%&  &  & (-1)^{n-1}I_{m}%
%\end{array}
%\right] O_{s}.
%\end{eqnarray*}
Thus, we have 
\begin{equation}
\mathrm{Ker}(C_{s}^{\dagger })=\mathrm{Ker}(O_{s}) .  \label{key_1}
\end{equation}

Now, we show that 
\begin{equation}
\mathrm{Ker}\left( O_{G}\right) =\mathrm{Ker} (O_{s}) .
\label{eq:ker_G_Sigma}
\end{equation}
Let $\mu \in \mathrm{Ker}(O_{s})$. Then, $C_{-}(\Omega _{-})^{j}\mu =0,\ \
j=0,1,\ldots$. As a result, $C_{-}(-i\Omega _{-}-\frac{1 }{2}C_{-}^{\dagger
}C_{-})^{j}\mu =0,\ \ j=0,1,\ldots $. That is, $\mu \in \mathrm{Ker}(O_{G})$%
. Hence, $\mathrm{Ker}(O_{s}) \subset \mathrm{Ker}(O_{G})$. The fact that, $%
\mathrm{Ker} (O_{G}) \subset \mathrm{Ker}(O_{s})$, can be proved similarly,
thus proving Eq. (\ref{eq:ker_G_Sigma}). We can establish the relation 
\begin{equation}
\mathrm{Ker}( C_{G}^{\dagger }) =\mathrm{Ker}( C_{s}^{\dagger }) .
\label{eq:ker_G_Sigma_2}
\end{equation}
similarly. Finally, Eq. (\ref{key_2}) follows from Eqs. (\ref{key_1})-(\ref{eq:ker_G_Sigma_2}). \hfill $\blacksquare $

%%%%%%%%%%%%%%%%%%%%

Theorem \ref{thm:passive_Kalman} demonstrates that the Kalman decomposition
of passive linear quantum systems can contain only $co$ and $\bar{c}\bar{o}$
subsystems. This property is due to the special structure of passive
systems, and does not hold for general linear quantum systems; see, e.g.,
Theorem \ref{thm:general_Kalman_1}. An immediate consequence of this result
is that, an uncontrollable mode is necessarily an unobservable mode. As was
discussed in Section \ref{sec:lqs}, only unitary coordinate transformations
preserve the quantum structure of passive linear quantum systems, and are
thus allowed to be used to achieve the Kalman decomposition. Although in the
case of general linear systems, it will require some effort to construct a
Bogoliubov or symplectic transformation for this purpose, the situation is
very simple in the passive case. Indeed, in the case of passive linear
quantum systems, a decomposition of the space of system variables into a
controllable subspace and an uncontrollable subspace will achieve the Kalman
decomposition. However, it is a well-known fact that this decomposition can
always be performed via a unitary matrix; e.g., see \cite{VD81} and the
references therein.

It is easily seen from Definition \ref{def:DFS} that $\mathrm{Ker}(O_{G})$
is the DFS of system (\ref{eq:passive_sys_a})-(\ref{eq:passive_sys_b}), if
it is non-trivial. In \cite[Lemma 2]{GZ15}, it is shown that for a passive
linear quantum system, Hurwitz stability, controllability and observability
are all equivalent. From this and Theorem \ref{thm:passive_Kalman}, we
conclude that if the passive linear quantum system (\ref{eq:passive_sys_a})-(%
\ref{eq:passive_sys_b}) is not Hurwitz stable, it must have a non-trivial
DFS. In what follows, we characterize the DFS of a passive linear quantum
system.

%%%%%%%%%%%%%%%%%%%%%

\begin{theorem}
\label{thm:passive_DFS_characterization} The DFS of the passive linear
quantum system (\ref{eq:passive_sys_a})-(\ref{eq:passive_sys_b}) is spanned
by the eigenvectors of the matrix $\mathcal{A}$ whose corresponding
eigenvalues are on the imaginary axis.
\end{theorem}

\textit{Proof}: It is a well-known fact that $\mathrm{Ker}(O _{G})$ is an
invariant subspace of $\mathcal{A}$. Hence, it is spanned by its
eigenvectors (including generalized ones). First, we show that all
eigenvectors of $\mathcal{A}$ with imaginary eigenvalues belong to $\mathrm{%
Ker}(O _{G})$. Let $\lambda $ be an eigenvalue of $\mathcal{A}$ with $%
\mathrm{Re}(\lambda )=0$, and let $\mu \neq 0$ be the corresponding
eigenvector. From the proof of Theorem \ref{thm:passive_Kalman}, it suffices
to show that $\mu \in \mathrm{Ker}(O_{s})$. That is, $C_{-}
\Omega_{-}^{k}\mu =0$ for $k=0,1,2,\ldots$ From 
\begin{equation}  \label{eq:feb1}
\mathcal{A}\mu =(-\imath\Omega_{-}-\frac{1}{2} C_{-}^{\dag}C_{-})\mu
=\lambda \mu,
\end{equation}
we have $\mu ^{\dagger }(-\imath\Omega _{-}-\frac{1}{2}C_{-}^{\dagger
}C_{-})\mu =\lambda \mu ^{\dagger }\mu$ %\begin{equation*}
%\mu ^{\dagger }(-\imath\Omega _{-}-\frac{1}{2}C_{-}^{\dagger }C_{-})\mu
%=\lambda \mu ^{\dagger }\mu
%\end{equation*}
and $\mu ^{\dagger }(\imath\Omega _{-}-\frac{1}{2}C_{-}^{\dagger }C_{-})\mu
= \lambda^\ast\mu ^{\dagger }\mu$. %\begin{equation*}
%\mu ^{\dagger }(\imath\Omega _{-}-\frac{1}{2}C_{-}^{\dagger }C_{-})\mu =
%\lambda^\ast\mu ^{\dagger }\mu .
%\end{equation*}
Adding these two equations, we get $-\mu ^{\dagger }C_{-}^{\dagger }C_{-}\mu
=2\mathrm{Re}(\lambda )\mu ^{\dagger }\mu =0$, %\begin{equation*}
%-\mu ^{\dagger }C_{-}^{\dagger }C_{-}\mu =2\mathrm{Re}(\lambda )\mu
%^{\dagger }\mu =0,
%\end{equation*}
which implies $C_{-}\mu =0$. Substituting $C_{-}\mu =0$ into Eq. (\ref%
{eq:feb1}) yields $\Omega_- \mu = \imath\lambda \mu$. As a result, $C_-
\Omega_-^k \mu = (\imath\lambda)^k C_{-}\mu =0, ~~ k=0,1,2,\ldots$. Thus we
have $\mu \in \mathrm{Ker}(O_{s}) = \mathrm{Ker}(O_{G})$. Next, we show that
if $\imath\omega$, $\omega \in \mathbb{R}$, is an eigenvalue of the matrix $%
\mathcal{A}$ for the passive linear quantum system (\ref{eq:passive_sys_a})-(%
\ref{eq:passive_sys_b}), then its geometric multiplicity is one. This way,
generalized eigenvectors for imaginary eigenvalues are excluded. To see
this, suppose that the geometric multiplicity is two. Then, in an
appropriate basis, the matrix $\mathcal{A}$ has a Jordan block $\bigl[ 
\begin{smallmatrix}
\imath\omega & 1 \\ 
0 & \imath\omega%
\end{smallmatrix}
\bigr]$. Clearly, the matrix 
\begin{equation*}
\left[ 
\begin{array}{cc}
\imath\omega & 1 \\ 
0 & \imath\omega%
\end{array}
\right] +\left[ 
\begin{array}{cc}
\imath\omega & 1 \\ 
0 & \imath\omega%
\end{array}
\right]^\dag = \left[ 
\begin{array}{cc}
0 & 1 \\ 
1 & 0%
\end{array}
\right]
\end{equation*}
is indefinite. However, from Eq. (\ref{eq:passive_PR}), we have $\mathcal{A}+%
\mathcal{A}^\dag = -C_{-}^{\dag} C_{-}$, %\begin{equation*}
%\mathcal{A}+\mathcal{A}^\dag = -C_{-}^{\dag} C_{-},
%\end{equation*}
which is negative semi-definite, a contradiction. A similar argument
excludes cases of higher geometric multiplicity. Finally, to complete the
proof, we need to show that $\mathrm{Ker}(O_{G})$ is spanned only by
eigenvectors of $\mathcal{A}$ with eigenvalues on the imaginary axis. Let $%
\mu \in \mathrm{Ker}(O _{G})$ be an eigenvector of $\mathcal{A}%
=-\imath\Omega _{-}-\frac{1}{2} C_{-}^{\dag}C_{-}$, with eigenvalue $\lambda$%
. Then, the equations $C_{-}\mu=0$, and $\mathcal{A}\mu = \lambda \mu$,
imply that $-\imath\Omega _{-}\mu = \lambda \mu$. However, $\Omega_{-}$ is a
Hermitian matrix, and this implies that $\lambda$ is imaginary. \hfill $%
\blacksquare $

We end this section with a simple example.

%%%%%%%%%%%%%%%%%%%%%

\begin{example}
Consider a passive linear quantum system with parameters $C_{-}=[1~1]$ and $%
\Omega _{-}=I_{2}$. The corresponding QSDEs are 
\begin{eqnarray*}
\boldsymbol{\dot{a}}_{1}(t) &=&-(\imath+\frac{1}{2})\boldsymbol{a}_{1}(t) -%
\frac{1 }{2}\boldsymbol{a}_{2}(t) -\boldsymbol{b}(t) , \\
\boldsymbol{\dot{a}}_{2}(t) &=&-\frac{1}{2}\boldsymbol{a}_{1}(t) -(\imath+%
\frac{1 }{2})\boldsymbol{a}_{2}(t) -\boldsymbol{b}(t) , \\
\boldsymbol{b}_{\mathrm{out}}(t) &=&\boldsymbol{a}_{1}(t) +\boldsymbol{a}
_{2}(t) +\boldsymbol{b}(t).
\end{eqnarray*}
If we let $T=\frac{1}{\sqrt{2}} \left[%
\begin{smallmatrix}
1 & 1 \\ 
-1 & 1%
\end{smallmatrix}%
\right]$, and $\left[%
\begin{smallmatrix}
\boldsymbol{a}_{DF} \\ 
\boldsymbol{a}_{D}%
\end{smallmatrix}%
\right]\triangleq T^{\dagger }\boldsymbol{a}$, the QSDEs for $\boldsymbol{a}%
_{DF}$ and $\boldsymbol{a}_{D}$ are the following: 
\begin{eqnarray*}
\boldsymbol{\dot{a}}_{DF}(t) &=&-\imath\boldsymbol{a}_{DF}(t) , \\
\boldsymbol{\dot{a}}_{D}(t) &=&-\left( 1+\imath\right) \boldsymbol{a}_{D}(t)
- \sqrt{2}\boldsymbol{b}(t) , \\
b_{\mathrm{out}}(t) &=&\sqrt{2}\boldsymbol{a}_{D}(t) +\boldsymbol{b}(t) .
\end{eqnarray*}
Clearly, $\boldsymbol{a}_{DF}$ is a DF mode.
\end{example}

%%%%%%%%%%%%%%%%%%%%%%%%%%
%%%%%%%%%%%%%%%%%%%%%%%%%%
%%%%%%%%%%%%%%%%%%%%%%%%%%

\section{The Kalman decomposition for general linear quantum systems}

\label{sec:general}

In this section, we construct the Kalman decomposition for a general linear
quantum system and uncover its special structure. In Subsection \ref%
{subsec:complex}, we derive the decomposition in the complex
annihilation-creation operator representation, and show that it can be
achieved with a unitary and Bogoliubov coordinate transformation. Then, we
translate the main results of this subsection, Theorems \ref%
{thm:general_Kalman_1} and \ref{thm:general_Kalman_2}, into the real
quadrature operator representation, Theorems \ref{thm:general_Kalman_3} and %
\ref{thm:general_Kalman_4} in Subsection \ref{subsec:real}. Finally, some
special cases of the Kalman decomposition are investigated in Subsection \ref%
{subsec:special_cases}.

\subsection{The Kalman decomposition in the complex annihilation-creation
operator representation}

\label{subsec:complex}

To make the presentation easy to follow, we first establish a series of
lemmas that are used to prove the main results of this subsection, Theorems %
\ref{thm:general_Kalman_1} and \ref{thm:general_Kalman_2}.

Define an auxiliary matrix \cite[Eq. (7)]{GZ15}: 
\begin{equation*}
O_{s} \triangleq \left[ 
\begin{array}{c}
\mathcal{C} \\ 
\mathcal{C}\,(J_{n}\Omega) \\ 
\vdots \\ 
\mathcal{C}(J_{n}\Omega )^{2n-1}%
\end{array}
\right] .
\end{equation*}
By \cite[Proposition 2]{GZ15}, we know that 
\begin{equation*}
\mathrm{Ker}\left( O_{G}\right) =\mathrm{Ker}\left( O _{s}\right) ,\ \ 
\mathrm{Ker}( C_{G}^{\dagger }) =\mathrm{Ker} \left( O_{s}J_{n}\right) .
\end{equation*}
Instead of working directly with $\mathrm{Ker}\left( O_{G}\right) $ and $%
\mathrm{Ker} ( C_{G}^{\dagger })$, it will be easier to work with $\mathrm{%
Ker}\left(O_{s}\right) $ and $\mathrm{Ker}\left( O _{s}J_{n}\right)$.

We start by characterizing the controllable subspace $\mathrm{Im}(C _{G})$,
the uncontrollable subspace $\mathrm{Ker}(C_{G}^{\dagger })$, the observable
subspace $\mathrm{Im}(O_{G}^{\dagger })$, and the unobservable subspace $%
\mathrm{Ker}\left( O_{G}\right)$. We first establish the following result.

%%%%%%%%%%%%%%%%%%%%%%%%%

\begin{lemma}
\label{lem:CO} The unobservable subspace $\mathrm{Ker}\left( O _{s}\right) $
and the uncontrollable subspace $\mathrm{Ker}\left( O _{s}J_{n}\right) $ are
related by 
\begin{equation}  \label{eq:f2a}
\mathrm{Ker}\left( O_{s}\right) =J_{n}\mathrm{Ker}\left( O _{s}J_{n}\right) .
\end{equation}
Similarly, the controllable subspace $\mathrm{Im}(C_{G})$ and the observable
subspace $\mathrm{Im}(O_{G}^{\dagger })$ are related by 
\begin{equation}  \label{eq:fc}
\mathrm{Im}(C_{G})=J_{n}\mathrm{Im}(O_{G}^{\dagger }).
\end{equation}
\end{lemma}

\textit{Proof:} Eq. (\ref{eq:f2a}) can be established in a straightforward
way. Hence, we concentrate on Eq. (\ref{eq:fc}). Noticing that $\mathrm{Im}%
(C_{G}) = \mathrm{Ker}(C_{G}^{\dagger })^{\perp }= \mathrm{Ker}\left(
O_{s}J_{n}\right) ^{\perp }$, %\begin{eqnarray*}
%\mathrm{Im}(C_{G}) = \mathrm{Ker}(C_{G}^{\dagger })^{\perp }= \mathrm{Ker}%
%\left( O_{s}J_{n}\right) ^{\perp },  %\label{eq:temp_c}
%\end{eqnarray*}
and $\mathrm{Im}(O_{G}^{\dagger }) = \mathrm{Ker}\left( O _{G}\right)
^{\perp }=\mathrm{Ker}\left( O_{s}\right) ^{\perp } =(J_{n} \mathrm{Ker}%
\left( O_{s}J_{n}\right) )^{\perp }$, %\begin{eqnarray*}
%\mathrm{Im}(O_{G}^{\dagger }) &=& \mathrm{Ker}\left( O _{G}\right) ^{\perp }=%
%\mathrm{Ker}\left( O_{s}\right) ^{\perp }  \nonumber \\
%&=&(J_{n} \mathrm{Ker}\left( O_{s}J_{n}\right) )^{\perp },  %\label{eq:temp_d}
%\end{eqnarray*}
where Eq. (\ref{eq:f2a}) is used in the last step, it suffices to show that 
\begin{equation}
\mathrm{Ker}\left( O_{s}J_{n}\right) ^{\perp }=J_{n}(J_{n}\mathrm{Ker }%
\left( O_{s}J_{n}\right) )^{\perp }.  \label{eq:f3}
\end{equation}
However, 
\begin{eqnarray*}
&&(J_{n}\mathrm{Ker}\left( O_{s}J_{n}\right) )^{\perp } \\
&=&\left\{ x:(J_{n}y)^{\dagger }x=0,\forall y\in \mathrm{Ker}%
\left(O_{s}J_{n}\right) \right\} \\
&=&J_{n}\left\{ w:w^{\dagger }y=0,\forall y\in \mathrm{Ker}\left( O
_{s}J_{n}\right) \right\} \\
&=&J_{n}(\mathrm{Ker}(O_{s}J_{n}))^{\perp }.
\end{eqnarray*}
Therefore, Eq. (\ref{eq:f3}) holds, and so does Eq. (\ref{eq:fc}). \hfill $%
\blacksquare $

Now, let us define the four subspaces used in the Kalman decomposition: 
\begin{eqnarray}
R_{c\bar{o}} &\triangleq &\mathrm{Im}(C_{G})\cap \mathrm{Ker}(O_{G}),
\label{Ra0} \\
R_{co} &\triangleq &\mathrm{Im}(C_{G})\cap \mathrm{Im}(O _{G}^{\dagger }),
\label{Rb0} \\
R_{\bar{c}\bar{o}} &\triangleq &\mathrm{Ker(}C_{G}^{\dagger })\cap \mathrm{%
Ker}(O_{G}),  \label{Rc0} \\
R_{\bar{c}o} &\triangleq &\mathrm{Ker}(C_{G}^{\dagger })\cap \mathrm{Im}%
(O_{G}^{\dagger }).  \label{Rd0}
\end{eqnarray}
That is, $R_{c\bar{o}}$, $R_{co}$, $R_{\bar{c}\bar{o}}$, and $R_{\bar{c}o}$
are respectively the controllable/unobservable ($c\bar{o}$),
controllable/observable ($co$), uncontrollable/unobservable ($\bar{c}\bar{o} 
$), and uncontrollable/observable ($\bar{c}o$) subspaces of system (\ref%
{eq:sys_a})-(\ref{eq:sys_b}).

The following lemma, which is an immediate consequence of Lemma \ref{lem:CO}, reveals relations among the subspaces $R_{c\bar{o}}$, $R_{co}$, $R_{\bar{c}%
\bar{o}}$, and $R_{\bar{c}o}$.

%%%%%%%%%%%%%%%%%%%%%

\begin{lemma}
\label{lem:subspaces} The subspaces $R_{c\bar{o}}$, $R_{co}$, $R_{\bar{c}%
\bar{o}}$, and $R_{\bar{c}o}$ can be expressed as 
\begin{eqnarray*}
R_{c\bar{o}} &=&\mathrm{Ker}\left( O_{s}J_{n}\right) ^{\perp }\cap \mathrm{%
Ker}(O_{s}),  \label{Ra} \\
R_{co} &=&\mathrm{Ker}\left( O_{s}J_{n}\right) ^{\perp }\cap \mathrm{Ker}%
\left( O_{s}\right) ^{\perp }, \\
R_{\bar{c}\bar{o}} &=&\mathrm{Ker}\left( O_{s}J_{n}\right) \cap \mathrm{Ker}%
\left( O_{s}\right) , \\
R_{\bar{c}o} &=&\mathrm{Ker}\left( O_{s}J_{n}\right) \cap \mathrm{Ker}\left(
O_{s}\right) ^{\perp }.
\end{eqnarray*}%
Moreover, they enjoy the following properties: $R_{c\bar{o}}\perp R_{co}\perp
R_{\bar{c}\bar{o}}\perp R_{\bar{c}o}$, %\begin{equation*}
%R_{c\bar{o}}\perp R_{co}\perp R_{\bar{c}\bar{o}}\perp R_{\bar{c}o},
%\end{equation*}
and 
\begin{equation}
R_{c\bar{o}}=J_{n}R_{\bar{c}o},~R_{co}=J_{n}R_{co},~R_{\bar{c}\bar{o}%
}=J_{n}R_{\bar{c}\bar{o}}.  \label{perp_4}
\end{equation}%
Furthermore, the vector space $\mathbb{C}^{2n}$ is the direct sum of these
orthogonal subspaces. That is, $\mathbb{C}^{2n}=R_{c\bar{o}}\oplus
R_{co}\oplus R_{\bar{c}\bar{o}}\oplus R_{\bar{c}o}$. %\begin{equation*}
%\mathbb{C}^{2n}=R_{c\bar{o}}\oplus R_{co}\oplus R_{\bar{c}\bar{o}}\oplus R_{%
%\bar{c}o}.  %\label{space}
%\end{equation*}
\end{lemma}

The next lemma shows that we can choose bases with a special structure for
the subspaces $R_{co}$ and $R_{\bar{c}\bar{o}}$.

%%%%%%%%%%%%%%%%%%%%%%

\begin{lemma}
\label{lem:basis_Rco} We have:

\begin{description}
\item[(i)] There exists a unitary and Bogoliubov matrix $T_{co}$ of the form 
\begin{equation}  \label{Tco}
T_{co}= \left[ 
\begin{array}{cc}
Z_{1} & 0 \\ 
0 & Z_{1}^{\#}%
\end{array}
\right],
\end{equation}%
where $Z_{1} \in \mathbb{C}^{n \times n_{1}}$ ($n_1\geq0$), such that its
columns form an orthonormal basis for $R_{co}$.

\item[(ii)] Similarly, there exists a unitary and Bogoliubov matrix $T_{\bar{%
c}\bar{o}}$ of the form 
\begin{equation*}
T_{\bar{c}\bar{o}}= \left[ 
\begin{array}{cc}
Z_{2} & 0 \\ 
0 & Z_{2}^{\#}%
\end{array}
\right],
\end{equation*}
where $Z_{2} \in \mathbb{C}^{n \times n_{2}}$ ($n_2\geq 0$), such that its
columns form an orthonormal basis for $R_{\bar{c}\bar{o}}$.
\end{description}
\end{lemma}

\textit{Proof:\ } We first establish Item (i). Let $\bigl[
\begin{smallmatrix}
e_{1} \\ 
f_{1}%
\end{smallmatrix}
\bigr]$ be a nonzero vector in the subspace $R_{co}$. Then, from the second
relation in Eq. (\ref{perp_4}), we have that $\bigl[
\begin{smallmatrix}
e_{1} \\ 
-f_{1}%
\end{smallmatrix}
\bigr] \in R_{co}$. Therefore, the vectors $\bigl[
\begin{smallmatrix}
e_{1} \\ 
0%
\end{smallmatrix}
\bigr]$, $\bigl[
\begin{smallmatrix}
0 \\ 
f_{1}%
\end{smallmatrix}
\bigr] \in R_{co}$. Moreover, due to the doubled-up structure of the system
matrices, it can be readily shown that $\bigl[
\begin{smallmatrix}
0 \\ 
e_{1}^{\#}%
\end{smallmatrix}
\bigr] $, $\bigl[
\begin{smallmatrix}
f_{1}^{\#} \\ 
0%
\end{smallmatrix}
\bigr] \in R_{co}$, as well. Because $\bigl[
\begin{smallmatrix}
e_{1} \\ 
f_{1}%
\end{smallmatrix}
\bigr] \neq 0$, $e_{1}$ and $f_{1}$ cannot both be zero. If $e_{1}\neq 0$,
define $z_{1}\triangleq \frac{1}{\left\Vert e_{1}\right\Vert }e_{1}$;
otherwise, define $z_{1}\triangleq \frac{1}{\left\Vert f_{1}^{\#}\right\Vert 
}f_{1}^{\#}$. Then, $\bigl[
\begin{smallmatrix}
z_{1} \\ 
0%
\end{smallmatrix}
\bigr] $ and $\bigl[
\begin{smallmatrix}
0 \\ 
z_{1}^{\#}%
\end{smallmatrix}
\bigr] $ are nonzero orthonormal vectors in $R_{co}$. Take another nonzero
vector $\bigl[
\begin{smallmatrix}
e_{2} \\ 
f_{2}%
\end{smallmatrix}
\bigr] \in R_{co}$ which is orthogonal to both $\bigl[
\begin{smallmatrix}
z_{1} \\ 
0%
\end{smallmatrix}
\bigr] $ and $\bigl[
\begin{smallmatrix}
0 \\ 
z_{1}^{\#}%
\end{smallmatrix}
\bigr] $. \ Then $z_{1}^{\dagger }e_{2}=0$ and $z_{1}^{\dag}f_{2}^{\#}=0$.
If $e_{2}\neq 0$, define $z_{2}\triangleq \frac{1}{\left\Vert
e_{2}\right\Vert }e_{2}$; otherwise, define $z_{2}\triangleq \frac{1}{%
\left\Vert f_{2}^{\#}\right\Vert }f_{2}^{\#}$. Then, $\bigl[
\begin{smallmatrix}
z_{1} \\ 
0%
\end{smallmatrix}
\bigr]$, $\bigl[
\begin{smallmatrix}
0 \\ 
z_{1}^{\#}%
\end{smallmatrix}
\bigr] $, $\bigl[
\begin{smallmatrix}
z_{2} \\ 
0%
\end{smallmatrix}
\bigr] $, $\bigl[
\begin{smallmatrix}
0 \\ 
z_{2}^{\#}%
\end{smallmatrix}
\bigr] $ are orthonormal vectors in $R_{co}$. Repeat this procedure to get
the matrix $T_{co}$ in Eq. (\ref{Tco}), with $Z_{1}=[z_{1} \ z_{2} \ldots
z_{n_{1}}]$. Clearly, the columns of $T_{co}$ form an orthonormal basis of $%
R_{co}$. Moreover, by the construction given above, $Z_{1}^{\dag} Z_{1} =
I_{n_{1}}$ holds. As a result, $T_{co}^{\dagger }T_{co}=I_{2n_{1}}$ and $%
T_{co}^{\dagger }J_{n}T_{co}=J_{n_{1}}$.

Item (ii) can be established in a similar way. \hfill\ $\blacksquare $

%%%%%%%%%%%%%%%%%%%

\begin{remark}
\label{rem:aug16} The above proof is more rigorous than that of \cite[Lemma 1%
]{GZ15}, which fails to discuss the case where $e_j=0$ or $f_j=0$.
\end{remark}

%%%%%%%%%%%%%%%%%%%

\begin{remark}
\label{rem:aug15_dim} It follows from Lemma \ref{lem:basis_Rco} that the
dimensions of the subspaces $R_{co}$ and $R_{\bar{c}\bar{o}}$ are both even (%
$2n_1$ and $2n_2$, respectively). Let the dimensions of the subspaces $R_{c%
\bar{o}}$ and $R_{\bar{c}o}$ be $n_3$ and $n_4$ respectively. Due to first
relation in Eq. (\ref{perp_4}), and the fact that $J_{n}$ is invertible, we
must have that $n_4 = n_3$. Hence, $2(n_1+ n_2+ n_3)=2n$.
\end{remark}

In order to construct special orthonormal bases for the subspaces $R_{c\bar{o%
}}$ and $R_{\bar{c}o}$, the following three lemmas are needed.

%%%%%%%%%%%%%%%%%%%%

\begin{lemma}
\label{lem:Delta} Let $M,N\in \mathbb{C}^{r\times k}$ and $x_{1},x_{2} \in 
\mathbb{C}^{k}$. If 
\begin{equation}
\Delta (M,N)\left[ 
\begin{array}{c}
x_{1} \\ 
x_{2}%
\end{array}%
\right] =0,  \label{eq:jan15_5}
\end{equation}
then 
\begin{equation}
\Delta (M,N)\left[ 
\begin{array}{c}
x_{1}+x_{2}^{\#} \\ 
x_{1}^{\#}+x_{2}%
\end{array}%
\right] =0.  \label{eq:jan15_4}
\end{equation}
\end{lemma}

\textit{Proof: \ } Eq. (\ref{eq:jan15_5}) is equivalent to 
\begin{eqnarray}
Mx_{1}+Nx_{2} &=&0,  \label{eq:temp_dec13_1a} \\
N^{\#}x_{1}+M^{\#}x_{2} &=&0.  \label{eq:temp_dec13_1b}
\end{eqnarray}%
Conjugating both sides of Eq. (\ref{eq:temp_dec13_1b}) yields 
\begin{equation}
Mx_{2}^{\#}+Nx_{1}^{\#}=0.  \label{eq:temp_dec13_1c}
\end{equation}%
Adding Eqs. (\ref{eq:temp_dec13_1a}) and (\ref{eq:temp_dec13_1c}) yields $%
M(x_{1}+x_{2}^{\#})+N(x_{1}^{\#}+x_{2})=0$. %\begin{eqnarray*}
%M(x_{1}+x_{2}^{\#})+N(x_{1}^{\#}+x_{2}) &=&0.
%\end{eqnarray*}
Conjugating both sides of the above equation gives $N^{\#}(x_{1}+x_{2}^{%
\#})+M^{\#}(x_{1}^{\#}+x_{2}) =0$. %\begin{eqnarray*}
%N^{\#}(x_{1}+x_{2}^{\#})+M^{\#}(x_{1}^{\#}+x_{2}) &=&0.
%\end{eqnarray*}%
In compact form, the above two equations become 
\begin{equation*}
\Delta (M,N)\left[ 
\begin{array}{c}
x_{1}+x_{2}^{\#} \\ 
x_{1}^{\#}+x_{2}%
\end{array}%
\right] =0,
\end{equation*}%
which is Eq. (\ref{eq:jan15_4}). \hfill\ $\blacksquare $

%%%%%%%%%%%%%%%%%%%%

\begin{lemma}
\label{lem:CG} If $\bigl[%
\begin{smallmatrix}
x_{1} \\ 
x_{2}%
\end{smallmatrix}%
\bigr] \in \mathrm{Im}(C_{G})$, then $\bigl[%
\begin{smallmatrix}
x_{1}+x_{2}^{\#} \\ 
x_{2}+x_{1}^{\#}%
\end{smallmatrix}%
\bigr] \in \mathrm{Im}(C_{G})$.
\end{lemma}

\textit{Proof: \ } The matrices $\mathcal{A}$ and $\mathcal{B}$ in Eq. (\ref%
{eq:sys_ABCD}) are doubled-up. Hence, $\mathcal{A}^{k}\mathcal{B}$ is also
doubled-up, for all $k=1,\ldots $. That is, each block column of the
controllability matrix $C_{G}$ is doubled-up. As a result, upon a column
permutation, $C_{G}$ is of the form $\Delta \left(C_{G,+},C_{G,-}\right)$.
Then, given $\bigl[%
\begin{smallmatrix}
x_{1} \\ 
x_{2}%
\end{smallmatrix}%
\bigr] \in \mathrm{Im}(C_{G})$, there exist vectors $z_{+}$ and $z_{-}$ such
that 
\begin{equation*}
\left[ 
\begin{array}{c}
x_{1} \\ 
x_{2}%
\end{array}%
\right] =\Delta \left( C_{G,+},C_{G,-}\right) \left[ 
\begin{array}{c}
z_{+} \\ 
z_{-}%
\end{array}%
\right] .
\end{equation*}%
Consequently, it can be easily shown that 
\begin{equation*}
\left[ 
\begin{array}{c}
x_{1}+x_{2}^{\#} \\ 
x_{2}+x_{1}^{\#}%
\end{array}%
\right] =\Delta \left( C_{G,+},C_{G,-}\right) \left[ 
\begin{array}{c}
z_{+}+z_{-}^{\#} \\ 
z_{-}+z_{+}^{\#}%
\end{array}%
\right] .
\end{equation*}%
That is, $\bigl[%
\begin{smallmatrix}
x_{1}+x_{2}^{\#} \\ 
x_{2}+x_{1}^{\#}%
\end{smallmatrix}
\bigr] \in \mathrm{Im}(C_{G})$. \hfill\ $\blacksquare $

Lemmas \ref{lem:Delta} and \ref{lem:CG} can be used to establish the
following result.

%%%%%%%%%%%%%%%%%%%%

\begin{lemma}
\label{lem:Rcbaro} If $\bigl[%
\begin{smallmatrix}
x_{1} \\ 
x_{2}%
\end{smallmatrix}%
\bigr] \in R_{c\bar{o}}$, then $\bigl[%
\begin{smallmatrix}
x_{1}+x_{2}^{\#} \\ 
x_{1}^{\#}+x_{2}%
\end{smallmatrix}%
\bigr] \in R_{c\bar{o}}$.
\end{lemma}

\textit{Proof: \ } Consider a vector $\bigl[%
\begin{smallmatrix}
x_{1} \\ 
x_{2}%
\end{smallmatrix}%
\bigr] \in R_{c\bar{o}}$. From Eq. (\ref{Ra0}), $\bigl[%
\begin{smallmatrix}
x_{1} \\ 
x_{2}%
\end{smallmatrix}%
\bigr] \in \mathrm{Im}(C_{G})\cap \mathrm{Ker}(O_{G})$. According to Lemma %
\ref{lem:CG}, 
\begin{equation}
\left[ 
\begin{array}{c}
x_{1}+x_{2}^{\#} \\ 
x_{1}^{\#}+x_{2}%
\end{array}%
\right] \in \mathrm{Im}(C_{G}).  \label{july27_2}
\end{equation}%
Also, since $\bigl[%
\begin{smallmatrix}
x_{1} \\ 
x_{2}%
\end{smallmatrix}%
\bigr] \in \mathrm{Ker}(O_{G})$, by Lemma \ref{lem:Delta}, 
\begin{equation}
\left[ 
\begin{array}{c}
x_{1}+x_{2}^{\#} \\ 
x_{1}^{\#}+x_{2}%
\end{array}%
\right] \in \mathrm{Ker}(O_{G}).  \label{eq:jan15_7}
\end{equation}%
Eqs. (\ref{july27_2}) and (\ref{eq:jan15_7}) yield%
\begin{equation*}
\left[ 
\begin{array}{c}
x_{1}+x_{2}^{\#} \\ 
x_{1}^{\#}+x_{2}%
\end{array}%
\right] \in \mathrm{Im}(C_{G})\cap \mathrm{Ker}(O_{G})=R_{c\bar{o}}.
\end{equation*}%
$\hfill\ \blacksquare $

%%%%%%%%%%%%%%%%%%%%

\begin{remark}
\label{rem: july28} Lemma \ref{lem:Rcbaro} also holds for the subspaces $R_{%
\bar{c}o}$, $R_{co}$, and $R_{\bar{c}\bar{o}}$.
\end{remark}

We are ready to construct special orthonormal bases for the subspaces $R_{c%
\bar{o}}$ and $R_{\bar{c}o}$ in the following lemma.

%%%%%%%%%%%%%%%%%%%%

\begin{lemma}
\label{lem:basis_Rh} There exists a matrix $T_{c\bar{o}}$ of the form 
\begin{equation*}
T_{c\bar{o}}\triangleq \frac{1}{\sqrt{2}}\left[ 
\begin{array}{cc}
X & Y \\ 
X^{\#} & -Y^{\#}%
\end{array}%
\right] ,
\end{equation*}%
where $X\in \mathbb{C}^{n\times n_{a}}$ and $Y\in \mathbb{C} ^{n\times
n_{b}} $ ($n_{a}\geq 0$, $n_{b}\geq 0$, $n_{a} + n_{b}=n_{3}$) satisfy $%
X^{\dagger }X=I_{n_{a}}$, $Y^{\dagger }Y=I_{n_{b}}$ and $X^{\dag} Y = 0$,
such that its columns form an orthonormal basis of $R_{c\bar{o}}$. Also, the
columns of $T_{\bar{c}o}\triangleq J_{n}T_{c\bar{o}}$ form an orthonormal
basis of $R_{\bar{c}o}$.
\end{lemma}

\textit{Proof:} Let $X=[x_{1},~\cdots,~x_{n_{a}}]$ and $Y=[y_{1},~\cdots
,~y_{n_{b}}]$, for some non-negative integers $n_a, n_b\geq 0$ such that $%
n_a+n_b=n_3$. We use the following algorithm to construct the vectors $%
x_1,\ldots,x_{n_a}$ and $y_1,\ldots,y_{n_b}$ sequentially.

\textit{Step 0.} Set indices $j=k=0$.

\textit{Step 1.} Pick a nonzero vector $\bigl[%
\begin{smallmatrix}
u \\ 
v%
\end{smallmatrix}%
\bigr] \in R_{c\bar{o}}$ . By Lemma \ref{lem:Rcbaro}, $\bigl[%
\begin{smallmatrix}
u+v^{\#} \\ 
u^{\#}+v%
\end{smallmatrix}%
\bigr] \in R_{c\bar{o}}$. There are two possibilities:

Case (I). $u+v^{\#}\neq 0 $. In this case, define $x_{1}\triangleq \frac{1}{%
\Vert u+v^{\#}\Vert }(u+v^{\#})$, where $\Vert \cdot \Vert $ denotes the
vector Euclidean norm. Clearly, $x_{1}^{\dag }x_{1}=1$, and $\bigl[%
\begin{smallmatrix}
x_{1} \\ 
x_{1}^{\#}%
\end{smallmatrix}%
\bigr] \in R_{c\bar{o}}$. Set $j\to j+1$.

Case (II). $u+v^{\#}=0 $. In this case, $v=-u^{\#}$. Define $y_{1}\triangleq 
\frac{u}{\left\Vert u\right\Vert }$. We have $y_{1}^{\dagger }y_{1}=1$ and $%
\bigl[%
\begin{smallmatrix}
y_{1} \\ 
-y_{1}^{\#}%
\end{smallmatrix}%
\bigr] \in R_{c\bar{o}}$. Set $k\to k+1$.

According to the above, in the first step of the algorithm we generate
either $x_1$ or $y_1$.

\textit{Step} $p=j+k$. Up to this step, we have generated $x_1, \ldots, x_j$%
, and $y_1, \ldots, y_k$. Now, let us take a nonzero vector $\bigl[%
\begin{smallmatrix}
x \\ 
y%
\end{smallmatrix}%
\bigr] \in R_{c\bar{o}}$ which satisfies 
\begin{equation}
x_{i}^{\dag }x+x_{i}^{\top}y=0,\ \ i=1,\ldots ,j  \label{july27_3a}
\end{equation}%
and%
\begin{equation}
y_{l}^{\dag }x-y_{l}^{\top}y=0,\ \ l=1,\ldots ,k.  \label{july27_3a2}
\end{equation}%
Complex conjugating both sides of Eqs. (\ref{july27_3a}) and (\ref%
{july27_3a2}) yields%
\begin{equation}
x_{i}^{\top}x^{\#}+x_{i}^{\dag }y^{\#}=0,\ \ i=1,\ldots ,j  \label{july27_3b}
\end{equation}%
and%
\begin{equation}
-y_{l}^{\top}x^{\#}+y_{l}^{\dagger }y^{\#}=0,\ \ l=1,\ldots ,k,
\label{july27_3b2}
\end{equation}%
respectively. Adding Eqs. (\ref{july27_3a}) and (\ref{july27_3b}) we get 
\begin{equation}
x_{i}^{\dag }(x+y^{\#})+x_{i}^{\top}(x^{\#}+y)=0.  \label{july27_3c}
\end{equation}%
That is, $\bigl[%
\begin{smallmatrix}
x+y^{\#} \\ 
x^{\#}+y%
\end{smallmatrix}%
\bigr] $ is orthogonal to $\bigl[%
\begin{smallmatrix}
x_{i} \\ 
x_{i}^{\#}%
\end{smallmatrix}%
\bigr] $, $\ i=1,\ldots ,j$. Similarly, using Eqs. (\ref{july27_3a2}) and (%
\ref{july27_3b2}) we get $y_{l}^{\dag }(x+y^{\#})-y_{l}^{\top}(x^{\#}+y)=0,\
l=1,\ldots ,k$. %\begin{equation*}
%y_{l}^{\dag }(x+y^{\#})-y_{l}^{\top}(x^{\#}+y)=0,\ \ l=1,\ldots ,k.
%\end{equation*}%
That is, $\bigl[%
\begin{smallmatrix}
x+y^{\#} \\ 
x^{\#}+y%
\end{smallmatrix}%
\bigr] $ is orthogonal to $\bigl[%
\begin{smallmatrix}
y_{j} \\ 
-y_{j}^{\#}%
\end{smallmatrix}%
\bigr] $, for all $l=1,\ldots ,k$. Again, there are two possibilities:

Case (I). $x+y^{\#}\neq 0$. In this case, define $x_{j+1}\triangleq \frac{1}{%
\Vert x+y^{\#}\Vert }(x+y^{\#})$. Clearly, $\bigl[%
\begin{smallmatrix}
x_{j+1} \\ 
x_{j+1}^{\#}%
\end{smallmatrix}%
\bigr] $ is orthogonal to all of the vectors $\bigl[%
\begin{smallmatrix}
x_{i} \\ 
x_{i}^{\#}%
\end{smallmatrix}%
\bigr] $ and $\bigl[%
\begin{smallmatrix}
y_{j} \\ 
-y_{j}^{\#}%
\end{smallmatrix}%
\bigr] $ for $i=1,\ldots ,j$ and $l=1,\ldots,k$. Set $j\rightarrow j+1 $.

Case (II). $x+y^{\#}=0$. In this case, define $y_{k+1}\triangleq \frac{1}{%
\Vert x\Vert }x$. Then we have, 
\begin{equation*}
x_{i}^{\dag }y_{k+1}+x_{i}^{\top}(-y_{k+1}^{\#}) = \frac{1}{\Vert x\Vert }%
(x_{i}^{\dag }x+x_{i}^{\top}y)=0
\end{equation*}
for all $i=1,\ldots ,j$, %\begin{eqnarray*}
%\lefteqn{x_{i}^{\dag }y_{k+1}+x_{i}^{\top}(-y_{k+1}^{\#})} \\
%%
%&=&\frac{1}{\Vert x\Vert }(x_{i}^{\dag }x+x_{i}^{\top}y)=0,\ \ i=1,\ldots ,j
%\end{eqnarray*}%
and%
\begin{equation*}
y_{l}^{\dag }y_{k+1}-y_{l}^{\top}(-y_{k+1}^{\#}) = \frac{1}{\Vert x\Vert }%
\left( y_{l}^{\dag }x-y_{l}^{\top}y\right) =0
\end{equation*}
for all $l=1,\ldots ,k$. %\begin{eqnarray*}
%\lefteqn{y_{l}^{\dag }y_{k+1}-y_{l}^{\top}(-y_{k+1}^{\#})}  \nonumber \\
%%&=&\frac{1}{\Vert x\Vert }\left( y_{l}^{\dag }x+y_{l}^{\top}x^{\#}\right) \\
%&=&\frac{1}{\Vert x\Vert }\left( y_{l}^{\dag }x-y_{l}^{\top}y\right) =0,\ \
%l=1,\ldots ,k.
%\end{eqnarray*}%
That is, $\bigl[%
\begin{smallmatrix}
y_{k+1} \\ 
-y_{k+1}^{\#}%
\end{smallmatrix}%
\bigr] $ is orthogonal to all of the vectors $\bigl[%
\begin{smallmatrix}
x_{i} \\ 
x_{i}^{\#}%
\end{smallmatrix}%
\bigr] $ and $\bigl[%
\begin{smallmatrix}
y_{j} \\ 
-y_{j}^{\#}%
\end{smallmatrix}%
\bigr] $ for $=1,\ldots ,j$ and $l=1,\ldots,k$. Set $k\rightarrow k+1 $.

\textit{Step} $n_3$. The algorithm terminates.

When the above algorithm terminates, we will have constructed the matrices $%
X $, and $Y$ in the definition of $T_{c\bar{o}}$. It is clear from the above
construction that the columns of $T_{c\bar{o}}$ form an orthonormal basis
for the subspace $R_{c\bar{o}}$. From the first relation in Eq. (\ref{perp_4}%
), it follows that the columns of $T_{\bar{c}o}\triangleq J_{n}T_{c\bar{o}}$
form an orthonormal basis of $R_{\bar{c}o}$. Finally, we prove that $X$ and $%
Y$ satisfy the relations $X^{\dagger }X=I_{n_{a}}$, $Y^{\dagger }Y=I_{n_{b}}$%
, and $X^{\dag} Y = 0$. Indeed, from the fact that the columns of $T_{c\bar{o%
}}$ form an orthonormal basis of $R_{c\bar{o}}$, we have that $T_{c\bar{o}%
}^{\dag}T_{c\bar{o}}=I_{n_{3}}$, from which it follows that $X^{\dag}X +
X^{\top}X^{\#}=2I_{n_a}$. Similarly, since $R_{c\bar{o}}\perp R_{\bar{c}o}$,
we have that $T_{\bar{c}o}^{\dag}T_{c\bar{o}}=T_{c\bar{o}}^{\dag} J_{n} T_{c%
\bar{o}}=0$, which implies the relation $X^{\dag}X - X^{\top}X^{\#}=0_{n_a}$%
. Adding these equations gives $X^{\dagger }X=I_{n_{a}}$. The other two
relations can be proved similarly.\hfill\ $\blacksquare$

We are now ready to construct a unitary and Bogoliubov transformation matrix
that achieves the Kalman decomposition of the system (\ref{eq:sys_a})-(\ref%
{eq:sys_b}). From now on, we will use the notation $R_{h}=R_{c\bar{o}}
\oplus R_{\bar{c}o}$.

\begin{theorem}
\label{thm:general_Kalman_1} Let 
\begin{eqnarray*}
T_{co} &=&\left[ 
\begin{array}{cc}
Z_{1} & 0 \\ 
0 & Z_{1}^{\#}%
\end{array}%
\right] ,\ \ T_{\bar{c}\bar{o}}=\left[ 
\begin{array}{cc}
Z_{2} & 0 \\ 
0 & Z_{2}^{\#}%
\end{array}%
\right] , \\
T_{c\bar{o}} &=&\frac{1}{\sqrt{2}}\left[ 
\begin{array}{cc}
X & Y \\ 
X^{\#} & -Y^{\#}%
\end{array}%
\right] ,\ \ T_{\bar{c}o}=\frac{1}{\sqrt{2}}\left[ 
\begin{array}{cc}
X & Y \\ 
-X^{\#} & Y^{\#}%
\end{array}%
\right]
\end{eqnarray*}%
be constructed as in Lemmas \ref{lem:basis_Rco} and \ref{lem:basis_Rh}, and
let $Z_{3}\triangleq \lbrack X~~Y]$. %\begin{equation*}%\label{sept2_Z3}
%Z_{3} \triangleq [X ~~ Y].
%\end{equation*}
Then the matrix 
\begin{equation}
\tilde{T}\triangleq \left[ 
\begin{array}{cccccc}
Z_{3} & Z_{1} & Z_{2} & 0 & 0 & 0 \\ 
0 & 0 & 0 & Z_{3}^{\#} & Z_{1}^{\#} & Z_{2}^{\#}%
\end{array}%
\right]  \label{tilde_T}
\end{equation}%
is a unitary and Bogoliubov matrix (i.e., it satisfies $\tilde{T}^{\dag }%
\tilde{T}=I_{2n}$ and $\tilde{T}^{\dag }J_{n}\tilde{T}=J_{n}$), and
decomposes the system variables of the linear quantum system (\ref{eq:sys_a}%
)-(\ref{eq:sys_b}) as follows: 
\begin{equation}
\left[ 
\begin{array}{cccccc}
\boldsymbol{a}_{h}^{\top } & \boldsymbol{a}_{co}^{\top } & \boldsymbol{a}_{%
\bar{c}\bar{o}}^{\top } & \boldsymbol{a}_{h}^{\dag } & \boldsymbol{a}%
_{co}^{\dag } & \boldsymbol{a}_{\bar{c}\bar{o}}^{\dag }%
\end{array}%
\right] ^{\top }=\tilde{T}^{\dag }\boldsymbol{\breve{a}}.
\label{complex_Kalman_1}
\end{equation}
\end{theorem}

\textit{Proof}: From Lemmas \ref{lem:basis_Rco}, and \ref{lem:basis_Rh}, we
have that $Z_{1}^{\dag }Z_{1}=I_{n_{1}}$, $Z_{2}^{\dag }Z_{2}=I_{n_{2}}$,
and $X^{\dagger }X=I_{n_{a}}$, $Y^{\dagger }Y=I_{n_{b}}$, and $X^{\dag }Y=0$%
. From the last three equations, we deduce that $Z_{3}^{\dag
}Z_{3}=I_{n_{3}} $. Also, from the orthogonality of the subspaces $R_{co}$, $%
R_{\bar{c}\bar{o}}$, and $R_{h}=R_{c\bar{o}}\oplus R_{\bar{c}o}$, we have
that $Z_{i}^{\dag }Z_{j}=0$, $i,j=1,2,3,i\neq j$. Then, the equations $%
\tilde{T}^{\dag }\tilde{T}=I_{2n}$ and $\tilde{T}^{\dag }J_{n}\tilde{T}%
=J_{n} $ follow immediately. From Lemma \ref{lem:basis_Rco}, we have that 
\begin{equation}\label{sept2_co}
\boldsymbol{\breve{a}}_{co}=\left[ 
\begin{array}{c}
\boldsymbol{a}_{co} \\ 
\boldsymbol{a}_{co}^{\#}%
\end{array}%
\right] \triangleq T_{co}^{\dag }\boldsymbol{a}=\left[ 
\begin{array}{c}
Z_{1}^{\dag }\boldsymbol{a} \\ 
(Z_{1}^{\dag }\boldsymbol{a})^{\#}%
\end{array}%
\right]
\end{equation}%
are the $co$ variables. Similarly, 
\begin{equation}\label{sept2_barco}
\boldsymbol{\breve{a}}_{\bar{c}\bar{o}}=\left[ 
\begin{array}{c}
\boldsymbol{a}_{\bar{c}\bar{o}} \\ 
\boldsymbol{a}_{\bar{c}\bar{o}}^{\#}%
\end{array}%
\right] \triangleq T_{\bar{c}\bar{o}}^{\dag }\boldsymbol{a}=\left[ 
\begin{array}{c}
Z_{2}^{\dag }\boldsymbol{a} \\ 
(Z_{2}^{\dag }\boldsymbol{a})^{\#}%
\end{array}%
\right]
\end{equation}%
are the $\bar{c}\bar{o}$ variables. Finally, from Lemma \ref{lem:basis_Rh}
we have that the columns of 
\begin{equation*}
\frac{1}{\sqrt{2}}\left[ 
\begin{array}{cccc}
X & Y & X & Y \\ 
X^{\#} & -Y^{\#} & -X^{\#} & Y^{\#}%
\end{array}%
\right]
\end{equation*}%
form an orthonormal basis for $R_{h}=R_{c\bar{o}}\oplus R_{\bar{c}o}$. Using
simple manipulations, it is easy to see that the same is true for 
\begin{equation}\label{aug31_1}
\left[ 
\begin{array}{cccc}
X & Y & 0 & 0 \\ 
0 & 0 & X^{\#} & Y^{\#}%
\end{array}%
\right] =\left[ 
\begin{array}{cc}
Z_{3} & 0 \\ 
0 & Z_{3}^{\#}%
\end{array}%
\right] \triangleq T_{h}.
\end{equation}%
Hence, 
\begin{equation}\label{sept2_h}
\boldsymbol{\breve{a}}_{h}=\left[ 
\begin{array}{c}
\boldsymbol{a}_{h} \\ 
\boldsymbol{a}_{h}^{\#}%
\end{array}%
\right] \triangleq T_{h}^{\dag }\boldsymbol{a}=\left[ 
\begin{array}{c}
Z_{3}^{\dag }\boldsymbol{a} \\ 
(Z_{3}^{\dag }\boldsymbol{a})^{\#}%
\end{array}%
\right]
\end{equation}%
are the $h=c\bar{o}\cup \bar{c}o$ variables. Hence, (\ref{complex_Kalman_1})
follows. \hfill $\blacksquare $

Although $\tilde{T}$ is useful in decomposing the space of variables of the
system (\ref{eq:sys_a})-(\ref{eq:sys_b}) into its $R_{co}$, $R_{\bar{c}\bar{o%
}}$, and $R_{h}=R_{c\bar{o}}\oplus R_{\bar{c}o}$ subspaces, it is not
directly useful in putting (\ref{eq:sys_a})-(\ref{eq:sys_b}) into the Kalman
canonical form. The reason is that the evolution equation for $\tilde{T}%
^{\dag }\boldsymbol{\breve{a}}$ mixes the evolution of variables in
different subspaces in a non-obvious way. To put (\ref{eq:sys_a})-(\ref%
{eq:sys_b}) into a Kalman-like canonical form, we introduce the following
transformation: 
\begin{eqnarray}
T &\triangleq &\left[ 
\begin{array}{c|c|c}
T_{h} & T_{co} & T_{\bar{c}\bar{o}}%
\end{array}%
\right]  \notag \\
&=&\left[ 
\begin{array}{cc|cc|cc}
Z_{3} & 0 & Z_{1} & 0 & Z_{2} & 0 \\ 
0 & Z_{3}^{\#} & 0 & Z_{1}^{\#} & 0 & Z_{2}^{\#}%
\end{array}%
\right] ,  \label{T}
\end{eqnarray}%
where $T_{h}$ was defined in Eq. (\ref{aug31_1}). Similarly to $T_{co}$ and $%
T_{\bar{c}\bar{o}}$, $T_{h}$ satisfies $T_{h}^{\dagger }T_{h}=I_{2n_{3}}$,
and $T_{h}^{\dagger }J_{n}T_{h}=J_{n_{3}}$. From the identities $Z_{i}^{\dag
}Z_{j}=\delta _{ij}I_{n_{i}}$, $i,j=1,2,3$, established in Lemmas \ref%
{lem:basis_Rco} and \ref{lem:basis_Rh}, and Theorem \ref%
{thm:general_Kalman_1}, it follows that $T^{\dagger }T=I_{2n}$, that is, $T$
is unitary, and also that 
\begin{equation*}
T^{\dagger }J_{n}T=\left[ 
\begin{array}{ccc}
J_{n_{3}} & 0 & 0 \\ 
0 & J_{n_{1}} & 0 \\ 
0 & 0 & J_{n_{2}}%
\end{array}%
\right] .
\end{equation*}%
That is, $T$ is \emph{blockwise Bogoliubov}. From this, we have the
following theorem.

%%%%%%%%%%%%%%%%%%%%

\begin{theorem}
\label{thm:general_Kalman_2} The unitary and blockwise Bogoliubov coordinate
transformation 
\begin{equation}  \label{complex_Kalman_2}
\left[%
\begin{array}{c}
\boldsymbol{\breve{a}}_{h} \\ 
\boldsymbol{\breve{a}}_{co} \\ 
\boldsymbol{\breve{a}}_{\bar{c}\bar{o}}%
\end{array}%
\right] = T^{\dagger }\boldsymbol{\breve{a}}
\end{equation}%
transforms the linear quantum system (\ref{eq:sys_a})-(\ref{eq:sys_b}) into
the form 
\begin{eqnarray}
\left[%
\begin{array}{c}
\boldsymbol{\dot{\breve{a}}}_{h}(t) \\ 
\boldsymbol{\dot{\breve{a}}}_{co}(t) \\ 
\boldsymbol{\dot{\breve{a}}}_{\bar{c}\bar{o}}(t)%
\end{array}%
\right] &=& \bar{\mathcal{A}} \left[%
\begin{array}{c}
\boldsymbol{\breve{a}}_{h}(t) \\ 
\boldsymbol{\breve{a}}_{co}(t) \\ 
\boldsymbol{\breve{a}}_{\bar{c}\bar{o}}(t)%
\end{array}%
\right] + \bar{\mathcal{B}}\, \boldsymbol{\breve{b}}(t),
\label{eq:complex_Kalman_ss} \\
\boldsymbol{\breve{b}}_{\mathrm{out}}(t) &=& \bar{\mathcal{C}} \left[%
\begin{array}{c}
\boldsymbol{\breve{a}}_{h}(t) \\ 
\boldsymbol{\breve{a}}_{co}(t) \\ 
\boldsymbol{\breve{a}}_{\bar{c}\bar{o}}(t)%
\end{array}%
\right] +\boldsymbol{\breve{b}}(t),  \label{eq:complex_Kalman_io}
\end{eqnarray}
where 
\begin{eqnarray}
\bar{\mathcal{A}} &\triangleq& T^{\dag}\mathcal{A}T= \left[%
\begin{array}{ccc}
\mathcal{A}_{h} & \mathcal{A}_{12} & \mathcal{A}_{13} \\ 
\mathcal{A}_{21} & \mathcal{A}_{co} & 0 \\ 
\mathcal{A}_{31} & 0 & \mathcal{A}_{\bar{c}\bar{o}}%
\end{array}%
\right],  \notag \\
\bar{\mathcal{B}} &\triangleq& T^{\dag}\mathcal{B}= \left[%
\begin{array}{c}
\mathcal{B}_{h} \\ 
\mathcal{B}_{co} \\ 
0%
\end{array}%
\right],  \notag \\
\bar{\mathcal{C}} &\triangleq& \mathcal{C}T= \left[%
\begin{array}{ccc}
\mathcal{C}_{h} & \mathcal{C}_{co} & 0%
\end{array}%
\right].  \label{eq:complex_Kalman_sys_ABCD}
\end{eqnarray}
\end{theorem}

\textit{Proof}: The proof follows from the following well-known invariance
properties of linear systems; e.g., see \cite[Chapter 2]{HK97}: 
\begin{equation}
\mathcal{A}R_{c\bar{o}}\subset R_{c\bar{o}},~\mathcal{A}R_{co}\subset R_{c%
\bar{o}} \oplus R_{co}, ~\mathcal{A}R_{\bar{c}\bar{o}}\subset R_{c\bar{o}}
\oplus R_{\bar{c}\bar{o}}  \label{subset_a}
\end{equation}
and 
\begin{eqnarray}
&& \mathrm{Im}(\mathcal{B}) \subset \mathrm{Im}(C_{G})=R_{c\bar{o}} \oplus
R_{co},  \notag \\
&& \mathrm{Ker}(O_{G}) = R_{c\bar{o}} \oplus R_{\bar{c}\bar{o}} \subset 
\mathrm{Ker}(C),  \label{subset_b}
\end{eqnarray}
which imply 
\begin{eqnarray*}
&&\mathcal{A}R_{co}\subset R_{h} \oplus R_{co}, \ \ \mathcal{A}R_{\bar{c}%
\bar{o}}\subset R_{h} \oplus R_{\bar{c}\bar{o}}, \\
&&\mathrm{Im}(\mathcal{B}) \subset R_{h} \oplus R_{co}, \ \ R_{\bar{c}\bar{o}%
} \subset \mathrm{Ker}(C).
\end{eqnarray*}
\hfill $\blacksquare$

%%%%%%%%%%%%%%%%%%%%%%%%%%%%%%%%

\begin{remark}
\label{rem:sept4} From a physics perspective, one expects that the $co$
subsystem 
\begin{eqnarray*}
\boldsymbol{\dot{\breve{a}}}_{co}(t) &=&\mathcal{A}_{co}\boldsymbol{\breve{a}%
}_{co}(t)+\mathcal{B}_{co}\boldsymbol{\breve{b}}(t), \\
\boldsymbol{\breve{b}}_{\mathrm{out}}(t) &=&\mathcal{C}_{co}\boldsymbol{%
\breve{a}}_{co}(t)+\boldsymbol{\breve{b}}(t),
\end{eqnarray*}%
and the $\bar{c}\bar{o}$ subsystem 
\begin{equation*}
\boldsymbol{\dot{\breve{a}}}_{\bar{c}\bar{o}}(t)=\mathcal{A}_{\bar{c}\bar{o}}%
\boldsymbol{\breve{a}}_{\bar{c}\bar{o}}(t),
\end{equation*}%
are respectively linear quantum systems in their own right. The proof is as
follows: From the second of Eqs. (\ref{eq:complex_Kalman_sys_ABCD}), we have
that 
\begin{eqnarray*}
&&\left[ 
\begin{array}{c}
\mathcal{B}_{h} \\ 
\mathcal{B}_{co} \\ 
0%
\end{array}%
\right] =\bar{\mathcal{B}}=T^{\dag }\mathcal{B}=-T^{\dag }\mathcal{C}^{\flat
}=-T^{\dag }J_{n}C^{\dag }J_{m} \\
&=&-(T^{\dag }J_{n}T)\,(CT)^{\dag }J_{m}=-\left[ 
\begin{array}{ccc}
J_{n_{3}} & 0 & 0 \\ 
0 & J_{n_{1}} & 0 \\ 
0 & 0 & J_{n_{2}}%
\end{array}%
\right] \bar{\mathcal{C}}^{\dag }J_{m} \\
&=&-\left[ 
\begin{array}{c}
J_{n_{3}}\mathcal{C}_{h}^{\dag }J_{m} \\ 
J_{n_{1}}\mathcal{C}_{co}^{\dag }J_{m} \\ 
0%
\end{array}%
\right] =-\left[ 
\begin{array}{c}
\mathcal{C}_{h}^{\flat } \\ 
\mathcal{C}_{co}^{\flat } \\ 
0%
\end{array}%
\right] ,
\end{eqnarray*}%
from which follows that 
\begin{equation}
\mathcal{B}_{co}=-\mathcal{C}_{co}^{\flat }.  \label{eq:rem:sept4_a}
\end{equation}%
From this, we also conclude that 
\begin{equation*}
\mathcal{B}^{\flat }T=-\mathcal{C}T=-\bar{\mathcal{C}}=\left[ 
\begin{array}{ccc}
\mathcal{B}_{h}^{\flat } & \mathcal{B}_{co}^{\flat } & 0%
\end{array}%
\right] ,
\end{equation*}%
and, hence, 
\begin{eqnarray}
&&T^{\dag }\mathcal{B}\mathcal{B}^{\flat }T=\left[ 
\begin{array}{c}
\mathcal{B}_{h} \\ 
\mathcal{B}_{co} \\ 
0%
\end{array}%
\right] \left[ 
\begin{array}{ccc}
\mathcal{B}_{h}^{\flat } & \mathcal{B}_{co}^{\flat } & 0%
\end{array}%
\right]  \notag \\
&=&\left[ 
\begin{array}{ccc}
\mathcal{B}_{h}\mathcal{B}_{h}^{\flat } & \mathcal{B}_{h}\mathcal{B}%
_{co}^{\flat } & 0 \\ 
\mathcal{B}_{co}\mathcal{B}_{h}^{\flat } & \mathcal{B}_{co}\mathcal{B}%
_{co}^{\flat } & 0 \\ 
0 & 0 & 0%
\end{array}%
\right] .  \label{eq:rem:sept4_b}
\end{eqnarray}%
Also, 
\begin{eqnarray}
&&T^{\dag }\mathcal{A}^{\flat }T=T^{\dag }J_{n}\mathcal{A}^{\dag }J_{n}T 
\notag \\
&=&(T^{\dag }J_{n}T)\,(T^{\dag }\mathcal{A}T)^{\dag }(T^{\dag }J_{n}T) 
\notag \\
&=&\left[ 
\begin{array}{ccc}
J_{n_{3}} & 0 & 0 \\ 
0 & J_{n_{1}} & 0 \\ 
0 & 0 & J_{n_{2}}%
\end{array}%
\right] \bar{\mathcal{A}}^{\dag }\left[ 
\begin{array}{ccc}
J_{n_{3}} & 0 & 0 \\ 
0 & J_{n_{1}} & 0 \\ 
0 & 0 & J_{n_{2}}%
\end{array}%
\right]  \notag \\
&=&\left[ 
\begin{array}{ccc}
\mathcal{A}_{h}^{\flat } & \mathcal{A}_{21}^{\flat } & \mathcal{A}%
_{31}^{\flat } \\ 
\mathcal{A}_{12}^{\flat } & \mathcal{A}_{co}^{\flat } & 0 \\ 
\mathcal{A}_{13}^{\flat } & 0 & \mathcal{A}_{\bar{c}\bar{o}}^{\flat }%
\end{array}%
\right] .  \label{eq:rem:sept4_c}
\end{eqnarray}%
Now, we multiply both sides of the first of the Eqs. (\ref{eq:PR}) by $%
T^{\dag }$ from the left, and $T$ from the right: 
\begin{equation*}
T^{\dag }\mathcal{A}T+T^{\dag }\mathcal{A}^{\flat }T+T^{\dag }\mathcal{B}%
\mathcal{B}^{\flat }T=0.
\end{equation*}%
Using Eqs. (\ref{eq:rem:sept4_b}) and (\ref{eq:rem:sept4_c}), the $(2,2)$
and $(3,3)$ blocks of the resulting block-matrix equation are, respectively, 
\begin{eqnarray}
\mathcal{A}_{co}+\mathcal{A}_{co}^{\flat }+\mathcal{B}_{co}\mathcal{B}%
_{co}^{\flat } &=&0,  \label{eq:rem:sept4_d} \\
\mathcal{A}_{\bar{c}\bar{o}}+\mathcal{A}_{\bar{c}\bar{o}}^{\flat } &=&0.
\label{eq:rem:sept4_e}
\end{eqnarray}%
Eqs. (\ref{eq:rem:sept4_a}) and (\ref{eq:rem:sept4_d}) are the physical
realizability conditions for the $co$ subsystem, while (\ref{eq:rem:sept4_e}%
) is the physical realizability condition for the $\bar{c}\bar{o}$ subsystem
(which has no inputs/outputs).
\end{remark}

\begin{remark}
\label{rem:unitary} We emphasize the fact that the transformation matrices $%
\tilde{T}$ in Eq. (\ref{complex_Kalman_1}) and $T$ in Eq. (\ref%
{complex_Kalman_2}) are unitary, in addition to being Bogoliubov or
blockwise Bogoliubov, respectively. This property is due to the special
structure of linear quantum systems and does not hold in general for linear
systems. A consequence of this is that these transformations can be applied
in a numerically stable way. Also, similarly to the classical case, they are
not unique.
\end{remark}

\begin{remark}
\label{rem:tf_complex} The sub-matrices of the matrix $\bar{\mathcal{A}}$
defined in Eq. (\ref{eq:complex_Kalman_sys_ABCD}) satisfy the following
identity: 
\begin{eqnarray}  \label{tf_complex}
\left[%
\begin{array}{c}
\mathcal{A}_{21} \\ 
\mathcal{A}_{31}%
\end{array}%
\right]\left( sI-\mathcal{A}_{h} \right) ^{-1} \left[%
\begin{array}{cc}
\mathcal{A}_{12} & \mathcal{A}_{13}%
\end{array}%
\right] = 0.
\end{eqnarray}
This result is established in Remark \ref{rem:tf_BAE}, in the next
subsection. It follows from (\ref{tf_complex}) that 
\begin{equation}  \label{eigenvalues}
\sigma(\bar{\mathcal{A}})=\sigma(\mathcal{A}_{co})\cup \sigma(\mathcal{A}_{%
\bar{c}\bar{o}}) \cup \sigma(\mathcal{A}_{h}).
\end{equation}
% From Theorem \ref{thm:general_Kalman_2}, we can see that the transfer function from $\boldsymbol{\breve{a}}_{co}$ to $\boldsymbol{\breve{a}}_{h}$ is $ \Xi_{\boldsymbol{\breve{a}}_{co} \to \boldsymbol{\breve{a}}_{h}} [s]=(sI-\mathcal{A}_{h})^{-1} \mathcal{A}_{12}$, and similarly,  the transfer function from $\boldsymbol{\breve{a}}_{h}$ to $\boldsymbol{\breve{a}}_{co}$ is $\Xi_{\boldsymbol{\breve{a}}_{h} \to \boldsymbol{\breve{a}}_{co}} [s]=(sI-\mathcal{A}_{co})^{-1} \mathcal{A}_{21}$. Thus, the transfer function from $\boldsymbol{\breve{a}}_{co}$ to itself {\it through} $\boldsymbol{\breve{a}}_{h}$ is $\Xi_{\boldsymbol{\breve{a}}_{h} \to \boldsymbol{\breve{a}}_{co}} [s]\Xi_{\boldsymbol{\breve{a}}_{co} \to \boldsymbol{\breve{a}}_{h}} [s] =(sI-\mathcal{A}_{co})^{-1} \mathcal{A}_{21} (sI-\mathcal{A}_{h})^{-1} \mathcal{A}_{12}$.  In fact, the transfer function $\Xi_{\boldsymbol{\breve{a}}_{h} \to \boldsymbol{\breve{a}}_{co}} [s]\Xi_{\boldsymbol{\breve{a}}_{co} \to \boldsymbol{\breve{a}}_{h}} [s]=0$, as demonstrated in Example \ref{ex:july28_complex}. Unfortunately, this can not be seen immediately from Theorem \ref{thm:general_Kalman_2}. In Subsection \ref{subsec:real}, we switch to the real quadrature  operator representation and prove this fact. A similar statement can be made for the transfer function from  $\boldsymbol{\breve{a}}_{\bar{c}\bar{o}}$ to itself {\it through} $\boldsymbol{\breve{a}}_{h}$; see Remark \ref{rem:tf_BAE}.
\end{remark}

We end this subsection with an illustrative example. 
%%%%%%%%%%%%%%%%%%%%%%%%%

\begin{example}
\label{ex:july28_complex} Consider the linear quantum system (\ref{eq:sys_a}%
)-(\ref{eq:sys_b}) with parameters 
\begin{equation*}
C_{-}=[1~0],~C_{+}=[0~0],\ \Omega _{-}=\Omega _{+}=\left[ 
\begin{array}{cc}
0 & 1 \\ 
1 & 0%
\end{array}%
\right] .
\end{equation*}%
Then, 
\begin{eqnarray}
\boldsymbol{H} &=&(\boldsymbol{a}_{1}+\boldsymbol{a}_{1}^{\ast })(%
\boldsymbol{a}_{2}+\boldsymbol{a}_{2}^{\ast }),  \label{sept3_H} \\
\boldsymbol{L} &=&\boldsymbol{a}_{1}.  \label{sept3_L}
\end{eqnarray}%
The transformation matrix $T$ in Eq. (\ref{T}) is computed to be%
\begin{equation*}
T=\left[ 
\begin{array}{c|c}
T_{h} & T_{co}%
\end{array}%
\right] =\left[ 
\begin{array}{cccc}
0 & 0 & 1 & 0 \\ 
1 & 0 & 0 & 0 \\ 
0 & 0 & 0 & 1 \\ 
0 & 1 & 0 & 0%
\end{array}%
\right]
\end{equation*}%
Hence, using Eq. (\ref{complex_Kalman_2}) we have 
\begin{equation*}
\boldsymbol{\breve{a}}_{h}=T_{h}^{\dagger }\boldsymbol{\breve{a}=}\left[ 
\begin{array}{c}
a_{2} \\ 
a_{2}^{\ast }%
\end{array}%
\right] ,\ \ \boldsymbol{\breve{a}}_{co}=T_{co}^{\dagger }\boldsymbol{\breve{%
a}=}\left[ 
\begin{array}{c}
a_{1} \\ 
a_{1}^{\ast }%
\end{array}%
\right] .
\end{equation*}%
The corresponding Kalman canonical form is%
\begin{eqnarray*}
\boldsymbol{\dot{\breve{a}}}_{h}(t) &=&\imath\left[ 
\begin{array}{cc}
-1 & -1 \\ 
1 & 1%
\end{array}%
\right] \boldsymbol{\breve{a}}_{co}(t), \\
\boldsymbol{\dot{\breve{a}}}_{co}(t) &=&\imath\left[ 
\begin{array}{cc}
-1 & -1 \\ 
1 &1%
\end{array}%
\right] \boldsymbol{\breve{a}}_{h}(t)--\frac{1}{2}\boldsymbol{\breve{a}}%
_{co}(t)-\boldsymbol{\breve{b}}(t), \\
\boldsymbol{\breve{b}}_{\mathrm{out}}(t) &=&\boldsymbol{\breve{a}}_{co}(t)+%
\boldsymbol{\breve{b}}(t).
\end{eqnarray*}
It can be easily seen that the transfer function 
\begin{equation*}
\mathcal{A}_{21}\left( sI-\mathcal{A}_{h}\right) ^{-1}\mathcal{A}_{12}=-\frac{%
1}{s}\left[ 
\begin{array}{cc}
-1 & -1 \\ 
1 & 1%
\end{array}%
\right] \left[ 
\begin{array}{cc}
-1 & -1 \\ 
1 & 1%
\end{array}%
\right] =0,
\end{equation*}%
as required by (\ref{tf_complex}). %%
%\begin{eqnarray*}
%\boldsymbol{\dot{\breve{a}}}_{2}(t) &=&\left[
%\begin{array}{cc}
%-\imath  & -\imath  \\
%\imath  & \imath
%\end{array}%
%\right] \boldsymbol{\breve{a}}_{1}(t), \\
%\boldsymbol{\dot{\breve{a}}}_{1}(t) &=&-\frac{1}{2}\boldsymbol{\breve{a}}%
%_{1}(t)+\left[
%\begin{array}{cc}
%-\imath  & -\imath  \\
%\imath  & \imath
%\end{array}%
%\right] \boldsymbol{\breve{a}}_{2}(t)-\boldsymbol{\breve{b}}(t) \\
%\boldsymbol{\breve{b}}_{\mathrm{out}}(t) &=&\boldsymbol{\breve{a}}_{1}(t)+%
%\boldsymbol{\breve{b}}(t),
%\end{eqnarray*}
\end{example}

%%%%%%%%%%%%%%%%%%%%%%%%
%%%%%%%%%%%%%%%%%%%%%%%%

\subsection{The Kalman decomposition in the real quadrature operator
representation}

\label{subsec:real}

In this subsection, we present the Kalman decomposition for a linear quantum
system in the real quadrature operator representation, namely a system of
the form (\ref{eq:real_sys_a})-(\ref{eq:real_sys_b}).

First, let us introduce the following system variables for $R_{co}$, $R_{%
\bar{c}\bar{o}}$, and $R_h$, in the real quadrature operator representation: 
\begin{eqnarray}
\boldsymbol{x}_{co} &\equiv& \left[ 
\begin{array}{c}
\boldsymbol{q}_{co} \\ 
\boldsymbol{p}_{co}%
\end{array}
\right] \triangleq V_{n_{1}}\boldsymbol{\breve{a}}_{co}, \ \boldsymbol{x}_{%
\bar{c}\bar{o}} \equiv \left[ 
\begin{array}{c}
\boldsymbol{q}_{\bar{c}\bar{o}} \\ 
\boldsymbol{p}_{\bar{c}\bar{o}}%
\end{array}
\right] \triangleq V_{n_{2}} \boldsymbol{\breve{a}}_{\bar{c}\bar{o}},  \notag
\\
\boldsymbol{\tilde{x}}_{h} &\equiv& \left[ 
\begin{array}{c}
\boldsymbol{\tilde{q}}_{h} \\ 
\boldsymbol{\tilde{p}}_{h}%
\end{array}
\right] \triangleq V_{n_{3}}\boldsymbol{\breve{a}}_{h}.  \label{sept1_2}
\end{eqnarray}
Then, the following result is a direct consequence of Theorem \ref%
{thm:general_Kalman_1}, which gives the Kalman decomposition for the linear
quantum system (\ref{eq:real_sys_a})-(\ref{eq:real_sys_b}):

%%%%%%%%%%%%%%%%%%

\begin{theorem}
\label{thm:general_Kalman_3} Let $\tilde{S} \triangleq V_{n} \tilde{T}
V_{n}^{\dag}$, %%
%\begin{equation*}%\label{tilde_S}
%\tilde{S} \triangleq V_{n} \tilde{T} V_{n}^{\dag},
%\end{equation*}
%%
where $\tilde{T}$ is given by Eq. (\ref{tilde_T}). $\tilde{S}$ is a real
orthogonal and symplectic coordinate transformation that decomposes the
space of variables of the linear quantum system (\ref{eq:real_sys_a})-(\ref%
{eq:real_sys_b}), as follows: 
\begin{equation}  \label{real_Kalman_1}
\left[%
\begin{array}{cccccc}
\boldsymbol{\tilde{q}}_{h}^{\top} & \boldsymbol{q}_{co}^{\top} & \boldsymbol{%
q}_{\bar{c}\bar{o}}^{\top} & \boldsymbol{\tilde{p}}_{h}^{\top} & \boldsymbol{%
p}_{co}^{\top} & \boldsymbol{p}_{\bar{c}\bar{o}}^{\top}%
\end{array}%
\right]^{\top} = \tilde{S}^{\top} \boldsymbol{x}.
\end{equation}
\end{theorem}

\textit{Proof}: Firstly, since $\tilde{T}$ is Bogoliubov and $V_{n}$ is
unitary, $\tilde{S}= V_{n} \tilde{T} V_{n}^{\dag}$ in Eq. (\ref%
{real_Kalman_1}) is real symplectic. Secondly, $\tilde{S}$ is unitary
because it is a product of three unitary matrices. A real unitary matrix is
orthogonal. Thus, $\tilde{S}$ is a real orthogonal and symplectic coordinate
transformation. %\begin{equation} \label{spet2_TT}
%\tilde{T}=T\Lambda,
%\end{equation}
%where
%\begin{equation} \label{spet2_Lambda}
%\Lambda \triangleq \left[
%\begin{array}{cccccc}
%I_{n_{3}} & 0 & 0 & 0 & 0 & 0 \\
%0 & 0 & 0 & I_{n_{3}} & 0 & 0 \\
%0 & I_{n_{1}} & 0 & 0 & 0 & 0 \\
%0 & 0 & 0 & 0 & I_{n_{1}} & 0 \\
%0 & 0 & I_{n_{2}} & 0 & 0 & 0 \\
%0 & 0 & 0 & 0 & 0 & I_{n_{2}}%
%\end{array}%
%\right]
%\end{equation}
Finally, by Eqs. (\ref{complex_to_real_trans}), (\ref{complex_Kalman_1}), (%
\ref{sept2_co}), (\ref{sept2_barco}), and (\ref{sept2_h}), we get 
\begin{eqnarray*}
&&\tilde{S}^{\top} \boldsymbol{x} =\tilde{S}^\dag \boldsymbol{x} = V_n 
\tilde{T}^\dag V_n^\dag \boldsymbol{x} = V_n \tilde{T}^\dag \boldsymbol{%
\breve{a}} \\
&=& V_n \left[%
\begin{array}{c}
\boldsymbol{a}_{h} \\ 
\boldsymbol{a}_{co} \\ 
\boldsymbol{a}_{\bar{c}\bar{o}} \\ 
\boldsymbol{a}_{h}^{\#} \\ 
\boldsymbol{a}_{co}^{\#} \\ 
\boldsymbol{a}_{\bar{c}\bar{o}}^{\#} \\ 
\end{array}%
\right] = \left[ 
\begin{array}{c}
\boldsymbol{\tilde{q}}_{h} \\ 
\boldsymbol{q}_{co} \\ 
\boldsymbol{q}_{\bar{c}\bar{o}} \\ 
\boldsymbol{\tilde{p}}_{h} \\ 
\boldsymbol{p}_{co} \\ 
\boldsymbol{p}_{\bar{c}\bar{o}}%
\end{array}
\right],
\end{eqnarray*}
which is Eq. (\ref{real_Kalman_1}). \hfill $\blacksquare$

Now, we proceed to prove the analog of Theorem \ref{thm:general_Kalman_2},
namely Theorem \ref{thm:general_Kalman_4}. However, before we do this, we
introduce a new set of variables for the $h=c\bar{o} \cup \bar{c}o$
subspace, in the real quadrature operator representation. The reason for
this is that using these new variables, we can reveal more structure in the
real quadrature operator representation of Kalman canonical form system
matrices (\ref{eq:real_Kalman_sys_ABCD}) than in the creation-annihilation
operator representation (\ref{eq:complex_Kalman_sys_ABCD}).

To this end, let us define the matrix $\Pi \in \mathbb{C}^{2n_{3}\times
2n_{3}}$ by 
\begin{equation*}
\Pi \triangleq \left[ 
\begin{array}{cccc}
I_{n_{a}} & 0 & 0 & 0 \\ 
0 & 0 & 0 & -I_{n_{b}} \\ 
0 & 0 & I_{n_{a}} & 0 \\ 
0 & I_{n_{b}} & 0 & 0%
\end{array}%
\right].
\end{equation*}
It is easy to verify that $\Pi \Pi^{\top} = \Pi^{\top} \Pi = I_{2n_3}$, and $%
\Pi \mathbb{J}_{n_3} \Pi^{\top} = \Pi^{\top} \mathbb{J}_{n_3} \Pi = \mathbb{J%
}_{n_3}$, that is $\Pi$ is orthogonal and symplectic. Now, let 
\begin{equation}
\tilde{V}_{n_{3}} \triangleq \Pi V_{n_{3}},  \label{eq:tilde_V_n3}
\end{equation}
and define a new set of system variables for $R_{h}$ by 
\begin{equation}
\boldsymbol{x}_{h} \equiv \left[ 
\begin{array}{c}
\boldsymbol{q}_{h} \\ \hline
\boldsymbol{p}_{h}%
\end{array}%
\right] \triangleq \tilde{V}_{n_{3}} \boldsymbol{\breve{a}}_{h} = \Pi \left[ 
\begin{array}{c}
\boldsymbol{\tilde{q}}_{h} \\ 
\boldsymbol{\tilde{p}}_{h}%
\end{array}
\right],  \label{eq:QP}
\end{equation}
using Eqs. (\ref{sept1_2}) and (\ref{eq:tilde_V_n3}). Since $\Pi$ is real
symplectic, it follows that $\boldsymbol{q}_{h}$ and $\boldsymbol{p}_{h}$
are self-adjoint operators, and that $[\boldsymbol{q}_{h},\boldsymbol{q}%
_{h}^{\top}]=0$, $[\boldsymbol{p}_{h},\boldsymbol{p}_{h}^{\top}]=0$, and $[%
\boldsymbol{q}_{h},\boldsymbol{p}_{h}^{\top}]=\imath I_{n_{3}}$. Hence, $%
\boldsymbol{q}_{h,i}$ and $\boldsymbol{p}_{h,i}$ are conjugate observables
for $i=1,\ldots,n_{3}$. We find it preferable to work with $\boldsymbol{q}%
_{h}$ and $\boldsymbol{p}_{h}$, rather than $\boldsymbol{\tilde{q}}_{h}$ and 
$\boldsymbol{\tilde{p}}_{h}$, because they allow us to transform the linear
quantum system (\ref{eq:real_sys_a})-(\ref{eq:real_sys_b}) to the standard
Kalman canonical form, as to be given in Theorem \ref{thm:general_Kalman_4}.

To prove the analog of Theorem \ref{thm:general_Kalman_2}, namely Theorem %
\ref{thm:general_Kalman_4}, we need two lemmas. Lemma \ref%
{lem:structure_transfer} transforms the structure of the system matrices in
Eqs. (\ref{eq:complex_Kalman_ss})-(\ref{eq:complex_Kalman_io}) to the real
quadrature representation with variables $(\boldsymbol{q}_{h}, \boldsymbol{p}%
_{h}, \boldsymbol{x}_{co}, \boldsymbol{x}_{\bar{c}\bar{o}})$, and Lemma \ref%
{lem:S} establishes properties of the matrix that transforms the system to
this representation.

%%%%%%%%%%%%%%%%%%%%%%%%%

\begin{lemma}
\label{lem:structure_transfer} Let 
\begin{equation*}
\tilde{V}_{n} \triangleq 
\mathrm{diag}\left(\tilde{V}_{n_{3}}, V_{n_{1}}, V_{n_{2}}\right).
% \left[%
% \begin{array}{ccc}
% \tilde{V}_{n_{3}} &  & \parbox{12pt}{\Huge 0} \\ 
% & V_{n_{1}} &  \\ 
% \parbox{12pt}{\Huge 0} &  & V_{n_{2}}%
% \end{array}%
% \right].
\end{equation*}
Then, 
\begin{eqnarray}
\bar{A} &\triangleq& \tilde{V}_{n}\bar{\mathcal{A}}\tilde{V}_{n}^{\dag} = %
\left[%
\begin{array}{cc|c|c}
A_{h}^{11} & A_{h}^{12} & A_{12} & A_{13} \\ 
0 & A_{h}^{22} & 0 & 0 \\ \hline
0 & A_{21} & A_{co} & 0 \\ \hline
0 & A_{31} & 0 & A_{\bar{c}\bar{o}}%
\end{array}%
\right],  \notag \\
\bar{B} &\triangleq& \tilde{V}_{n}\bar{\mathcal{B}}V_{m}^{\dag} = \left[%
\begin{array}{c}
B_{h} \\ 
0 \\ \hline
B_{co} \\ \hline
0%
\end{array}%
\right],  \notag \\
\bar{C} &\triangleq& V_{m}\bar{\mathcal{C}} \tilde{V}_{n}^{\dag} = \left[%
\begin{array}{cc|c|c}
0 & C_{h} & C_{co} & 0%
\end{array}%
\right].  \label{eq:real_Kalman_sys_ABCD}
\end{eqnarray}
\end{lemma}

\textit{Proof}: It follows from the definitions of $\bar{A}$, $\bar{B}$, and 
$\bar{C}$ in Eq. (\ref{eq:real_Kalman_sys_ABCD}), along with Eq. (\ref%
{eq:complex_Kalman_sys_ABCD}), that $\bar{A}=\hat{T}^{\dag }\mathcal{A}\hat{T%
}$, $\bar{B}=\hat{T}^{\dag }\mathcal{B}V_{m}^{\dag }$, and $\bar{C}=V_{m}%
\mathcal{C}\hat{T}$, where 
\begin{eqnarray*}
\hat{T} &\triangleq &T\tilde{V}_{n}^{\dag }=\left[ 
\begin{array}{ccc}
T_{h}\tilde{V}_{n_{3}}^{\dag } & T_{co}V_{n_{1}}^{\dag } & T_{\bar{c}\bar{o}%
}V_{n_{2}}^{\dag } 
\end{array}%
\right]  \\
&\equiv &\left[ 
\begin{array}{ccc}
\hat{T}_{h} & \hat{T}_{co} & \hat{T}_{\bar{c}\bar{o}}
\end{array}%
\right] .
\end{eqnarray*}%
Since the columns of $T_{h}$, $T_{co}$ and $T_{\bar{c}\bar{o}}$ are
orthonormal bases of $R_{h}$, $R_{co}$ and $R_{\bar{c}\bar{o}}$,
respectively, and $\tilde{V}_{n_{3}}$, $V_{n_{1}}$, and $V_{n_{2}}$ are
unitary, the same is true for $\hat{T}_{h}$, $\hat{T}_{co}$ and $\hat{T}_{%
\bar{c}\bar{o}}$. Using the definitions of $T_{h}$ and $\tilde{V}_{n_{3}}$,
we can show that 
\begin{eqnarray*}
\hat{T}_{h} &=&T_{h}\tilde{V}_{n_{3}}^{\dag }=\frac{1}{\sqrt{2}}\left[ 
\begin{array}{cc|cc}
X & -\imath Y & \imath X & Y \\ 
X^{\#} & \imath Y^{\#} & -\imath X^{\#} & Y^{\#}%
\end{array}%
\right]  \\
&=: &\left[ 
\begin{array}{c|c}
\hat{T}_{c\bar{o}} & \hat{T}_{\bar{c}o}%
\end{array}%
\right] .
\end{eqnarray*}%
That is, the columns of $\hat{T}_{h}$ are the union of a basis for $R_{c\bar{%
o}}$, namely the columns of $\hat{T}_{c\bar{o}}=\bigl[%
\begin{smallmatrix}
X & -\imath Y \\ 
X^{\#} & \imath Y^{\#}%
\end{smallmatrix}%
\bigr]$, and a basis for $R_{c\bar{o}}$, namely the columns of $\hat{T}_{%
\bar{c}o}=\bigl[%
\begin{smallmatrix}
\imath X & Y \\ 
-\imath X^{\#} & Y^{\#}%
\end{smallmatrix}%
\bigr]$; see also Lemma \ref{lem:basis_Rh}. The structure of $\bar{A}$, $%
\bar{B}$, and $\bar{C}$ in Eq. (\ref{eq:real_Kalman_sys_ABCD}) then follows
from the invariance properties Eqs. (\ref{subset_a})-(\ref{subset_b}). For
example, $\mathcal{A}R_{c\bar{o}}\subset R_{c\bar{o}}$ implies $\hat{T}_{%
\bar{c}o}^{\dag }\mathcal{A}\hat{T}_{c\bar{o}}=0$. Hence, the $(2,1)$ block
of $\bar{A}$ is zero. Similarly, $\mathcal{A}R_{co}\subset R_{c\bar{o}%
}+R_{co}$ implies $\hat{T}_{\bar{c}o}^{\dag }\mathcal{A}\hat{T}_{co}=0$.
Hence, the $(2,3)$ block of $\bar{A}$ is zero. The rest of the block-zero
entries of $\bar{\mathcal{A}}$, $\bar{\mathcal{B}}$, and $\bar{\mathcal{C}}$
can be obtained similarly. \hfill $\blacksquare $

%%%%%%%%%%%%%%%%%%%

\begin{remark}
\label{rem:tf_BAE} The structure of the matrix $\bar A$ given in (\ref%
{eq:real_Kalman_sys_ABCD}) implies 
\begin{eqnarray}  \label{tf_real}
\lefteqn{\left[ \begin{array}{cc} 0 & A_{21} \\ 0 & A_{31} \end{array}\right]%
\left( sI-\left[ \begin{array}{cc} A_{h}^{11} & A_{h}^{12} \\ 0 & A_{h}^{22}
\\ \end{array}\right] \right) ^{-1}\left[ \begin{array}{cc} A_{12} &
A_{13}\\ 0 & 0 \end{array}\right]}  
\notag \\
&&=0. \hspace{6cm}
\end{eqnarray}
Also, it follows from Eqs. (\ref{eq:complex_Kalman_sys_ABCD}) and (\ref%
{eq:real_Kalman_sys_ABCD}) that 
\begin{eqnarray*}
\left[%
\begin{array}{cc}
A_{h}^{11} & A_{h}^{12} \\ 
0 & A_{h}^{22}%
\end{array}%
\right] &=& \tilde{V}_{n_{3}}\mathcal{A}_{h}\tilde{V}_{n_{3}}^\dagger; 
\notag \\
\left[%
\begin{array}{cc}
A_{12} & A_{13} \\ 
0 & 0%
\end{array}%
\right] &=& \tilde{V}_{n_{3}} \left[%
\begin{array}{cc}
\mathcal{A}_{12} & \mathcal{A}_{13}%
\end{array}%
\right] \left[%
\begin{array}{cc}
V_{n_{1}}^\dagger & 0 \\ 
0 & V_{n_{2}}^\dagger%
\end{array}%
\right];  \notag \\
\left[%
\begin{array}{cc}
0 & A_{21} \\ 
0 & A_{31}%
\end{array}%
\right] &=& \left[%
\begin{array}{cc}
V_{n_{1}} & 0 \\ 
0 & V_{n_{2}}%
\end{array}%
\right] \left[%
\begin{array}{c}
\mathcal{A}_{21} \\ 
\mathcal{A}_{31}%
\end{array}%
\right]\tilde{V}_{n_{3}}^\dagger.
\end{eqnarray*}
Then, since the matrices $\tilde{V}_{n_{3}}$, $V_{n_{1}}$ and $V_{n_{2}}$
are unitary, Eq. (\ref{tf_real}) implies that the condition (\ref{tf_complex}%
) is satisfied.
\end{remark}

%%%%%%%%%%%%%%%%%%%%

\begin{lemma}
\label{lem:S} Define $S \triangleq V_{n} T \tilde{V}_{n}^{\dag}$, %%
%\begin{equation*}%\label{S}
%S \triangleq V_{n} T \tilde{V}_{n}^{\dag},
%\end{equation*}
%%
where $T$ is defined in Eq. (\ref{T}). Then, $S$ is real, orthogonal and 
\emph{blockwise symplectic}; i.e., it satisfies 
\begin{equation}  \label{eq:block_sym}
S^{\top}\mathbb{J}_{n}S= 
\mathrm{diag} \left(\mathbb{J}_{n_{3}},\mathbb{J}_{n_{1}},\mathbb{J}_{n_{2}}\right).
% \left[%
% \begin{array}{ccc}
% \mathbb{J}_{n_{3}} &  & \parbox{12pt}{\Huge 0} \\ 
% & \mathbb{J}_{n_{1}} &  \\ 
% \parbox{12pt}{\Huge 0} &  & \mathbb{J}_{n_{2}}%
% \end{array}%
% \right].
\end{equation}
\end{lemma}

\textit{Proof:} First, notice that, 
\begin{eqnarray}
S^{\dag} \left[%
\begin{array}{c}
\boldsymbol{q} \\ 
\boldsymbol{p}%
\end{array}%
\right] = \tilde{V}_{n} T^{\dag} \boldsymbol{\breve{a}} = \tilde{V}_{n} %
\left[%
\begin{array}{c}
\boldsymbol{\breve{a}}_{h} \\ \hline
\boldsymbol{\breve{a}}_{co} \\ \hline
\boldsymbol{\breve{a}}_{\bar{c}\bar{o}}%
\end{array}%
\right] = \left[%
\begin{array}{c}
\boldsymbol{q}_{h} \\ 
\boldsymbol{p}_{h} \\ \hline
\boldsymbol{x}_{co} \\ \hline
\boldsymbol{x}_{\bar{c}\bar{o}}%
\end{array}%
\right].  \label{aug31_2}
\end{eqnarray}
Since $\boldsymbol{q}$, $\boldsymbol{p}$, $\boldsymbol{q}_{h}$, $\boldsymbol{%
p}_{h}$, $\boldsymbol{x}_{co}$, and $\boldsymbol{x}_{\bar{c}\bar{o}}$ are
all self-adjoint, $S$ is real. $S$ is also unitary, as a product of unitary
matrices, and hence it is orthogonal. Finally, using the equations $%
V_{n}^{\dag} \mathbb{J}_{n} V_{n}= -\imath J_{n} \Leftrightarrow V_{n} J_{n}
V_{n}^{\dag} = \imath \mathbb{J}_{n}$, and $T^{\dag} J_{n} T = \mathrm{diag}%
(J_{n_{3}},J_{n_{1}},J_{n_{3}})$, we have that 
\begin{eqnarray*}
S^{\top}\mathbb{J}_{n}S &=& - \imath 
\mathrm{diag}\left(\tilde{V}_{n_{3}}J_{n_{3}}\tilde{V}_{n_{3}}^{\dag}, V_{n_{1}}J_{n_{1}}V_{n_{1}}^{\dag},
V_{n_{2}}J_{n_{2}}V_{n_{2}}^{\dag}\right)
% \left[%
% \begin{array}{ccc}
% \tilde{V}_{n_{3}}J_{n_{3}}\tilde{V}_{n_{3}}^{\dag} &  & 
% \parbox{12pt}{\Huge
% 0} \\ 
% & V_{n_{1}}J_{n_{1}}V_{n_{1}}^{\dag} &  \\ 
% \parbox{12pt}{\Huge 0} &  & V_{n_{2}}J_{n_{2}}V_{n_{1}}^{\dag}%
% \end{array}%
% \right] 
\\
&=& 
\mathrm{diag}\left(\Pi\mathbb{J}_{n_{3}}\Pi^{\top},\mathbb{J}_{n_{1}},\mathbb{J}_{n_{2}}\right),
% \left[%
% \begin{array}{ccc}
% \Pi\mathbb{J}_{n_{3}}\Pi^{\top} &  & \parbox{12pt}{\Huge 0} \\ 
% & \mathbb{J}_{n_{1}} &  \\ 
% \parbox{12pt}{\Huge 0} &  & \mathbb{J}_{n_{2}}%
% \end{array}%
% \right],
\end{eqnarray*}
from which Eq. (\ref{eq:block_sym}) follows, because $\Pi\mathbb{J}%
_{n_{3}}\Pi^{\top} = \mathbb{J}_{n_{3}}$. \hfill $\blacksquare$

Now we can state the analog of Theorem \ref{thm:general_Kalman_2} in the
real quadrature operator representation.

\begin{figure}[tbph]
\centering
\includegraphics[width=0.4\textwidth]{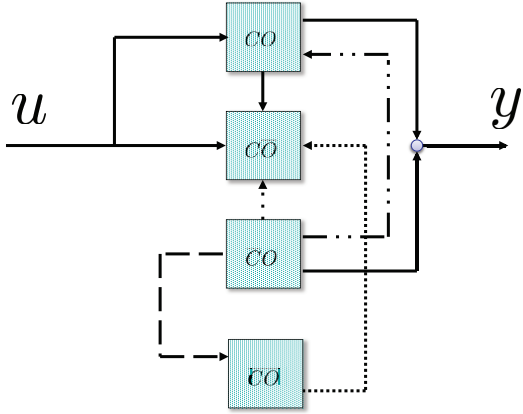}
\caption{Kalman decomposition of a linear quantum system. The solid lines
indicate that the blocks can be either controlled by the input $\boldsymbol{u%
}$ or observed via the output $\boldsymbol{y}$.}
\label{fig:kalman}
\end{figure}

\begin{theorem}
\label{thm:general_Kalman_4} The real orthogonal and blockwise symplectic
coordinate transformation 
\begin{eqnarray}
\left[%
\begin{array}{c}
\boldsymbol{q}_{h} \\ 
\boldsymbol{p}_{h} \\ \hline
\boldsymbol{x}_{co} \\ \hline
\boldsymbol{x}_{\bar{c}\bar{o}}%
\end{array}%
\right] = S^{\top} \boldsymbol{x}  \label{aug31_3}
\end{eqnarray}
transforms the linear quantum system (\ref{eq:real_sys_a})-(\ref%
{eq:real_sys_b}) into the form 
\begin{eqnarray}
\left[%
\begin{array}{c}
\boldsymbol{\dot{q}}_{h}(t) \\ 
\boldsymbol{\dot{p}}_{h}(t) \\ \hline
\boldsymbol{\dot{x}}_{co}(t) \\ \hline
\boldsymbol{\dot{x}}_{\bar{c}\bar{o}}(t)%
\end{array}%
\right] &=& \bar{A} \left[%
\begin{array}{c}
\boldsymbol{q}_{h}(t) \\ 
\boldsymbol{p}_{h}(t) \\ \hline
\boldsymbol{x}_{co}(t) \\ \hline
\boldsymbol{x}_{\bar{c}\bar{o}}(t)%
\end{array}%
\right] + \bar{B} \boldsymbol{u}(t),  \label{real_Kalman_ss} \\
\boldsymbol{y}(t) &=& \bar{C} \left[%
\begin{array}{c}
\boldsymbol{q}_{h}(t) \\ 
\boldsymbol{p}_{h}(t) \\ \hline
\boldsymbol{x}_{co}(t) \\ \hline
\boldsymbol{x}_{\bar{c}\bar{o}}(t)%
\end{array}%
\right] +\boldsymbol{u}(t),  \label{real_Kalman_io}
\end{eqnarray}
where matrices $\bar{A},\bar{B},\bar{C}$ were given in Eq. (\ref%
{eq:real_Kalman_sys_ABCD}). After a re-arrangement, the system (\ref%
{real_Kalman_ss})-(\ref{real_Kalman_io}) becomes 
\begin{eqnarray}
\left[%
\begin{array}{c}
\boldsymbol{\dot{q}}_{h}(t) \\ 
\boldsymbol{\dot{x}}_{co}(t) \\ 
\boldsymbol{\dot{x}}_{\bar{c}\bar{o}}(t) \\ 
\boldsymbol{\dot{p}}_{h}(t)%
\end{array}%
\right] &=& \left[ 
\begin{array}{cccc}
A_{h}^{11} & A_{12} & A_{13} & A_{h}^{12} \\ 
0 & A_{co} & 0 & A_{21} \\ 
0 & 0 & A_{\bar{c}\bar{o}} & A_{31} \\ 
0 & 0 & 0 & A_{h}^{22}%
\end{array}%
\right] \left[%
\begin{array}{c}
\boldsymbol{q}_{h}(t) \\ 
\boldsymbol{x}_{co}(t) \\ 
\boldsymbol{x}_{\bar{c}\bar{o}}(t) \\ 
\boldsymbol{p}_{h}(t)%
\end{array}%
\right]  \notag \\
&+& \left[ 
\begin{array}{c}
B_{h} \\ 
B_{co} \\ 
0 \\ 
0%
\end{array}%
\right] \boldsymbol{u}(t),  \label{real_Kalman_ss_2} \\
\boldsymbol{y}(t)&=& [0 \ C_{co} \ 0 \ C_{h}] \left[%
\begin{array}{c}
\boldsymbol{q}_{h}(t) \\ 
\boldsymbol{x}_{co}(t) \\ 
\boldsymbol{x}_{\bar{c}\bar{o}}(t) \\ 
\boldsymbol{p}_{h}(t)%
\end{array}%
\right] +\boldsymbol{u}(t).  \label{real_Kalman_io_2}
\end{eqnarray}%
A block diagram for the system (\ref{real_Kalman_ss})-(\ref{real_Kalman_io})
is given in Fig.~\ref{fig:kalman}.
\end{theorem}

\textit{Proof:} By Lemma \ref{lem:S}, $S$ is real. Therefore, Eq. (\ref%
{aug31_3}) is a re-statement of Eq. (\ref{aug31_2}). As a result, Theorem %
\ref{thm:general_Kalman_4} follows from the transformation (\ref%
{complex_to_real_trans}), the transformation (\ref{aug31_2}), Theorem \ref%
{thm:general_Kalman_2}, and Lemma \ref{lem:structure_transfer}. \hfill $%
\blacksquare$

%%%%%%%%%%%%%%%%%%%%

%%%%%%%%%%%%%%%%%%%%

\begin{remark}
\label{rem:eigenvalues} From Eq. (\ref{real_Kalman_ss_2}), we conclude that $%
\sigma(A)=\sigma(\bar{A})=\sigma(A_{co}) \cup \sigma(A_{\bar{c}\bar{o}})
\cup \sigma(A_{h}^{11}) \cup \sigma(A_{h}^{22})$. The definitions of $A_{co}$%
, $A_{\bar{c}\bar{o}}$, $A_{h}^{11}$, and $A_{h}^{22}$ in Eq. (\ref%
{eq:real_Kalman_sys_ABCD}), also imply that (\ref{eigenvalues}) is satisfied.
\end{remark}

%%%%%%%%%%%%%%%%%%%%

\begin{remark}
\label{rem:sept22_1} It can be seen from (\ref{real_Kalman_ss})-(\ref%
{real_Kalman_io}) or (\ref{real_Kalman_ss_2})-(\ref{real_Kalman_io_2}) that, 
$\boldsymbol{q}_{h,i}, i=1,\ldots,n_{3}$, are controllable but unobservable,
while $\boldsymbol{p}_{h,i}, i=1,\ldots,n_{3}$, are observable but
uncontrollable. We see that every $c\bar{o}$ variable must have an
associated $\bar{c}o$ variable. That is, they appear in conjugate pairs.
Notice that the variables $\boldsymbol{p}_{h,i}$ commute with each other at
equal times. Also, as seen from Eq. (\ref{real_Kalman_ss_2}), they evolve
without any influence from the inputs or other system variables. As shown in 
\cite{TC12}, the set of $\boldsymbol{p}_{h,i}$, $i=1,...,n_3$, is a QMFS
satisfying Eq. (\ref{eq:QMFS}), see Definition \ref{QMFS} and References 
\cite{TC12} and \cite{WC13}. This implies that each $\boldsymbol{p}_{h,i}$
satisfies Eq. (\ref{eq:QND}), hence, each $\boldsymbol{p}_{h,i}$ is a QND
variable, see Definition \ref{def:QND} and References \cite{HMW95} and \cite%
{NY14}. Moreover, $\boldsymbol{x}_{\bar{c}\bar{o},i}$, $i=1,\ldots,n_{2}$,
are DF modes, see Definition \ref{def:DFS} and References \cite{DFK+12},  \cite{NY13}, \cite{NY14}, and \cite{GZ15}. Finally, we
emphasize the fact that, not all linear quantum systems contain QND
variables and DF modes. Indeed, as in the classical case, for a specific
system, some of the subsystems may not be present; see Examples \ref%
{ex:july28_complex}, \ref{ex:july28} and Section \ref{sec:apps} for more
details.
\end{remark}

\begin{remark}
\label{rem:BAE} In (\ref{real_Kalman_ss_2})-(\ref{real_Kalman_io_2}), recall
that $\boldsymbol{u}=\bigl[%
\begin{smallmatrix}
\boldsymbol{q}_{\mathrm{in}} \\ 
\boldsymbol{p}_{\mathrm{in}}%
\end{smallmatrix}%
\bigr]$ and $\boldsymbol{y}=\bigl[%
\begin{smallmatrix}
\boldsymbol{q}_{\mathrm{out}} \\ 
\boldsymbol{p}_{\mathrm{out}}%
\end{smallmatrix}%
\bigr]$, as defined in Eq. (\ref{complex_to_real_trans}). Partition the
matrices $B_{co}$ and $C_{co}$ accordingly as 
\begin{equation*}
B_{co}=\left[ 
\begin{array}{cc}
B_{co,q} & B_{co,p}%
\end{array}%
\right] ,~C_{co}=\left[ 
\begin{array}{c}
C_{co,q} \\ 
C_{co,p}%
\end{array}%
\right] .
\end{equation*}%
If the transfer function $%
\begin{smallmatrix}
\Xi _{\boldsymbol{p}_{\mathrm{in}}\rightarrow \boldsymbol{q}_{\mathrm{out}%
}}(s)=C_{co,q}(sI-A_{co})^{-1}B_{co,p}=0%
\end{smallmatrix}%
$, then the input noise quadrature $\boldsymbol{p}_{\mathrm{in}}$ has no
influence on the output quadrature $\boldsymbol{q}_{\mathrm{out}}$. In this
case, the system (\ref{real_Kalman_ss_2})-(\ref{real_Kalman_io_2}) realizes
the BAE measurement of the output $\boldsymbol{q}_{\mathrm{out}}$ with
respect to the input $\boldsymbol{p}_{\mathrm{in}}$. Similarly, if the
transfer function $%
\begin{smallmatrix}
\Xi _{\boldsymbol{q}_{\mathrm{in}}\rightarrow \boldsymbol{p}_{\mathrm{out}%
}}(s)=C_{co,p}(sI-A_{co})^{-1}B_{co,q}=0%
\end{smallmatrix}%
$, then the system (\ref{real_Kalman_ss_2})-(\ref{real_Kalman_io_2})
realizes the BAE measurement of the output $\boldsymbol{p}_{\mathrm{out}}$
with respect to the input $\boldsymbol{q}_{\mathrm{in}}$. These properties
will be demonstrated in Example \ref{ex:july28} and  Example \ref{ex:woolley}. In the special case when there is no mode $\boldsymbol{x}%
_{co}$ in the system, we have 
\begin{equation*}
\left[ 
\begin{array}{c}
\boldsymbol{q}_{\mathrm{out}} \\ 
\boldsymbol{p}_{\mathrm{out}}%
\end{array}%
\right] =C_{h}\boldsymbol{p}_{h}+\left[ 
\begin{array}{c}
\boldsymbol{q}_{\mathrm{in}} \\ 
\boldsymbol{p}_{\mathrm{in}}%
\end{array}%
\right] .
\end{equation*}%
Since $\boldsymbol{p}_{h}$ is uncontrollable, it is clear that in this case,
BAE measurements are naturally achieved.
\end{remark}

Finally, we have the following result as a corollary of Theorems \ref%
{thm:general_Kalman_4} and \ref{thm:passive_DFS_characterization}.

%%%%%%%%%%%%%%%%%%%%%

\begin{corollary}
\label{cor:passive_Kalman} The Kalman canonical form of a passive linear
quantum system in the real quadrature operator representation can be
achieved by a real orthogonal transformation, and is as follows: 
\begin{eqnarray}
\left[ 
\begin{array}{c}
\boldsymbol{\dot{x}}_{co}(t) \\ 
\boldsymbol{\dot{x}}_{\bar{c}\bar{o}}(t)%
\end{array}%
\right]  &=&\left[ 
\begin{array}{cc}
A_{co} & 0 \\ 
0 & A_{\bar{c}\bar{o}}%
\end{array}%
\right] \left[ 
\begin{array}{c}
\boldsymbol{x}_{co}(t) \\ 
\boldsymbol{x}_{\bar{c}\bar{o}}(t)%
\end{array}%
\right]   \notag \\
&+&\left[ 
\begin{array}{c}
B_{co} \\ 
0%
\end{array}%
\right] \boldsymbol{u}(t),  \label{Kal_ss_feb15b} \\
\boldsymbol{y}(t) &=&\left[ 
\begin{array}{cc}
C_{co} & 0%
\end{array}%
\right] \left[ 
\begin{array}{c}
\boldsymbol{x}_{co}(t) \\ 
\boldsymbol{x}_{\bar{c}\bar{o}}(t)%
\end{array}%
\right] +\boldsymbol{u}(t).  \label{Kal_io_feb15b}
\end{eqnarray}%
Here, all eigenvalues of the matrix $A_{\bar{c}\bar{o}}$ are located on the
imaginary axis, and have geometric multiplicity one. Also, the real parts of
the eigenvalues of the matrix $\tilde{A}_{co}$ are strictly negative.
\end{corollary}

We end this subsection with an illustrative example.

\begin{example}
\label{ex:july28} For the system in Example \ref{ex:july28_complex}, the
Hamiltonian $\boldsymbol{H}$ and the coupling operator $\boldsymbol{L}$ in
Eqs. (\ref{sept3_H})-(\ref{sept3_L}) are given in the real quadrature
representation of the system by $\boldsymbol{H}=2\boldsymbol{q}_{1}%
\boldsymbol{q}_{2}$ and $\boldsymbol{L}=\frac{1}{\sqrt{2}}(\boldsymbol{q}%
_{1}+\imath \boldsymbol{p}_{1})$, respectively. %\begin{eqnarray*}
%\boldsymbol{H} &=&2\boldsymbol{%
%q}_{1}\boldsymbol{q}_{2}-\imath(\boldsymbol{p}_{1}\boldsymbol{q}_{2}+\boldsymbol{q%
%}_{1}\boldsymbol{p}_{2}), \\
%\boldsymbol{L} &=&\frac{\boldsymbol{q}_{1}+\imath \boldsymbol{p}_{1}}{\sqrt{2}}.
%\end{eqnarray*}
By applying Theorem \ref{thm:general_Kalman_4}, we find that the system
variables in the real quadrature representation form of the Kalman
decomposition are given by $\boldsymbol{q}_{h}=-\boldsymbol{p}_{2},\ 
\boldsymbol{p}_{h}=\boldsymbol{q}_{2},\ \boldsymbol{q}_{co}=\boldsymbol{q}%
_{1},\ \boldsymbol{p}_{co}=\boldsymbol{p}_{1}$. %\[
%\boldsymbol{q}_{h} = -\boldsymbol{p}_{2}, \   \boldsymbol{p}_{h} = \boldsymbol{q}_{2}, \  \boldsymbol{q}_{co} =  \boldsymbol{q}_{1},  \ \boldsymbol{p}_{co} = \boldsymbol{p}_{1}.
%\]
%\begin{equation*}
%\left[
%\begin{array}{c}
%\boldsymbol{q}_{h} \\
%\boldsymbol{p}_{h} \\
%\boldsymbol{q}_{co} \\
%\boldsymbol{p}_{co}%
%\end{array}%
%\right] =\left[
%\begin{array}{c}
%-\boldsymbol{p}_{2} \\
%\boldsymbol{q}_{2} \\
%\boldsymbol{q}_{1} \\
%\boldsymbol{p}_{1}%
%\end{array}%
%\right].
%\end{equation*}%
Also, the corresponding QSDEs are as follows: %\begin{eqnarray*}
\begin{eqnarray*}
\dot{\boldsymbol{p}}_{2}(t) &=&-2\boldsymbol{q}_{1}(t), \\
\dot{\boldsymbol{q}}_{1}(t) &=&-0.5\boldsymbol{q}_{1}(t)-\boldsymbol{q}_{%
\mathrm{in}}(t), \\
\dot{\boldsymbol{p}}_{1}(t) &=&-0.5\boldsymbol{p}_{1}(t)-2\boldsymbol{q}%
_{2}(t)-\boldsymbol{p}_{\mathrm{in}}(t), \\
\dot{\boldsymbol{q}}_{2}(t) &=&0, \\
\boldsymbol{q}_{\mathrm{out}}(t) &=&\boldsymbol{q}_{1}(t)+\boldsymbol{q}_{%
\mathrm{in}}(t), \\
\boldsymbol{p}_{\mathrm{out}}(t) &=&\boldsymbol{p}_{1}(t)+\boldsymbol{p}_{%
\mathrm{in}}(t).
\end{eqnarray*}%
It can be readily shown that

\begin{description}
\item[(i)] $\boldsymbol{p}_{2}$ is controllable but unobservable, while $%
\boldsymbol{q}_{2}$ is observable but uncontrollable. So, $\boldsymbol{q}%
_{2} $ is a QND variable.

\item[(ii)] Because the transfer function $\Xi_{\boldsymbol{q}_{\mathrm{in}}
\to \boldsymbol{p}_{\mathrm{out}}}(s)=0$, the system realizes a BAE measurement
of $\boldsymbol{p}_{\mathrm{out}}$ with respect to $\boldsymbol{q}_{\mathrm{%
in}}$.

\item[(iii)] Similarly, the system realizes a BAE measurement of $%
\boldsymbol{q}_{\mathrm{out}}$ with respect to $\boldsymbol{p}_{\mathrm{in}}$%
.
\end{description}
\end{example}

\subsection{Some special cases of the Kalman decomposition}

\label{subsec:special_cases}

In this subsection, we study two special cases of the Kalman decomposition.

\begin{proposition}
\label{prop:Rc} If $\mathrm{Ker}(O_{s})$ is an invariant space of $\Omega$,
then $\mathcal{A}_{13}=0$ and $\mathcal{A}_{31}=0$ in Eq. (\ref%
{eq:complex_Kalman_sys_ABCD}).
\end{proposition}

\textit{Proof:} Suppose $x\in $ $R_{\bar{c}\bar{o}}$. Then $O _{s}x=0 $. As
a result, $O_{s}\mathcal{A}x=-\imath O_{s}J_{n}\Omega x-\frac{1}{2}O _{s}%
\mathcal{C}^{\flat }\mathcal{C}x=0$. %\begin{equation*}
%O_{s}{\cal A}x=-\imath O_{s}J_{n}\Omega x-\frac{1}{2}O _{s}{\cal C}^{\flat }{\cal C}x=0.
%\end{equation*}
That is, $\mathcal{A}x\in \mathrm{Ker}(O_{s})$. On the other hand, if $%
\mathrm{Ker}(O_{s})$ is an invariant space of $\Omega $, then $\Omega x\in $ 
$\mathrm{Ker}(O_{s})$ for all $x\in $ $\mathrm{Ker}( O_{s})$. As a result, $%
O_{s}J_{n}\mathcal{A}x=-\imath O_{s}J_{n}J_{n}\Omega x-\frac{1}{2}O_{s}J_{n}%
\mathcal{C}^{\flat }\mathcal{C}x=-\imath O_{s}\Omega x=0$. 
%\begin{equation*}
%O_{s}J_{n}{\cal A}x=-\imath O_{s}J_{n}J_{n}\Omega x-\frac{1}{2}%
%O_{s}J_{n}{\cal C}^{\flat }{\cal C}x=-\imath O_{s}\Omega x=0.
%\end{equation*}
That is, $\mathcal{A}x\in \mathrm{Ker}(O_{s}J_{n})$. Consequently, $\mathcal{%
A}R_{\bar{c }\bar{o}}\subset R_{\bar{c}\bar{o}}$. Hence, $\mathcal{A}_{13}=0$%
. Next we show that $\mathcal{A}_{31}=0$. For $x\in $ $R_{\bar{c}\bar{o}}$,
by Eq. (\ref{perp_4}) we have $J_{n}x\in $ $J_{n}R_{\bar{c}\bar{o}}=R_{\bar{c%
}\bar{o} }$. Consequently, $\mathcal{A}^\dag x =\imath\Omega J_n x -\frac{1}{%
2}\mathcal{C}^\dag J_m \mathcal{C} J_n x = \imath \Omega J_n x \in R_{\bar{c}%
\bar{o}}$. %\begin{equation*}
%{\cal A}^\dag x =\imath\Omega J_n x -\frac{1}{2}{\cal C}^\dag J_m {\cal C} J_n x = \imath \Omega J_n x \in R_{\bar{c}\bar{o}}.
%\end{equation*}
So we have 
\begin{equation}
\mathcal{A}^{\dagger }R_{\bar{c}\bar{o}}\subset R_{\bar{c}\bar{o}}.
\label{eq:temp_b}
\end{equation}
Given $x\in R_{\bar{c}o}+R_{c\bar{o}}+R_{co}$, let $\mathcal{A}x=y_{1}+y_{2}$
where $y_{1}\in R_{\bar{c}\bar{o}}$ while $y_{2}\in (R_{\bar{c}\bar{o}%
})^{\bot }=R_{ \bar{c}o}+R_{c\bar{o}}+R_{co}$. Then $y_{1}^{\dagger }%
\mathcal{A}x=y_{1}^{\dagger }y_{1}+y_{1}^{\dagger }y_{2}=y_{1}^{\dagger
}y_{1}.$ However, by Eq. (\ref{eq:temp_b}), we have $\mathcal{A}^{\dagger
}y_{1}\in R_{\bar{c}\bar{o}}$, and hence $y_{1}^{\dagger }\mathcal{A}x=(%
\mathcal{A}^{\dagger }y_{1})^{\dagger }x=0$. As a result, $y_{1}^{\dagger
}y_{1}=0$, i.e., $y_{1}=0$ and $\mathcal{A}x \in R_{ \bar{c}o}+R_{c\bar{o}%
}+R_{co}$. Thus, we have 
\begin{equation*}
\mathcal{A}(R_{\bar{c}o}+R_{c\bar{o}})\subset \mathcal{A}(R_{\bar{c}o}+R_{c%
\bar{o}}+R_{co})\subset (R_{ \bar{c}o}+R_{c\bar{o}}+R_{co}).
\end{equation*}
This implies $\mathcal{A}_{31}=0$. \hfill $\blacksquare $

\begin{remark}
In some sense, Proposition \ref{prop:Rc} slightly relaxes the condition that 
$\mathrm{Ker}(O_{s})\cap \mathrm{Ker}(O_{s}J_{n})$ is an invariant subspace
of $\Omega $ in \cite[Lemma 1]{GZ15}.
\end{remark}

%%%%%%%%%%%%%%%%%%%%%

\begin{proposition}
\label{prop:co} If 
\begin{equation}
\mathrm{Ker}(\mathcal{C})^{\perp } \perp \ \mathrm{Ker}\left( O
_{s}J_{n}\right) ,  \label{eq:temp3}
\end{equation}
then $\mathcal{B}_{h}=0$ and $\mathcal{C}_{h}=0$ in Eq. (\ref%
{eq:complex_Kalman_sys_ABCD}).
\end{proposition}

\textit{Proof:} Eq. (\ref{eq:temp3}) can be restated as $\mathrm{Ker}(%
\mathcal{C})^{\perp } \perp R_{\bar{c}o} \oplus R_{\bar{c}\bar{o}}$, which
implies $\mathrm{Ker}(\mathcal{C})^{\perp } \perp R_{\bar{c}o}$. Since $%
\mathrm{Im}(\mathcal{B}) =\mathrm{Im}(-J_{n}\mathcal{C}^{\dagger
}J_{m})=J_{n}\mathrm{Im} (\mathcal{C}^{\dagger })=J_{n}\mathrm{Ker}(\mathcal{%
C})^{\perp }$, we have that $\mathrm{Im}(\mathcal{B}) \perp J_{n} R_{\bar{c}%
o}= R_{c\bar{o}}$. Hence, $\mathcal{B}_{h}=0$. Also, Eq. (\ref{eq:temp3})
implies $\mathrm{Ker}(\mathcal{C})^{\perp} \subseteq (R_{\bar{c}o} \oplus R_{%
\bar{c}\bar{o}})^{\perp}$, and equivalently $\mathrm{Ker}(\mathcal{C})
\supseteq R_{\bar{c}o} \oplus R_{\bar{c}\bar{o}}$. Then, $\mathrm{Ker}(%
\mathcal{C}) \supseteq R_{\bar{c}o}$, which implies $\mathcal{C}_{h}=0$.
\hfill $\blacksquare$

\section{Applications}

\label{sec:apps}

In this section, we apply the Kalman decomposition theory developed to two
physical systems.

\begin{figure}[tbph]
\centering
\includegraphics[width=0.33\textwidth]{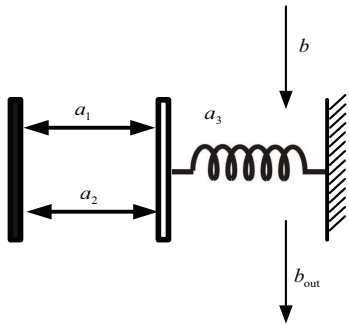}
\caption{An opto-mechanical system}
\label{fig:opto}
\end{figure}

\begin{example}
\label{ex:science_1} In this example, we investigate an opto-mechanical
system, as shown in Fig. \ref{fig:opto}. The optical cavity has two optical
modes, $\boldsymbol{a}_{1}$ and $\boldsymbol{a}_{2}$. The cavity is coupled
to a mechanical oscillator with mode $\boldsymbol{a}_{3}$, whose resonant
frequency is $\omega _{m}$. We ignore the optical damping, but keep the
mechanical damping as represented by $\boldsymbol{b}$ in  Fig. \ref{fig:opto}.  (Although the external mode $\boldsymbol{b}$ is thermal noise \cite{DFK+12}, we treat it here as a general quantum input, because our purpose is only to demonstrate our results.)  The coupling operator of the system is $\boldsymbol{L}=%
\sqrt{\kappa} \boldsymbol{a}_{3}$, where $\kappa >0$ is a coupling constant. 
Denote the optical detunings for $\boldsymbol{a}_{1}$ and $\boldsymbol{a}%
_{2} $ as $\Delta _{1}$ and $\Delta _{2}$, respectively. The Hamiltonian of
the system is given by 
\begin{eqnarray}
\boldsymbol{H} &=&\lambda _{1}\frac{\boldsymbol{a}_{1}+\boldsymbol{a}%
_{1}^{\ast }}{\sqrt{2}} \frac{\boldsymbol{a}_{3}+\boldsymbol{a}_{3}^{\ast }}{%
\sqrt{2}}+\lambda _{2} \frac{\boldsymbol{a}_{2}+\boldsymbol{a}_{2}^{\ast }}{%
\sqrt{2}}\frac{ \boldsymbol{a}_{3}+\boldsymbol{a}_{3}^{\ast }}{\sqrt{2}} 
\notag \\
&& -\Delta _{1}\boldsymbol{a}_{1}^{\ast }\boldsymbol{a}_{1}-\Delta _{2} 
\boldsymbol{a}_{2}^{\ast }\boldsymbol{a}_{2}+\omega _{m}\boldsymbol{a}
_{3}^{\ast }\boldsymbol{a}_{3},  \label{aug20_1}
\end{eqnarray}
where $\lambda _{1},\lambda _{2}  >0$ are the opto-mechanical
couplings. In the following, we discuss three cases of opto-mechanical
couplings, \cite[Sec. III]{AKM14}. Also, we let 
\begin{equation}  \label{rho12}
\lambda=\sqrt{\lambda _{1}^{2}+\lambda _{2}^{2}},~~\rho_1=\lambda_{1}/%
\lambda, ~~ \rho_2=\lambda_{2}/\lambda.
\end{equation}

\textit{Case 1: Red-detuned regime} In this case, the detuning between the laser frequency and  both cavity modes is negative. Moreover, we assume   $\Delta _{1}=\Delta_{2}=-\omega _{m}$. In this regime, the existence of an opto-mechanical dark
mode has been experimentally demonstrated in \cite{DFK+12}. The
opto-mechanical dark mode is a coherent superposition of the two optical
modes $\boldsymbol{a}_1$ and $\boldsymbol{a}_2$, and is decoupled from the
mechanical mode $\boldsymbol{a}_3$. Therefore, it is immune to thermal
noise, the major source of decoherence in this type of opto-mechanical
systems. In what follows, we apply the theory proposed in this paper to
derive the opto-mechanical dark mode in \cite{DFK+12}. In the rotating frame 
$\boldsymbol{a} _{1}(t)\rightarrow \boldsymbol{a}_{1}(t)e^{\imath\omega
_{m}t}$, $\boldsymbol{a}_{2}(t)\rightarrow \boldsymbol{a}_{2}(t)e^{\imath%
\omega _{m}t}$, and $\boldsymbol{a}_{3}(t)\rightarrow \boldsymbol{a}%
_{3}(t)e^{ \imath\omega _{m}t}$ (see, e.g., \cite[Eq. (31)]{AKM14}), the
Hamiltonian (\ref{aug20_1}) can be approximated by 
\begin{eqnarray*}
\boldsymbol{H}_{R} &=&\omega _{m}(\boldsymbol{a}_{1}^{\ast }\boldsymbol{a}%
_{1}+ \boldsymbol{a}_{2}^{\ast }\boldsymbol{a}_{2}+\boldsymbol{a}_{3}^{\ast
} \boldsymbol{a}_{3}) \\
&& +\frac{\lambda _{1}}{2}(\boldsymbol{a}_{1}\boldsymbol{a}_{3}^{\ast }+ 
\boldsymbol{a}_{1}^{\ast }\boldsymbol{a}_{3})+\frac{\lambda _{2}}{2}( 
\boldsymbol{a}_{2}\boldsymbol{a}_{3}^{\ast }+\boldsymbol{a}_{2}^{\ast } 
\boldsymbol{a}_{3}).  \notag
\end{eqnarray*}
In this case, the system is passive.  The
coordinate transformation 
\begin{equation*}
\left[ 
\begin{array}{c}
\boldsymbol{a}_{DF} \\ \hline
\boldsymbol{a}_{D}%
\end{array}
\right] =T^{\dag }\boldsymbol{\breve{a}}=\left[ 
\begin{array}{c}
\rho _{2}\boldsymbol{a}_{1}-\rho _{1}\boldsymbol{a}_{2} \\ \hline
\rho _{1}\boldsymbol{a}_{1}+\rho _{2}\boldsymbol{a}_{2} \\ 
\boldsymbol{a}_{3}%
\end{array}
\right] 
\end{equation*}
yields the following Kalman decomposition: 
\begin{eqnarray*}
\boldsymbol{\dot{a}}_{DF}(t) &=&-\imath\omega _{m}\boldsymbol{a}_{DF}(t), \\
\boldsymbol{\dot{a}}_{D}(t) &=& -\left[ 
\begin{array}{cc}
\imath\omega _{m} & \imath\frac{\lambda}{2} \\ 
\imath\frac{\lambda}{2} & \frac{\kappa}{2} + \imath\omega _{m}%
\end{array}
\right] \boldsymbol{a}_{D}(t)- \left[ 
\begin{array}{c}
0 \\ 
\sqrt{\kappa }%
\end{array}
\right] \boldsymbol{b}(t) \\
\boldsymbol{b}_{\mathrm{out}}(t) &=& \left[ 
\begin{array}{cc}
0 & \sqrt{\kappa }%
\end{array}
\right] \boldsymbol{a}_{D}(t)+\boldsymbol{b}(t).
\end{eqnarray*}
Clearly, $\boldsymbol{a}_{DF}$ is a DF mode (which is denoted $\boldsymbol{%
\hat{a}}_{D}$ in \cite{DFK+12}). It is a linear combination of the two cavity modes and is decoupled from the mechanical mode, thus being immune from the mechanical damping. This phenomenon has been observed in \cite{WC12a}, where the mode has been called ``mechanically dark''. Finally, in the real quadrature operator
representation, the DF mode is 
\begin{equation}
V_{1}\left[ 
\begin{array}{c}
\boldsymbol{a}_{DF} \\ 
\boldsymbol{a}_{DF}^{\ast }%
\end{array}
\right] = \left[ 
\begin{array}{c}
\rho _{2}\boldsymbol{q}_{1}-\rho _{1}\boldsymbol{q}_{2} \\ 
\rho _{2}\boldsymbol{p}_{1}-\rho _{1}\boldsymbol{p}_{2}%
\end{array}
\right] .  \label{aug17_1}
\end{equation}

\textit{Case 2: Blue-detuned regime} In this case,  the detuning between the laser frequency and both cavity modes is positive.  Moreover, we assume  $\Delta _{1}=\Delta
_{2}=\omega _{m}$. Under the rotating frame approximation $\boldsymbol{a}
_{1}(t)\rightarrow \boldsymbol{a}_{1}(t)e^{-\imath\omega _{m}t}$, $%
\boldsymbol{a}_{2}(t)\rightarrow \boldsymbol{a}_{2}(t)e^{-\imath\omega
_{m}t} $, and $\boldsymbol{a}_{3}(t)\rightarrow \boldsymbol{a}_{3}(t)e^{
\imath\omega _{m}t}$ (see, e.g., \cite[Eq. (32)]{AKM14}), the Hamiltonian (%
\ref{aug20_1}) can be approximated by 
\begin{eqnarray*}
\boldsymbol{H}_{B}&=&\lambda _{1}\frac{\boldsymbol{a}_{1}\boldsymbol{a}_{3}+%
\boldsymbol{a} _{1}^{\ast }\boldsymbol{a}_{3}^{\ast }}{2}+\lambda _{2}\frac{%
\boldsymbol{a} _{2}\boldsymbol{a}_{3}+\boldsymbol{a}_{2}^{\ast }\boldsymbol{a%
}_{3}^{\ast }}{ 2}  \notag \\
&& -\omega _{m}\boldsymbol{a}_{1}^{\ast }\boldsymbol{a}_{1}-\omega _{m} 
\boldsymbol{a}_{2}^{\ast }\boldsymbol{a}_{2}+\omega _{m}\boldsymbol{a}
_{3}^{\ast }\boldsymbol{a}_{3}.
\end{eqnarray*}
In this case, we find that there are no $\bar{c}o$ or $c\bar{o}$ subsystems.
%The unitary and blockwise Bogoliubov transformation matrix $T$ in Eq. (\ref%
%{T}) is given by 
%\begin{eqnarray*}
%T = [T_{co}\ |T_{\bar{c}\bar{o}}] = \left[ 
%\begin{array}{cccc|cc}
%0 & \rho_{1} & 0 & 0 & \rho_{2} & 0 \\ 
%0 & \rho_{2} & 0 & 0 & - \rho_{1} & 0 \\ 
%1 & 0 & 0 & 0 & 0 & 0 \\ 
%0 & 0 & 0 & \rho_{1} & 0 & \rho_{2} \\ 
%0 & 0 & 0 & \rho_{2} & 0 & -\rho_{1} \\ 
%0 & 0 & 1 & 0 & 0 & 0%
%\end{array}
%\right],
%\end{eqnarray*}
%where $\rho_{1}$ and $\rho_{2}$ are defined as in (\ref{rho12}). Thus using
%Eq. (\ref{aug31_3}), we obtain 
By Theorem \ref{thm:general_Kalman_4},
\begin{equation*}
\boldsymbol{x}_{co} = \left[ 
\begin{array}{c}
\boldsymbol{q}_{3} \\ 
\rho _{1}\boldsymbol{q}_{1}+\rho _{2}\boldsymbol{q}_{2} \\ 
\boldsymbol{p}_{3} \\ 
\rho _{1}\boldsymbol{p}_{1}+\rho _{2}\boldsymbol{p}_{2}%
\end{array}
\right], \ \ \boldsymbol{x}_{\bar{c}\bar{o}} = \left[ 
\begin{array}{c}
\rho _{2}\boldsymbol{q}_{1}-\rho _{1}\boldsymbol{q}_{2} \\ 
\rho _{2}\boldsymbol{p}_{1}-\rho _{1}\boldsymbol{p}_{2}%
\end{array}
\right].
\end{equation*}
%\begin{equation*}
%\left[
%\begin{array}{c}
%\boldsymbol{x}_{co} \\ \hline
%\boldsymbol{x}_{\bar{c}\bar{o}}
%\end{array}
%\right] =\left[
%\begin{array}{c}
%\boldsymbol{q}_{3} \\
%\rho _{1}\boldsymbol{q}_{1}+\rho _{2}\boldsymbol{q}_{2} \\
%\boldsymbol{p}_{3} \\
%\rho _{1}\boldsymbol{p}_{1}+\rho _{2}\boldsymbol{p}_{2} \\ \hline
%\rho _{2}\boldsymbol{q}_{1}-\rho _{1}\boldsymbol{q}_{2} \\
%\rho _{2}\boldsymbol{p}_{1}-\rho _{1}\boldsymbol{p}_{2}
%\end{array}
%\right].
%\end{equation*}
Also, Eqs. (\ref{real_Kalman_ss_2})-(\ref{real_Kalman_io_2}) take the form 
\begin{eqnarray*}
\boldsymbol{\dot{x}}_{co}(t) &=&A_{co}\boldsymbol{x}_{co}(t)- \left[ 
\begin{array}{cc}
\sqrt{\kappa} & 0 \\ 
0 & 0 \\ 
0 & \sqrt{\kappa} \\ 
0 & 0%
\end{array}
\right] \left[ 
\begin{array}{c}
\boldsymbol{q}_{\mathrm{in}}(t) \\ 
\boldsymbol{p}_{\mathrm{in}}(t)%
\end{array}
\right] , \\
\boldsymbol{\dot{x}}_{\bar{c}\bar{o}}(t) &=& \left[ 
\begin{array}{cr}
0 & -\omega _{m} \\ 
\omega _{m} & 0%
\end{array}
\right] \boldsymbol{x}_{\bar{c}\bar{o}}(t), \\
\left[ 
\begin{array}{c}
\boldsymbol{q}_{\mathrm{out}}(t) \\ 
\boldsymbol{p}_{\mathrm{out}}(t)%
\end{array}
\right] &=& \left[ 
\begin{array}{cccc}
\sqrt{\kappa} & 0 & 0 & 0 \\ 
0 & 0 & \sqrt{\kappa} & 0%
\end{array}
\right] \boldsymbol{x}_{co}(t) + \left[ 
\begin{array}{c}
\boldsymbol{q}_{\mathrm{in}}(t) \\ 
\boldsymbol{p}_{\mathrm{in}}(t)%
\end{array}
\right] ,
\end{eqnarray*}
where 
\begin{eqnarray*}
A_{co} &=& \left[ 
\begin{array}{cccc}
-\kappa & 0 & \omega _{m} & -\frac{\lambda}{2} \\ 
0 & 0 & -\frac{\lambda}{2} & -\omega _{m} \\ 
-\omega _{m} & -\frac{\lambda}{2} & -\kappa & 0 \\ 
-\frac{\lambda}{2} & \omega _{m} & 0 & 0%
\end{array}
\right] .
\end{eqnarray*}
Clearly, $\boldsymbol{x}_{\bar{c}\bar{o}}$ is a DF mode. Indeed, it is
exactly the same as that in Eq. (\ref{aug17_1}) for the red-detuned regime
case.

\textit{Case 3: Phase-shift regime} In this case, the two cavity modes are resonant with their respective driving lasers). Moreover, $\Delta _{1}=\Delta
_{2}=0 $ (see, e.g., \cite[Eq. (33)]{AKM14}). 
%The unitary and blockwise Bogoliubov transformation matrix $T$ in Eq. (\ref%
%{T}) is given by 
%\begin{eqnarray*}
%T&=& \left[ 
%\begin{array}{c|c|c}
%T_{h} & T_{co} & T_{\bar{c}\bar{o}}%
%\end{array}
%\right] \\
%&=& \left[ 
%\begin{array}{cc|cc|cc}
%\rho_{1} & 0 & 0 & 0 & \rho_{2} & 0 \\ 
%\rho_{2} & 0 & 0 & 0 & - \rho_{1} & 0 \\ 
%0 & 0 & 1 & 0 & 0 & 0 \\ 
%0 & \rho_{1} & 0 & 0 & 0 & \rho_{2} \\ 
%0 & \rho_{2} & 0 & 0 & 0 & -\rho_{1} \\ 
%0 & 0 & 0 & 1 & 0 & 0%
%\end{array}
%\right],
%\end{eqnarray*}
%where $\rho_{1}$ and $\rho_{2}$ are defined as in (\ref{rho12}). Thus by Eq.
%(\ref{aug31_3}), 
By Theorem \ref{thm:general_Kalman_4},
\begin{equation*}
\left[ 
\begin{array}{c}
\boldsymbol{q}_{h}(t) \\ 
\boldsymbol{p}_{h}(t) \\ \hline
\boldsymbol{x}_{co}(t) \\ \hline
\boldsymbol{x}_{\bar{c}\bar{o}}(t)%
\end{array}
\right] =\left[ 
\begin{array}{c}
-\rho _{1}\boldsymbol{p}_{1}-\rho _{2}\boldsymbol{p}_{2} \\ 
\rho _{1}\boldsymbol{q}_{1}+\rho _{2}\boldsymbol{q}_{2} \\ \hline
\boldsymbol{q}_{3} \\ 
\boldsymbol{p}_{3} \\ \hline
\rho _{2}\boldsymbol{q}_{1}-\rho _{1}\boldsymbol{q}_{2} \\ 
\rho _{2}\boldsymbol{p}_{1}-\rho _{1}\boldsymbol{p}_{2}%
\end{array}
\right].
\end{equation*}
Then, Eqs. (\ref{real_Kalman_ss_2})-(\ref{real_Kalman_io_2}) take the form 
\begin{eqnarray*}
\left[ 
\begin{array}{c}
\boldsymbol{\dot{q}}_{h}(t) \\ 
\boldsymbol{\dot{p}}_{h}(t)%
\end{array}
\right] &=& \left[ 
\begin{array}{cc}
\lambda & 0 \\ 
0 & 0%
\end{array}
\right] \boldsymbol{x}_{co}(t), \\
\boldsymbol{\dot{x}}_{co}(t) &=& \left[ 
\begin{array}{cc}
-\kappa/2 & \omega _{m} \\ 
-\omega _{m} & -\kappa/2%
\end{array}
\right] \boldsymbol{x}_{co}(t) \\
&& -\lambda \left[ 
\begin{array}{c}
0 \\ 
\boldsymbol{p}_{h}(t)%
\end{array}
\right] -\sqrt{\kappa }\left[ 
\begin{array}{c}
\boldsymbol{q}_{\mathrm{in}}(t) \\ 
\boldsymbol{p}_{\mathrm{in}}(t)%
\end{array}
\right] , \\
\boldsymbol{\dot{x}}_{\bar{c}\bar{o}}(t) &=&0, 
\\
\left[ 
\begin{array}{c}
\boldsymbol{q}_{\mathrm{out}}(t) \\ 
\boldsymbol{p}_{\mathrm{out}}(t)%
\end{array}
\right] &=&\sqrt{\kappa }\boldsymbol{x}_{co}(t)+ \left[ 
\begin{array}{c}
\boldsymbol{q}_{\mathrm{in}}(t) \\ 
\boldsymbol{p}_{\mathrm{in}}(t)%
\end{array}
\right] .
\end{eqnarray*}
Clearly, $\boldsymbol{x}_{\bar{c}\bar{o}}(t)$ is a DF mode (which is the
same as Cases 1 and 2 above). On the other hand,  $\boldsymbol{p} _{h}(t)$ is a constant for all $t\geq 0$, thus is a QND
variable. Actually,  $\boldsymbol{p} _{h}$ could be measured continuously with no quantum limit on the predictability of these measurements as the measurement back-action only drives its conjugate operator $\boldsymbol{q} _{h}$. 
\end{example}

\begin{figure}[tbph]
\centering
\includegraphics[width=0.33\textwidth]{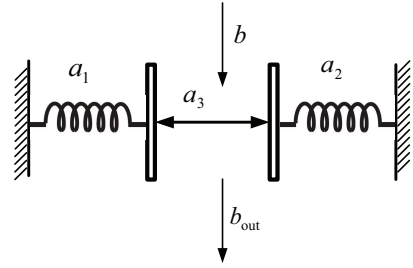}
\caption{Schematic diagram of an opto-mechanical system studied in 
\protect\cite{ODP+16} and \protect\cite{WC13}.}
\label{fig:prl}
\end{figure}

%The rough idea of BAE measurement is: When an observable $X$ is being
%continuously measured, the measurement affects its conjugate observable $P$,
%but which is decoupled from $X$. As a result, $[X(t),X(r)]=0$ for all time
%constants $t,r$. That is, $X$ is a QND variable, see Definition \ref{def:QND}.

\begin{example}
\label{ex:woolley} The opto-mechanical system, as shown in Fig. \ref{fig:prl}%
, has been studied theoretically in \cite{WC13}, and implemented
experimentally in \cite{ODP+16}. Back-action evading measurements of
collective quadratures of the two mechanical oscillators were demonstrated
in this system. 
%(The basic idea of BAE measurement is: When we measure an observable, say $X$, the measurement affects its conjugate observable, say $P$, but which is decoupled from $X$. Therefore, BAE measurement is closely related to QND variables.)
Here, the two mechanical oscillators with modes $\boldsymbol{a}_{1}$ and $%
\boldsymbol{a}_{2}$, are not directly coupled. Instead, they are coupled to
a microwave cavity, with mode $\boldsymbol{a}_{3}$. In this system, the
mechanical damping is much smaller than the optical damping. (In the
experimental paper \cite{ODP+16}, the mechanical damping is around $10^{-6}$
times that of the optical damping.) Therefore, in what follows we neglect
the mechanical damping. The system Hamiltonian (\cite[Eq. (1)]{ODP+16}, \cite%
[Eq. (A6)]{WC13}) is the following: 
\begin{eqnarray*}
\boldsymbol{H} &=&\Omega (\boldsymbol{a}_{1}^{\ast }\boldsymbol{a}_{1}-%
\boldsymbol{a}_{2}^{\ast }\boldsymbol{a}_{2})+g(\boldsymbol{a}_{1}+%
\boldsymbol{a}_{1}^{\ast })(\boldsymbol{a}_{3}+\boldsymbol{a}_{3}^{\ast }) 
\notag \\
&&+g(\boldsymbol{a}_{2}+\boldsymbol{a}_{2}^{\ast })(\boldsymbol{a}_{3}+%
\boldsymbol{a}_{3}^{\ast }).
\end{eqnarray*}
The $g$ used here is $G$ in \cite[Eq. (1)]{ODP+16} and equals $g_{a}\bar{c}%
=g_{b}\bar{c}$ in \cite[Eq. (A6)]{WC13}).
The optical coupling is $\boldsymbol{L}=\sqrt{\kappa }\boldsymbol{a}_{3}$.
% The unitary and blockwise Bogoliubov transformation
%matrix of (\ref{T}) is given by 
%\begin{eqnarray*}
%T &=& \left[%
%\begin{array}{c|c}
%T_{h} & T_{co}%
%\end{array}%
%\right] \\
%&=& \left[%
%\begin{array}{cccc|cc}
%-1/\sqrt{2} & 1/\sqrt{2} & 0 & 0 & 0 & 0 \\ 
%1/\sqrt{2} & 1/\sqrt{2} & 0 & 0 & 0 & 0 \\ 
%0 & 0 & 0 & 0 & 1 & 0 \\ 
%0 & 0 & -1/\sqrt{2} & 1/\sqrt{2} & 0 & 0 \\ 
%0 & 0 & 1/\sqrt{2} & 1/\sqrt{2} & 0 & 0 \\ 
%0 & 0 & 0 & 0 & 0 & 1%
%\end{array}%
%\right],
%\end{eqnarray*}
%and thus by Eq. (\ref{aug31_3}), 
By Theorem \ref{thm:general_Kalman_4},
\begin{equation*}
\left[ 
\begin{array}{c}
\boldsymbol{q}_{h} \\ \hline
\boldsymbol{p}_{h}%
\end{array}
\right] =\frac{1}{\sqrt{2}}\left[ 
\begin{array}{c}
\boldsymbol{q}_{2}-\boldsymbol{q}_{1} \\ 
-\boldsymbol{p}_{1}-\boldsymbol{p}_{2} \\ \hline
\boldsymbol{p}_{2}-\boldsymbol{p}_{1} \\ 
\boldsymbol{q}_{1}+\boldsymbol{q}_{2}%
\end{array}
\right] , \ \ \boldsymbol{x}_{co}= \left[ 
\begin{array}{c}
\boldsymbol{q}_{3} \\ 
\boldsymbol{p}_{3}%
\end{array}
\right] .
\end{equation*}
Then, Eqs. (\ref{real_Kalman_ss_2})-(\ref{real_Kalman_io_2}) take the form 
\begin{eqnarray}
\boldsymbol{\dot{q}}_{h}(t) &=& \left[ 
\begin{array}{cr}
0 & \Omega \\ 
-\Omega & 0%
\end{array}
\right] \boldsymbol{q}_{h}(t)+ \left[ 
\begin{array}{cc}
0 & 0 \\ 
2\sqrt{2}g & 0%
\end{array}
\right] \boldsymbol{x}_{co}(t),  \notag \\
\boldsymbol{\dot{x}}_{co}(t) &=&-\frac{\kappa }{2}\boldsymbol{x}_{co}(t)- %
\left[ 
\begin{array}{cc}
0 & 0 \\ 
0 & 2\sqrt{2}g%
\end{array}
\right] \boldsymbol{p}_{h}(t)  \notag \\
&& -\sqrt{\kappa }\left[ 
\begin{array}{c}
\boldsymbol{q}_{\mathrm{in}}(t) \\ 
\boldsymbol{p}_{\mathrm{in}}(t)%
\end{array}
\right], 
\label{aug23_1} 
\\
\boldsymbol{\dot{p}}_{h}(t) &=& \left[ 
\begin{array}{cc}
0 & \Omega \\ 
-\Omega & 0%
\end{array}
\right] \boldsymbol{p}_{h}(t),  
\notag
 \\
\left[ 
\begin{array}{c}
\boldsymbol{q}_{\mathrm{out}}(t) \\ 
\boldsymbol{p}_{\mathrm{out}}(t)%
\end{array}
\right] &=&\sqrt{\kappa }\boldsymbol{x}_{co}(t)+ \left[ 
\begin{array}{c}
\boldsymbol{q}_{\mathrm{in}}(t) \\ 
\boldsymbol{p}_{\mathrm{in}}(t)%
\end{array}
\right].
\notag
\end{eqnarray}
The components of $\boldsymbol{p}_{h}$ are linear combinations of variables of the two mechanical oscillators, are immune from optical damping, and form a QMFS. Moreover, the second entry of $\boldsymbol{p}_{h}$, can be measured via a measurement on the optical cavity, and the back-action will only affect the dynamics of the mechanical quadratures in $\boldsymbol{q}_{h}$, which are conjugate to those in $\boldsymbol{p}_{h}$. 
It can be readily shown that the system
realizes a BAE measurement of $\boldsymbol{q}_{\mathrm{out}}$ with respect
to $\boldsymbol{p}_{\mathrm{in}}$, and a BAE measurement of $\boldsymbol{p}_{%
\mathrm{out}}$ with respect to $\boldsymbol{q}_{\mathrm{in}}$. Finally,
notice that $\frac{\boldsymbol{q}_{1}+\boldsymbol{q}_{2}}{\sqrt{2}}$, the
second entry of $\boldsymbol{p}_{h}$, is exactly $X_{+}$ in \cite{ODP+16}
and \cite{WC13}, which couples to the microwave cavity dynamics $\boldsymbol{%
x}_{co}$, as can be seen in Eq. (\ref{aug23_1}).
\end{example}

%%%%%%%%%%%%%%%%%%%%%%%%
%%%%%%%%%%%%%%%%%%%%%%%%
%%%%%%%%%%%%%%%%%%%%%%%%

\section{Conclusion}

\label{sec:con}

In this paper, we have studied the Kalman decomposition for linear quantum
systems. We have shown that it can always be performed with a unitary
Bogoliubov coordinate transformation in the complex annihilation-creation
operator representation. Alternatively, it can be performed with an
orthogonal symplectic coordinate transformation in the real quadrature
representation. These are the only coordinate transformations allowed by
quantum mechanics to preserve the physical realizability conditions for
linear quantum systems. Because the coordinate transformations are unitary,
they can be performed in a numerically stable way. Furthermore, the
decomposition is performed in a constructive way, as in the classical case.
We have shown that a system in the Kalman canonical form has an interesting
structure. For passive linear quantum systems, only $co$ and $\bar{c}\bar{o}$
subsystems may exist, because the uncontrollable and the unobservable
subspaces are identical; a characterization of these subspaces has also been
given. In the general case, $c\bar{o}$ and $\bar{c}o$ subsystems may be
present, but their respective system variables must be conjugates of each
other. The Kalman canonical decomposition naturally exposes the system's
decoherence-free modes, quantum-nondemolition variables,
quantum-mechanics-free-subspaces, and back-action evasion measurement, which
are important resources in quantum information science. The methodology
proposed in this paper should be helpful in the analysis and synthesis of
linear quantum control systems.

\bigskip

{\bf Acknowledgement.}  This paper was initialized from discussions at the quantum control engineering programme at the Isaac Newton Institute for Mathematical Sciences in 2014. Guofeng Zhang, Ian Petersen and John Gough are grateful to the kind support and hospitality of the Isaac Newton Institute  for Mathematical Sciences at the University of Cambridge.

%%%%%%%%%%%%%%%%%%%%%%%%%%%%%
%%%%%%%%%%%%%%%%%%%%%%%%%%%%%
%%%%%%%%%%%%%%%%%%%%%%%%%%%%%

\begin{IEEEbiography}[{\includegraphics[width=1in,height=1.25in,clip,keepaspectratio]{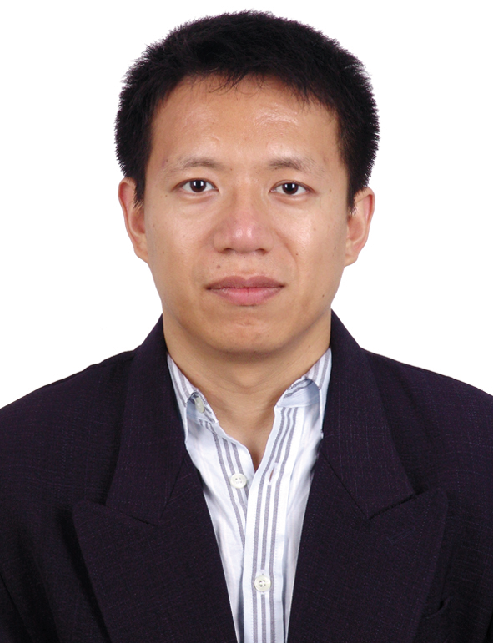}}]{Guofeng Zhang} received a Ph.D. degree in Applied Mathematics from the University of Alberta in 2005. During 2005-2006, he was a Postdoc Fellow  at the University of Windsor, Windsor, Canada. He joined  the University of Electronic Science and Technology of China in 2007. From April 2010 to December 2011 he was a Research Fellow in  the Australian National University. He joined the Hong Kong Polytechnic University in December 2011 and is currently an Assistant Professor.
\end{IEEEbiography}
\begin{IEEEbiography}[{\includegraphics[width=1in,height=1.25in,clip,keepaspectratio]{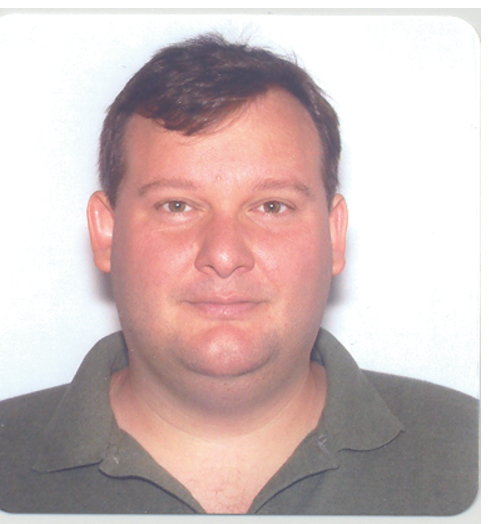}}]{Symeon Grivopoulos} received his Ph.D degree in Mechanical Engineering from the University of California at Santa Barbara in 2005. During 2006-2009, he was a Postdoctoral Researcher in the University of California at Santa Barbara. From 2014 to 2016, he was a Research Associate in the University of New South Wales at Canberra. 
\end{IEEEbiography}

\begin{IEEEbiography}[{\includegraphics[width=1in,height=1.25in,clip,keepaspectratio]{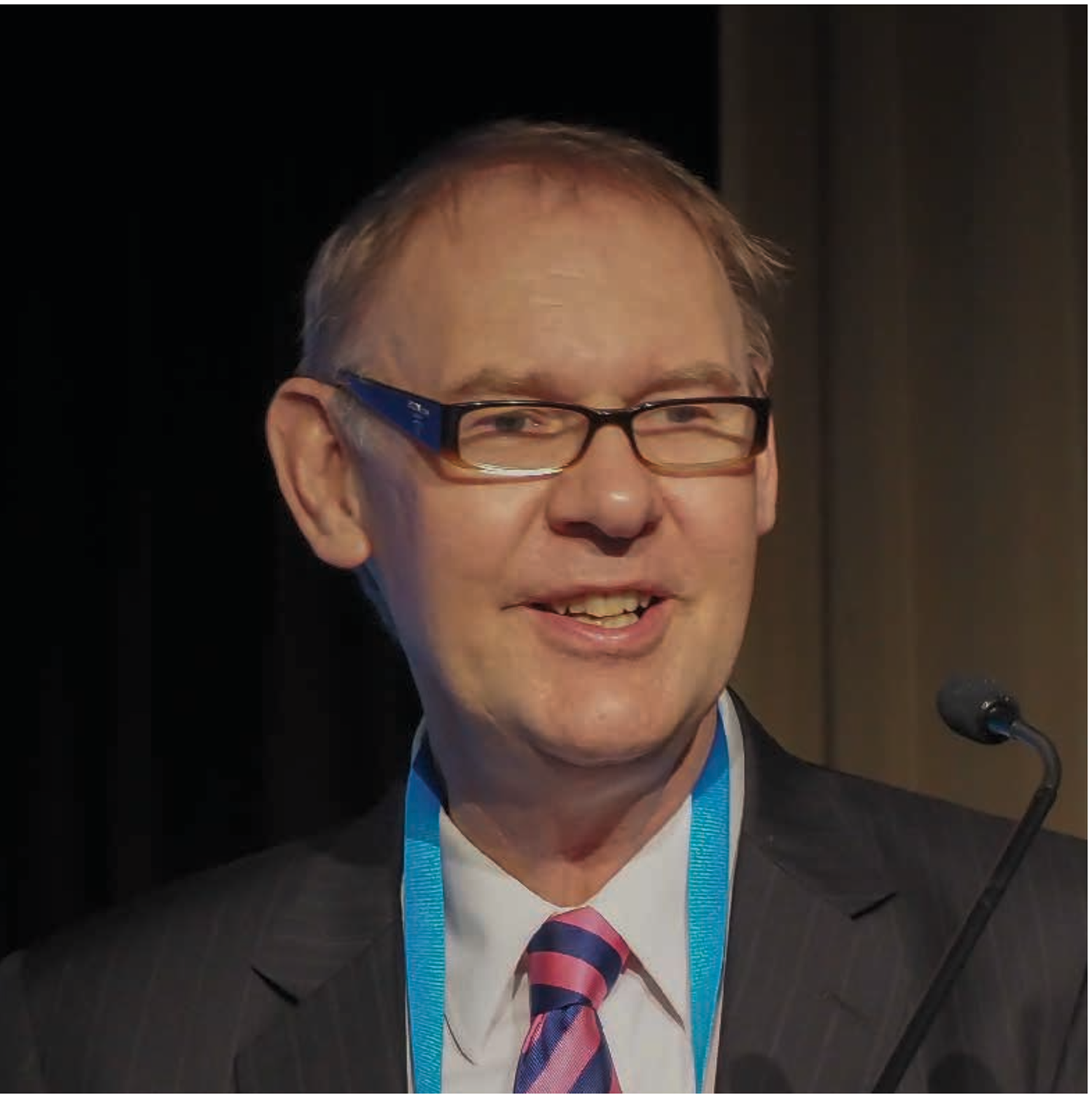}}]{Ian R. Petersen} received a Ph.D in Electrical Engineering in 1984 from the University of Rochester, USA. From 1983 to 1985 he was a Postdoctoral Fellow at the Australian National University. From 1985 to 2016 he was with the University of New South Wales Canberra. He is currently a Professor in the Research School of Engineering at the Australian National University.  

\end{IEEEbiography}

\begin{IEEEbiography}[{\includegraphics[width=1in,height=1.25in,clip,keepaspectratio]{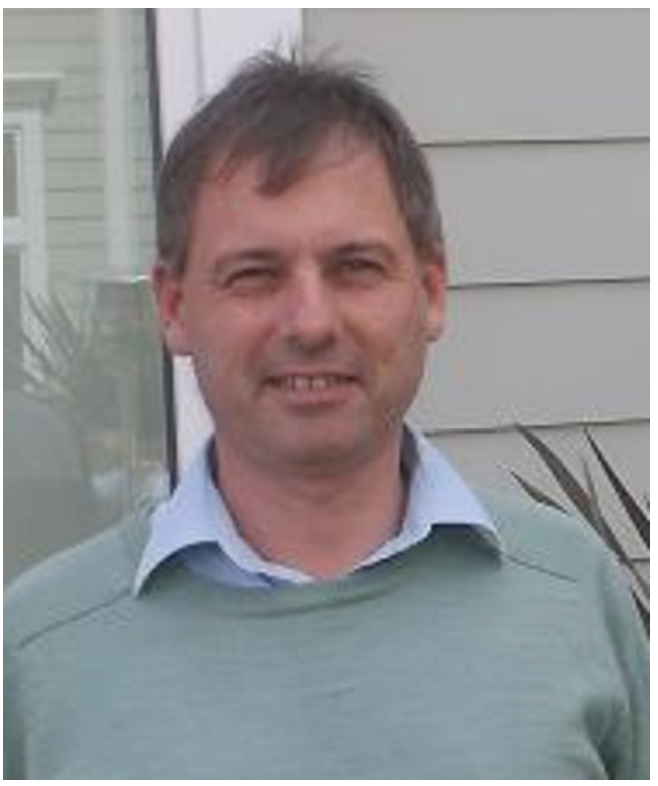}}]{John E. Gough}  received the Ph.D. degree in Mathematical Physics from the National University of Ireland, Dublin in 1992. He was reader in Mathematical Physics at  Nottingham- Trent University, up until 2007 when he joined the Institute of Mathematics and Physics at Aberystwyth University as established chair of Mathematics. 
\end{IEEEbiography}

%%%%%%%%%%%%%%%%%%
%%%%%%%%%%%%%%%%%%
%%%%%%%%%%%%%%%%%%

%%%%%%%%%%%%%%%%%%%%%%%%%%%%%%%%%%%%


\begin{thebibliography}{99}
\bibitem{AT12} C. Altafini and F. Ticozzi, ``Modeling and control of quantum
systems: an introduction,'' \textit{IEEE Trans. Automat. Contr.} vol. 57,
pp. 1898-1917, 2012.

\bibitem{AKM14} M. Aspelmeyer, T. J. Kippenberg, and F. Marquardt, ``Cavity
optomechanics,'' Rev. Mod. Phys., vol. 86, pp. 1391-1452, 2014.

%\bibitem{AV} B. D. O. Anderson and S. Vongpanitlerd, \textit{Network
%Analysis and Synthesis: A Modern Systems Theory Approach.} Prentice-Hall, Inc., Englewood Cliffs, NJ, 1973.

\bibitem{CF} M. J. Corless and A. Frazho, \textit{Linear Systems and
Control: an Operator Perspective}. Marcel Dekker, Inc., 2003. %?????

%\bibitem{DV09} C. A. Desoer and M. Vidyasagar, \textit{Feedback Systems:
%Input-Output Properties,} Society for Industrial and Applied Mathematics,
%Philadelphia, 2009.

\bibitem{DJ99} A. Doherty and K. Jacob, ``Feedback-control of quantum
systems using continuous state-estimation,'' \textit{Phys. Rev. A}, vol. 60,
pp. 2700-2711, 1999.

\bibitem{DP10} D. Dong and I. R. Petersen, ``Quantum control theory and
applications: a survey,'' \textit{IET Control Theory Appl.,} vol. 4, pp.
2651-2671, 2010.

\bibitem{DFK+12} C. Dong, V. Fiore, M. C. Kuzyk, and H. Wang,
``Optomechanical dark mode,'' \textit{Science}, 338(6114), pp. 1609-1613,
2012.


\bibitem{DLCZ01}
L.-M. Duan, M.~D. Lukin, J.~I. Cirac, and P.~Zoller.
\newblock Long-distance quantum communication with atomic ensembles and linear
  optics.
\newblock {\em Nature}, 414:413--418, 2001.


\bibitem{DEMPUJ16} L. A. Duffaut Espinosa, Z. Miao, I. R. Petersen, V.
Ugrinovskii, and M. R. James, "Physical realizability and preservation of
commutation and anticommutation relations for $n$-level quantum systems", 
\textit{SIAM J. Control Optim.,} vo. 54(2), 632-– 2016.

\bibitem{GZ00} C.W. Gardiner and P. Zoller, \emph{Quantum Noise.} \newblock
Springer, 2004.

\bibitem{GJ09} J. E. Gough and M. R. James, ``The series product and its
application to quantum feedforward and feedback networks,'' \emph{IEEE
Trans. Automat. Control}, vol. 54, no.11, pp. 2530-2544, 2009.

\bibitem{GJN10} J. E. Gough, M. R. James, and H. I. Nurdin, ``Squeezing
components in linear quantum feedback networks,'' \textit{Phys. Rev. A},
vol. 81, 023804, 2010.

\bibitem{GZ15} J. E. Gough and G. Zhang, ``On realization theory of quantum
linear systems'', \textit{Automatica}, vol. 59, pp. 139-151, 2015.

\bibitem{GY13} M. Guta and N. Yamamoto, ``Systems identification for passive
linear quantum systems,''  \emph{IEEE
Trans. Automat. Control}, vol. 61, no. 4, pp. 921-936, 2016.

\bibitem{HM12} R. Hamerly and H. Mabuchi, ``Advantages of coherent feedback
for cooling quantum oscillators,'' \textit{Phys. Rev. Lett.}, 109, 173602,
2012. % opto-mechanical

\bibitem{HCH+13} M. R. Hush, A. R. R. Carvalho, M. Hedges, and M. R. James,
``Analysis of the operation of gradient echo memories using a quantum
input-output model'', \textit{New Journal of Physics}, vol. 15, 085020, 2013.

%\bibitem{IYY+12} S. Iida, M. Yukawa, H. Yonezawa, N. Yamamoto, and A.
%Furusawa, ``Experimental demonstration of coherent feedback control on
%optical field squeezing'', \textit{IEEE Transactions on Automatic Control},
%vol. 57, no. 8, pp. 2045-2050, 2012. %????? experiment

\bibitem{KJ14} K. Jacobs, \textit{Quantum Measurement Theory and Its
Applications}. Cambridge University Press, Cambridge, New York, 2014.

\bibitem{JNP08} M. R. James and H. I. Nurdin and I. R. Petersen, ``$H^\infty$
control of linear quantum stochastic systems,'' 
\newblock  {\em IEEE Trans.
Automat. Control}, vol. 53, pp. 1787-1803, 2008.

\bibitem{JG10} M. R. James and J. E. Gough, \newblock ``Quantum dissipative
systems and feedback control design by interconnection,'' 
\newblock {\em
IEEE Trans. Automat. Contr.}, vol. 55, no. 8, pp. 1806-1821, 2010.

\bibitem{KAK13} J. Kerckhoff, R. W. Andrews, H. S. Ku, W. F. Kindel, K.
Cicak, R. W. Simmonds, and K. W. Lehnert, ``Tunable coupling to a mechanical
oscillator circuit using a coherent feedback network,'' \textit{Phys. Rev. X}%
, vol. 3, 021013, 2013. %????? experiment

\bibitem{HK97} H. Kimura, \textit{Chain-scattering Approach to $H^\infty$%
-Control}. Birkhauser, 1997.

\bibitem{Leo03} U. Leonhardt, ``Quantum physics of simple optical
instruments,'' \textit{Rep. Prog. Phys.}, vol. 66, pp. 1207-1249, 2003.

%\bibitem{MP11} A. Maalouf and I. R. Petersen, ``Coherent $H_{\infty} $
%control for a class of linear complex quantum systems,'' \newblock  \emph{%
%IEEE Trans. Automat. Contr.,} vol. 56, no. 2, pp. 309-319, 2011.

\bibitem{MP11a} A. Maalouf and I. R. Petersen, ``Bounded real properties for
a class of linear complex quantum systems,'' \emph{IEEE Trans. Automat.
Contr.,} vol. 56, no. 4, pp. 786-801, 2011.

\bibitem{M08} H. Mabuchi, ``Coherent-feedback control with a dynamic
compensator'', \textit{Phys. Rev. A}, vol. 78, 032323, 2008 
%????? experiment
%
%\bibitem{CTS+13} O. Crisafulli, N. Tezak, D. B. S. Soh, M. A. Armen, and H.
%Mabuchi, ``Squeezed light in an optical parametric oscillator network with
%coherent feedback quantum control,'' \textit{Opt. Express}, vol. 21, 18371,
%2013. %????? experiment

\bibitem{MHP+11} F. Massel, T. T. Heikkila, J. -M. Pirkkalainen, S. U. Cho,
H. Saloniemi, P. J. Hakonen, and M. A. Sillanpaa, ``Microwave amplification
with nanomechanical resonators,'' \textit{Nature}, vol. 480, pp. 351-354,
2011.

\bibitem{MCP+12} F. Massel, S. U. Cho, J.-M. Pirkkalainen, P. J. Hakonen, T.
T. Heikkila, M. A. Sillanpaa, ``Multimode circuit optomechanics near the
quantum limit,'' \textit{Nature Communications}, vol. 3, 987, 2012.

\bibitem{MJP+11} A. Matyas, C. Jirauschek, F. Peretti, P. Lugli, and G.
Csaba, ``Linear circuit models for on-chip quantum electrodynamics,'' 
\textit{IEEE Trans. Microwave Theory and Techniques}, vol. 59, pp. 65-71,
2011.

\bibitem{MvLGWRN06}
Nicolas~C. Menicucci, Peter van Loock, Mile Gu, Christian Weedbrook, Timothy~C.
  Ralph, and Michael~A. Nielsen.
\newblock Universal quantum computation with continuous-variable cluster
  states.
\newblock {\em Phys. Rev. Lett.}, 97:110501, Sep 2006.


\bibitem{NJP09} H. I. Nurdin, M. R. James, and I. R. Petersen, ``Coherent
quantum LQG control,'' \emph{Automatica}, vol. 45, pp. 1837-1846, 2009.

\bibitem{NJD09} H. I. Nurdin, M. R. James, and A. Doherty, ``Network
synthesis of linear dynamical quantum stochastic systems,'' \textit{SIAM J.
Contr. and Optim.}, vol. 48, pp. 2686-2718, 2009.

%\bibitem{HIN10a} H. I. Nurdin, ``Synthesis of linear quantum stochastic
%systems via quantum feedback networks'', \textit{IEEE Trans. Automat. Contr.}%
%, vol. 55, pp. 1008-1013, 2010.

%\bibitem{HIN10b} H. I. Nurdin, ``On synthesis of linear quantum stochastic
%systems by pure cascading,'' \textit{IEEE Trans. Automat. Contr.}, vol. 55,
%pp. 2439-2444, 2010. %?????

\bibitem{HIN13} H. I. Nurdin, ``Structures and transformations for model
reduction of linear quantum stochastic systems,'' \textit{IEEE Trans.
Automat. Contr.}, vol. 59, pp. 2413-2425, 2014.

%\bibitem{NGP15} H. I. Nurdin, S. Grivopoulos, and I. R. Petersen, ``The
%Transfer Function of Generic Linear Quantum Stochastic Systems Has a Pure
%Cascade Realization,'' arXiv:1509.05537v1 [quant-ph], 2015.

\bibitem{ODP+16} C. F. Ockeloen-Korppi, E. Damskagg, J.-M. Pirkkalainen, A.
A. Clerk, M. J. Woolley, M. A. Sillanpaa, ``Quantum back-action evading
measurement of collective mechanical modes,''  {\it Phys. Rev. Lett.}, vol. 117, 140401, 2016.

\bibitem{KRP92} K. R. Parthasarathy, \emph{An Introduction to Quantum
Stochastic Calculus}. Berlin, Germany: Birkhauser, 1992.

\bibitem{IRP11} I. R. Petersen, ``Cascade cavity realization for a class of
complex transfer functions arising in coherent quantum feedback control,'' 
\emph{Automatica}, vol. 47, no. 8, pp. 1757-1763, 2011.

%\bibitem{IRP13} I. R. Petersen, ``Singular perturbation approximations
%for a class of linear quantum systemsÕÕ, \textit{IEEE Trans. Automat. Contr.},
%vol. 58, pp. 193-198, 2013. %?????

\bibitem{PR14} P. Rouchon, ``Models and feedback stabilization of open
quantum systems,'' arXiv:1407.7810v3, 2015.

\bibitem{SW10} S. G. Schirmer and X. Wang, ``Stabilizing open quantum
systems by Markovian reservoir engineering,'' \textit{Phys. Rev. A}, vol.
81, 062306, 2010.

\bibitem{SP12} A. J. Shaiju and I. R. Petersen, ``A frequency domain
condition for the physical realizability of linear quantum systems'', \emph{%
IEEE Trans. Automat. Contr.}, vol. 57, pp. 2033-2044, 2012.

\bibitem{SvHM04} J. K. Stockton, R. van Handel, and H. Mabuchi,
``Deterministic dicke state preparation with continuous measurement and
control'', \textit{Physical Review A}, vol. 70, 022106, 2004.

\bibitem{TNP+11} N. Tezak, A. Niederberger, D. S. Pavlichin, G. Sarma, and
H. Mabuchi, ``Specification of photonic circuits using quantum hardware
description language,'' \textit{Philosophical Transactions of the Royal
Society A: Mathematical, Physical \& Engineering Sciences}, vol. 370, pp.
5270-5290, 2012.

\bibitem{T12} L. Tian, ``Adiabatic state conversion and pulse transmission
in optomechanical systems,'' \newblock {\it Phys. Rev. Lett.}, vol. 108,
153604, 2012.

\bibitem{TC10} M. Tsang and C. M. Caves, ``Coherent quantum-noise
cancellation for optomechanical sensors,'' \textit{Phys. Rev. Lett.}, vol.
105, 123601, 2010.

\bibitem{TC12} M. Tsang and C. M. Caves, ``Evading quantum mechanics:
engineering a classical subsystem within a quantum environment,'' {\it Phys. Rev.
Lett. } vol. 2, 031016, 2012.

\bibitem{WC12a} Y. Wang and A. A.  Clerk, ``Using Interference for High Fidelity Quantum State Transfer in Optomechanics,'' {\it Phys. Rev. Lett. }, vol. 108, 153603, 2012.

\bibitem{WC12b} Y. Wang and A. A.  Clerk, ``Using dark modes for high-fidelity optomechanical quantum state transfer,'' \textit{New J. Physics}, vol. 14, 105010, 2012.



\bibitem{VD81} P. M. Van Dooren, "The generalized eigenstructure problem in
linear system theory", \textit{IEEE Trans. Automat. Contr.,} vol. 26, no. 1,
1981.

\bibitem{WM08} D. F. Walls and G. J. Milburn, \emph{Quantum Optics}.
Springer, Berlin, 2008.

\bibitem{HMW95} H. M. Wiseman, ``Using feedback to eliminate back-action in
quantum measurements,'' \textit{Phys. Rev. A} vol. 51, pp. 2459-2468, 1995.

\bibitem{WD05} H. M. Wiseman and A. C. Doherty, ``Optimal unravellings for
feedback control in linear quantum systems,'' \textit{Phys. Rev. Lett.},
vol. 94, 070405, 2005.

\bibitem{WM10} H. W. Wiseman and G. J. Milburn, \emph{Quantum Measurement
and Control}. Cambridge University Press, Cambridge, UK, 2010.

\bibitem{WC13} M. J. Woolley and A. A. Clerk, ``Two-mode back-action-evading
measurements in cavity optomechanics,'' \textit{Phys. Rev. A,} vol. 87, no.
6, 063846, 2013.

\bibitem{NY13} N. Yamamoto, ``Decoherence-free linear quantum systems,'' 
\textit{IEEE Trans. Automat. Contr.}, vol. 59, pp.1845-1857, 2014.

\bibitem{NY14} N. Yamamoto, ``Coherent versus measurement feedback: Linear
systems theory for quantum information,'' \textit{Phys. Rev. X}, vol. 4,
041029, 2014.

\bibitem{YJ14} N. Yamamoto and M. R. James, ``Zero-dynamics principle for
perfect quantum memory in linear networks,'' \textit{New Journal of Physics,}
vol.16(7), 073032, 2014.

\bibitem{YK03a} M. Yanagisawa and H. Kimura, ``Transfer function approach to
quantum control-part I: Dynamics of quantum feedback systems, \emph{IEEE
Trans. Automat. Contr.}, vol. 48, pp. 2107-2120, 2003.

%\bibitem{YKS+12} H. Yuan, C. P Koch, P. Salamon, D. J Tannor,
%``Controllability on relaxation-free subspaces: On the relationship between
%adiabatic population transfer and optimal control,'' \textit{Physical Review
%A}, vol. 85, 033417, 2012.

\bibitem{Z14} G. Zhang, ``Analysis of quantum linear systems' response to
multi-photon states'', \textit{Automatica}, vol. 50, pp. 442-451, 2014,

%\bibitem{ZCB+10} K. Zhang, W. Chen, M. Bhattacharya, and P. Meystre,
%``Hamiltonian chaos in a coupled BEC-optomechanical cavity system,'' \textit{%
%Phys. Rev. A}, vol. 81, 013802, 2010.

\bibitem{ZLW+14} J. Zhang, Y. X. Liu, R.-B. Wu, K. Jacobs, and F. Nori,
``Quantum feedback: theory, experiments, and applications,''
ArXiv:1407.8536, 2014.

\bibitem{ZJ11} G. Zhang and M. R. James, ``Direct and indirect couplings in
coherent feedback control of linear quantum systems,'' \emph{IEEE Trans.
Automat. Contr.}, vol. 56, no. 7, pp. 1535-1550, 2011.

\bibitem{ZJ12} G. Zhang and M.R. James, ``Quantum feedback networks and
control: a brief survey,'' \textit{Chinese Science Bulletin}, vol. 57, no.
18, pp. 2200-2214, 2012, (arXiv:1201.6020v3 [quant-ph]).

\bibitem{ZLHZ12} G. Zhang, H. W. J. Lee, B. Huang, and H. Zhang, ``Coherent
feedback control of linear quantum optical systems via squeezing and phase
shift'', \textit{SIAM Journal on Control and Optimization}, vol. 50, pp.
2130-2150, 2012.

\bibitem{ZJ13} G. Zhang and M. R. James, ``On the response of quantum linear
systems to single photon input fields'', \textit{IEEE Trans. Automat. Contr.,%
} vol. 58, pp. 1221-1235, 2013.

%\bibitem{Z14} G. Zhang, ``Analysis of quantum linear systems' response to
%multi-photon states'', \textit{Automatica}, vol. 50, pp. 442-451, 2014,
%arXiv:1311.0357 [quant-ph]. %?????

\bibitem{ZZZ+15} X. Zhang, C.-L. Zou, N. Zhu, F. Marquardt, L. Jiang, and H.
X. Tang, ``Magnon dark modes and gradient memory,'' \textit{Nature
Communications,} vol. 6, 8914, 2015.

\bibitem{ZDG96} K. Zhou, J. Doyle, and K. Glover, \textit{Robust and Optimal
Control.} Prentice-Hall, Upper Saddle River, NJ, 1996.

%\bibitem{MZH91}
%P. Marte, P. Zoller, and J. L. Hall. Coherent atomic mirrors and beam
%splitters by adiabatic passage in multilevel systems. Phys. Rev. A,
%44:R4118ÐR4121, Oct 1991.

%
%\bibitem{SW10}
%S. G. Schirmer and X. Wang, ``Stabilizing open quantum systems by
%Markovian reservoir engineering,''  {\it Phys. Rev. A,} 81, 062306, 2010.

%\bibitem{Won85} W. M. Wonham, \textit{Linear Multivariable Control: A
%Geometric Approach,} third edition, Springer-Verlag, New York, 1985.
\end{thebibliography}
\end{document}